\documentclass[11pt,times,letter]{article}
\usepackage{amsmath, amsfonts,amsthm,amssymb,bm}
\usepackage{dsfont}

\usepackage[english]{babel}

\usepackage[normalem]{ulem}

\usepackage{latexsym,amssymb,amsfonts,graphicx}
\usepackage{amsmath,dsfont}
\usepackage{verbatim}
\usepackage{mathrsfs}
\usepackage{bm}
\usepackage{color}
\usepackage{epsfig}

\usepackage{url}
\usepackage{bm}
\usepackage{natbib}

\usepackage[colorlinks,linkcolor=blue,citecolor=blue,anchorcolor=blue]{hyperref}

\newcommand{\Px}{ \mathbb{P} }
\newcommand{\Qx}{ \mathbb{Q} }
\newcommand{\N}{ \mathds{N} }
\newcommand{\Ex}{ \mathbb{E} }

\def\esssup_#1{\underset{#1}{\Xi}}
\def\essinf_#1{\underset{#1}{\mathrm{ess\,inf\, }}}
\def\argmax_#1{\underset{#1}{\mathrm{arg\,max\, }}}
\def\argmin_#1{\underset{#1}{\mathrm{arg\,min\, }}}
\def\argXi_#1{\underset{#1}{\mathrm{arg\,\Xi\, }}}

\newcommand{\Hx}{\mathbb{H}}
\newcommand{\Gx}{\mathbb{G}}
\newcommand{\Fx}{\mathbb{F} }

\newcommand{\D}{\mathrm{d}}

\newcommand{\G}{\mathcal{G}}
\newcommand{\R}{\mathds{R}}

\newcommand{\Gam}{\mathnormal{\Gamma}}

\newtheorem{theorem}{Theorem}[section]
\newtheorem{definition}{Definition}[section]

\newtheorem{proposition}[theorem]{Proposition}
\newtheorem{remark}[theorem]{Remark}
\newtheorem{lemma}[theorem]{Lemma}

\newtheorem{example}{Example}[section]
\allowdisplaybreaks[4]

\definecolor{Red}{rgb}{1.00, 0.00, 0.00}

\definecolor{DRed}{rgb}{0.5, 0.00, 0.00}

\definecolor{Blue}{rgb}{0.00, 0.00, 1.00}

\definecolor{Green}{rgb}{0.0, 0.4, 0.0}

\title
{
Portfolio Choice with Market-Credit Risk Dependencies
}

\author{
Lijun Bo \thanks{Email: lijunbo@ustc.edu.cn, School of Mathematics and Statistics, Xidian University, Xi'an 710071, China and School of Mathematical Sciences, University of Science and Technology of China, Hefei 230026, China.} \and
Agostino Capponi \thanks{E-mail: ac3827@columbia.edu, Department of Industrial Engineering and Operations Research, Columbia University, New York, 10027, NY, USA. }}

\begin{document}
\maketitle
\begin{abstract}
We study an optimal investment/consumption problem in a model capturing market and credit risk dependencies. Stochastic factors drive both the default intensity and the volatility of the stocks in the portfolio. We use the martingale approach and analyze the recursive system of nonlinear Hamilton-Jacobi-Bellman equations associated with the dual problem. We transform such a system into an equivalent system of semi-linear PDEs, for which we establish existence and uniqueness of a bounded {global} classical solution. We obtain explicit representations for the optimal strategy, consumption path and wealth process, in terms of the solution to the recursive system of semi-linear PDEs. We numerically analyze the sensitivity of the optimal investment strategies to risk aversion, default risk and volatility.

\vspace{0.3 cm}

{\noindent{\textbf{AMS 2000 subject classifications.} 91G10, 91G40, 60J20}

\vspace{0.3 cm}

}
\end{abstract}

\section{Introduction}
Portfolio optimization problems, originating from the seminal work of \cite{Merton71}, have been the subject of considerable investigation. Important developments include the impact of trading constraints (see \cite{Cvitanic01} for a survey), the inclusion of stochastic volatility, see for instance \cite{FouqueThaleia}, and the forward utility approach (see \cite{Musiela}) to model the time changing preferences of an agent. 
The most direct extension of the log-normal assumption made by \cite{Merton71} is the stochastic factor model. Such a model is able to capture empirically observed features of price processes and has been successfully used in several contexts, including stochastic interest rates (e.g. \cite{Brennan}), {mean returns of individual stocks (e.g. \cite{BiePliska99})}, and stochastic volatility (e.g. \cite{NLHH,FouqueThaleia}). We also refer to \cite{Thaleia} for an excellent overview of the stochastic factor model. 

The objective of the present paper is to introduce a credit portfolio framework to assess the joint impact of systemic and macroeconomic factors on the optimal portfolio strategy of an investor. To the best of our knowledge, ours is the first model to incorporate in a tractable manner the impact of both stochastic volatility and systemic risk on optimal portfolio allocations. Prior work has focused on stochastic volatility models as surveyed above, direct contagion models based on interacting intensities (e.g. \cite{BoCappMF} and \cite{BoCappRC} and \cite{JY}), and default correlation through exposure to systematic factors (e.g. \cite{Callegaro}). None of the above studies take into account the joint impact of market and credit risk on optimal investment.
Empirical studies, however, suggest that the interplay of these two risks plays a critical role. It is well documented that stock return volatility is stochastic (e.g. \cite{Bollerslev,Ghysels}); moreover, there is evidence that credit spreads of a company are positively related to the equity return volatilities (e.g \cite{Campbell, Bakshi}). In the pricing space, model specifications featuring the dependence of default intensities on asset volatilities have been proposed by \cite{Bayar}, \cite{CarrWu}, \cite{CarrLinetsky} and \cite{MendozaMFa}, but only for a security underwritten by an individual entity. \cite{MendozaMF} extend the analysis to a multi-name credit-equity model. The importance of credit-market risk dependencies has also been highlighted by \cite{Basel}, which provides empirical evidence for their interaction both at the macro level, and at the micro level (sensitivity of individual bank risk to different risk factors). {A related branch of the literature has analyzed the optimal investment problem in a portfolio consisting of default risk sensitive assets. \cite{BielJang} consider a portfolio consisting a stock and a bond, but assume their price processes to be independent, thus ignoring market-credit risk dependencies. \cite{Pham10} and \cite{JiaoKP13} study optimal investment in a portfolio model consisting of multiple securities subject to default risk. They decompose the original control problem defined under the enlarged filtration, inclusive of default event information, into classical stochastic control problems under the reference filtration, using a finite backward induction procedure. \cite{IfJeanblanc} use the dual approach to solve the portfolio optimization problem in a market environment where the risk-free interest rate process can experience sudden jumps. \cite{DiNuSj13} consider optimal investment in defaultable assets when the investor has access to different sets of information. They find necessary and sufficient conditions for the existence of a portfolio which locally maximizes the expected investor's utility from terminal wealth.}

We consider a risk averse investor with power utility, who allocates his wealth across defaultable stocks and a bank account. The default intensity of a stock depends not only on its volatility, but also on common factors that influence the volatility processes of other stocks in the portfolio. These factors model the evolution of macro-economic variables which influence both the market and the credit risk of the portfolio. Moreover, the default intensity of each stock exhibit jumps when other stocks in the portfolio default. Empirically, it has been shown that for many financial sectors, e.g commercial banks, the default likelihood of an entity is likely to abruptly increase if some of its major counterparties default, see also \cite{Yu}.

There are several mathematical contributions in our efforts. {Our portfolio analysis employs the martingale approach to deal with market incompleteness, as in
\cite{KaSr98} (Chapter 5) and \cite{KrSc99}. 
Because of the default risk, we need to introduce an additional control process to establish the dual stochastic control problem. We show that the value function of the dual problem satisfies a recursive system of default-state dependent nonlinear PDEs. Using the power transform method developed by \cite{Thaleia01}, we transform the original system into a recursive system of semi-linear PDEs whose nonlinear coefficients are still non-Lipschitz continuous. We then employ a two-steps approach to establish existence and
uniqueness of a smooth solution to the system. {We first construct a system of truncated PDEs using stopping time arguments. Such a system
 falls within the class of semi-linear PDEs analyzed by \cite{BechererSchweizer}, for which existence and uniqueness of a bounded classical solution can be guaranteed. Using probabilistic representations of classical solutions to PDEs, we show the equivalence between the truncated system and the original system.
We further prove that the gradient of the solution to the recursive system is bounded, and obtain a closed-form representation
for the optimal admissible {investment} strategy, consumption path and wealth process. 
Our paper is also related to \cite{Pham02}, who studies an optimal investment problem under stochastic volatility. He discusses existence of classical solutions and provides gradient estimates for the solution to the HJB equation of the primal problem. His methodology cannot deal with the additional jump-to-default term appearing in our Hamiltonian.}
}



{We develop a numerical study for a special case of our model setup, in which the stochastic factor is constant and the portfolio consists of two stocks subject to credit risk. We find that the signs associated with the sensitivities of the investment strategies to the model parameters are in line with economic intuition. The investor reduces his holdings in the stock if (i) the default probability of the stock increases, (ii) his risk aversion increases, (iii) the planning horizon gets shorter, and (iv) the volatility parameter of the stocks' price processes increases.}

The rest of the paper is organized as follows. Section \ref{sec:modelinvestor} introduces the portfolio model and formulates the primal problem. {Section \ref{sec:dualver}
develops the dual formulation and provides a verification result. Section \ref{sec:optimal-ic} gives the optimal investment/consumption strategy. Section~\ref{sec:numerics} develops a numerical analysis. Section \ref{sec:conclusions} concludes.}

\section{The Model and Investor's Problem}\label{sec:modelinvestor}
We describe the model in Section \ref{sec:model}. We set up the primal problem of the investor in Section \ref{sec:primal}.

\subsection{The Model}\label{sec:model}

The portfolio model consists of $n\geq2$ defaultable stocks and a risk-free bank account $B_t$ with dynamics $dB_t= rB_tdt$, where $r>0$ is the constant interest rate.\footnote{Throughout the paper, we consider a constant interest rate since this is not the main focus of our analysis. Our results can be easily extended to the case of a stochastic interest rate $r(Y_t,H_t)$ as long as $r(y,z)$ is $C_b^1$ in $y\in D$ for each default state $z\in{\cal S}$.} We fix $T>0$ to be the finite target horizon and consider a complete filtered probability space $(\Omega,{\mathcal G},{\Gx},\Px)$, where $\Gx = \Fx \vee \Hx$.
Two independent $n$-dimensional standard Brownian motions $W=(W_t^i;\ i=1,\ldots,n)_{t\in[0,T]}^{\top}$ and $\bar{W}=(\bar{W}_t^i;\ i=1,\ldots,n)_{t\in[0,T]}^{\top}$ generate a filtration $\Fx:=({\mathcal{F}}_t)_{t\in[0,T]}$, where $\mathcal{F}_t=\sigma(W_s,\bar{W}_s;\ s\leq t)$.
We use $\top$ to denote the transpose operator.
The default state is described by a $n$-dimensional default indicator process $H=(H_t^i;\ i=1,\ldots,n)_{t\in[0,T]}$ with state space $\mathcal{S}=\{0,1\}^n$, where $H_t^i=1$ if the asset $i$ has defaulted by time $t$ and $H_t^i=0$ otherwise.
The default time of the $i$-th security is given by $\tau^i := \inf\{t\geq0;\ H_t^i=1\}$ for $i=1,\ldots,n$.
For $t\in[0,T]$, the sigma-algebra ${\cal H}_t:=\bigvee_{i=1}^n{\cal H}_t^i$, where ${\cal H}_t^i:=\sigma(H_s^i;\ s\leq t)$, contains information about default events of the stocks up to time $t$. The filtration $\Hx = ({\mathcal{H}}_t)_{t\in[0,T]}$ contains all information about default events till the target horizon $T$. Our model consists of three blocks: a stochastic factor, the price processes and the credit model. The stochastic factor influences not only the returns and volatility of the prices, but also the credit risk of the stocks.

{\bf Stochastic factor.} It is a reduced form model for the evolution of macroeconomic variables. Examples of these variables are interest rates, broad share price indices, and measures of economic activity or growth. Such a factor drives the drift, volatility, and the default intensities of the stock price processes. Consider a domain (open connected subset) $D \subseteq \R^m$. {The process $Y=(Y_t)_{t\in[0,T]}$ is referred to as the stochastic factor and has dynamics given by, $Y_0=y\in D$, and}
\begin{align}\label{eq:Yt}
dY_t &= \mu_0(Y_t)dt + \sigma_0(Y_t)\big[\rho dW_t+ \sqrt{1-\rho^2} d\bar{W}_t\big],
\end{align}
where the correlation coefficient $\rho\in(-1,1)$. The drift coefficient $\mu_0$ is an $\R^m$-valued column vector of functions and $\sigma_0$ is an $\R^{m\times n}$-valued matrix of
functions. For each $(t,y)\in[0,T]\times D$, let $Y_s^{t,y}$, $s\in[t,T]$, be the solution of Eq.~\eqref{eq:Yt}, with the constraint $Y_t^{t,y}=y\in D$ at time $t$.

{\bf Credit risk model.} We assume that the bivariate process $(Y,H)=(Y_t,H_t)_{t\in[0,T]}$ is Markovian with state space $D\times\mathcal{S}$. The default indicator process $H$ transits from a
state $H_t:=(H_t^1,\ldots,H_t^{i-1},H_t^i,H_t^{i+1},\ldots,H_t^n)$ in which the stock $i$ is alive ($H_t^i=0$) to the neighbouring state $\bar{H}_t^i:=(H_t^1,\ldots,H_t^{i-1},1-H_t^i,H_t^{i+1},\ldots,H_t^n)$
in which the stock $i$ has defaulted at a stochastic rate ${\bf1}_{H_t^i=0}\lambda_{i}(Y_t,H_t)={\bf1}_{H_t^i=0} \lambda_i(Y_t,(H_t^1,\ldots,H_t^{i-1},0,H_t^{i+1},\ldots,H_t^n))$.
Notice that the default intensity of the $i$-th stock may depend on the default state $H_t^j$ of other stocks $j\neq i$ in the portfolio, but is defined on the event that $H_t^i = 0$. By construction, simultaneous defaults are precluded in the model because transitions from $H_t$ can only occur to a state differing from $H_t$ in exactly one of the entries.
The intensity function $\lambda_i(y,z)$ is assumed to be strictly positive for all $z\in{\cal S}$. The default intensity of the $i$-th stock may change if (i) a stock in the
portfolio defaults (counterparty risk effect), and (ii) there are fluctuations in the macro-economic environment. Our default model thus belongs to the rich class of interacting
Markovian intensity models, introduced by \cite{FreyRung2010}. 

{\bf Price processes.} The vector of price processes of the $n$ stocks is denoted by
$\tilde{P}=(\tilde{P}_t^i;\ i=1,\ldots,n)_{t\in[0,T]}^{\top}$. For $t\in[0,T]$, the price process of the $i$-th defaultable stock is given by
\begin{equation}
\tilde{P}_t^i=(1-H_t^i)P_t^i, \qquad \; i = 1,\ldots,n.
\label{eq:pricedef}
\end{equation}
In other words, the price of the $i$-th stock is given by the {\it predefault price} $P_t^i$ up to $\tau_i-$, and  jumps to $0$ at time $\tau_i$, where it remains forever afterwards.
The dynamics of the pre-default price process $P=(P_t^i;\ i=1,\ldots,n)_{t\in[0,T]}^{\top}$ of the $n$ defaultable stocks is given by
\begin{align}\label{eq:Pt}
dP_t &= {\rm diag}(P_t)\left[\left(\mu(Y_t)+\lambda(Y_t,H_t)\right)dt+\sigma(Y_t)dW_t\right].
\end{align}
In the above expression, ${\rm diag}(P_t)$ is the diagonal $n$-dimensional square matrix whose $i$-th entry is $P_t^i$. The vector $\mu$ is an $\R^n$-valued function and the matrix $\sigma$ is an $\R^{n\times n}$-valued function. Further, $\sigma$ is assumed to be invertible and its inverse is denoted by $\sigma^{-1}$. The vector
$\lambda(y,z)=(\lambda_i(y,z);\ i=1,\ldots,n)^{\top}$ is the vector of {default intensities}.
Eq.~\eqref{eq:Pt} indicates that the investor holding the credit sensitive stock is compensated for the incurred default risk at the premium rate $\lambda(Y_t,H_t)$.
Using \eqref{eq:pricedef}, \eqref{eq:Pt} and integration by parts, we obtain the dynamics given by
\begin{align}\label{eq:barPt}
d\tilde{P}_t={\rm diag}(\tilde{P}_{t-})\left[\mu(Y_t)dt + \sigma(Y_t)dW_t-dM_t\right],
\end{align}
where $M_t=(M_t^i;\ i=1,\ldots,n)^{\top}$ is a pure jump $(\Px,\Gx)$-martingale given by, for $i=1,\ldots,n$,
\begin{align}\label{eq:taui}
M_t^i&:= H_t^i - \int_0^{t\wedge\tau_i}\lambda_i(Y_s,H_s)ds,\ \ \ \ \ \ t\in[0,T].
\end{align}
The joint process $(\tilde{P},H)$ satisfying \eqref{eq:barPt} and \eqref{eq:taui} can be constructed in an iterative manner following a procedure similar to that in Lemma A.1 of \cite{CappFrei}. For completeness, we next give the construction of the process $(\tilde{P},H)$: Let $({\xi}_{ij})_{i,j=1,\ldots,n}$ be independent standard exponentially distributed random variables. We assume that these are also independent of the Brownian motions $(W,\bar{W})$. We first consider the following SDE given by, for $i=1,\ldots,n$,
\begin{align}\label{eq:P0}
\frac{d\tilde{P}_t^{0,i}}{\tilde{P}_t^{0,i}}&=(\mu_i(Y_t)+\lambda_i(Y_t,0))dt + \sum_{j=1}^n\sigma_{ij}(Y_t)dW_t^{j},\quad \tilde{P}_0^{0,i}=\tilde{P}_0^i>0.
\end{align}
Then $\tilde{P}^{0,i}=(\tilde{P}^{0,i}_t)_{t\geq0}$ is a geometric Brownian motion whose coefficients depend on stochastic factors. Assume that no default has occurred at inception, i.e., $H_0={0}$. We define $\hat\tau_1$ as the first time that either of the $n$ stocks defaults, i.e.,
\begin{align*}
\hat\tau_1&:=\min_{i=1,\ldots,n}\tau_{1i},\quad
\tau_{1i}:=\inf\left\{t>0;\ \int_0^t\lambda_i(Y_s,0)ds\geq\xi_{1i}\right\},\ \ i=1,\ldots,n.
\end{align*}
For $i=1,\ldots,n$, we set $\tilde{P}_t^i=\tilde{P}_t^{0,i}$ and $H_t=H_0=0$ when $t\in[0,\hat\tau_1)$. Further, define $i_1:=\mathop{\arg\min}\limits_{i=1,\ldots,n}\tau_{1i}$
and define $\tilde{P}_{t}^{1,i_1}=0$ for $t\geq\hat{\tau}_1$. For $i\in\{1,\ldots,n\}\setminus\{i_1\}$, consider the following SDE: on $t\geq\hat{\tau}_1$,
\begin{align}\label{eq:SDE6}
\tilde{P}_{t}^{1,i}&= \tilde{P}_{\tau_1}^{0,i}+\int_{\hat{\tau}_1}^t\tilde{P}_{s}^{1,i}(\mu_i(Y_s)+\lambda_i(Y_s,0^{i_1}))ds
+ \sum_{j=1}^n\int_{\hat{\tau}_1}^t\tilde{P}_{s}^{1,i}\sigma_{ij}(Y_s)dW_s^{j}.
\end{align}
We use $0^{i_1}$ to denote the $n$-dimensional row vector whose entries are $0$ except for the $i_1$-th entry which is set to $1$. It can be easily seen that Eq.~\eqref{eq:SDE6} admits a unique positive strong solution $\tilde{P}_{t}^{1,i}$ on $t\geq\hat\tau_1$. Further, define the second default time
\begin{align*}
\hat\tau_2&:=\min_{i\in\{1,\ldots,n\}\setminus \{i_1\}}\tau_{2i},\quad
\tau_{2i}:=\inf\left\{t\geq\hat\tau_1;\ \int_{\hat\tau_1}^t\lambda_i(Y_s,0^{i_1})ds\geq\xi_{2i}\right\},\ \ i\in\{1,\ldots,n\}\setminus \{i_1\}.
\end{align*}
Proceeding similarly to the construction of the process $(\tilde{P}_t,H_t)$ for $t\in[0,\hat\tau_1)$, we set $\tilde{P}_t^i=\tilde{P}_t^{1,i}$ for all $i\in\{1,\ldots,n\}\setminus\{i_1\}$ and $H_t={H}_{\hat\tau_1}=0^{i_1}$ if $t\in[\hat\tau_1,\hat\tau_2)$.
Moreover, let $i_2:=\mathop{\arg\min}\limits_{i\in\{1,\ldots,n\}\setminus\{i_1\}}\tau_{2i}$. More generally, for $k=3,\ldots,n$, the $k$-th default time is specified by
\begin{align*}
\hat\tau_k&:=\min_{i\in\{1,\ldots,n\}\setminus \{i_1,\ldots,i_{k-1}\}}\tau_{ki},\nonumber\\
\tau_{ki}&:=\inf\left\{t\geq\hat\tau_{k-1};\ \int_{\hat\tau_{k-1}}^t\lambda_i(Y_s,0^{i_1,\ldots,i_{k-1}})ds\geq\xi_{ki}\right\},\ \ i\in\{1,\ldots,k\}\setminus \{i_1,\ldots,i_{k-1}\}.
\end{align*}
The defaulted names $i_1,\ldots,i_{k-1}$ in the equation above are defined following an iterative procedure, in a similar way to $i_1$ and $i_2$. For $i\in\{1,\ldots,n\}\setminus\{i_1,\ldots,i_{k-1}\}$, and for $t\geq\hat{\tau}_{k-1}$, consider the process
\begin{align*}
\tilde{P}_t^{k-1,i}&=\tilde{P}_{\hat{\tau}_{k-1}}^{k-2,i}+\int_{\hat{\tau}_{k-1}}^t\tilde{P}_s^{k-1,i}(\mu_i(Y_s)+\lambda_i(Y_s,0^{i_1,\ldots,i_{k-1}}))ds+ \sum_{j=1}^n\int_{\hat{\tau}_{k-1}}^t\tilde{P}_s^{k-1,i}\sigma_{ij}(Y_s)dW_s^{j}.
\end{align*}
In the expression above, we use $0^{i_1,\ldots,i_{k-1}}$ to denote the $n$-dimensional row vector whose $i_1, i_2, \ldots, i_{k-1}$-th entries are equal to $1$, and the remaining entries are set to $0$. Iterating the recursive procedure described above, we can construct the Markov process $(\tilde{P}_t,H_t)$ on $t\in[\hat\tau_{k-1},\hat\tau_k)$, until $k=n$. At time $t\geq\hat\tau_{n}$, all stocks in the portfolio have defaulted. If we set $\hat{\tau}_0=0$ and $\hat{\tau}_{n+1}=\infty$, then the process $N_t=\max\{i\leq n;\ \hat{\tau}_i\leq t\}$ counts the number of defaults in the interval $[0,t]$. For $k\geq1$, recall that the random variable $i_k$ denotes the identity of the stock defaulting at $\hat{\tau}_k$. Then, the default indicator process $H$ may be represented as a marked point process via the sequence $(\hat{\tau}_k,i_k)_{1\leq k\leq n}$, i.e., $H_t^i=\sum_{\hat{\tau}_k\leq t}{\bf1}_{i_k=i}$ for $i=1,\ldots,n$ (we use ${\bf1}_{i_k=i}$ to denote the indicator function of the event $\{i_k=i\}$). This concludes the construction of the process $(\tilde{P},H)$. Using the above construction of the credit risk model, the filtration $\Fx$ is included in the filtration $\hat{\Fx}:=\Fx\vee(\vee_{k,i}\sigma(\xi_{ki}))$ since the Brownian motions are independent of the random variables $(\xi_{ki})$. Thus $\Fx\subset\Gx\subset\hat{\Fx}$ implies that any $(\Px,\Fx)$-martingale is a $(\Px,\hat{\Fx})$-martingale and hence a $(\Px,\Gx)$-martingale. In particular,  $(W,\bar{W})$ is also a $(\Px,\Gx)$-Brownian motion.

In the special case of a portfolio consisting of one risk-free stock ($\lambda(y,z)\equiv0$) and a one-dimensional stochastic factor ($n=m=1$), our model reduces to the one considered by
\cite{NLHH}, see equations (2.1) and (2.2) therein. Before proceeding further, define the $n$-dimensional column vector $\xi:D\to\R^n$ by
\begin{align}\label{eq:thetai}
\xi(y) &:= \sigma^{-1}(y)\left(\mu(y)-re_n\right),\ \ \ \ \ y\in D.
\end{align}
{In the above expression, $e_n$ denotes the $n$-dimensional column vector with entries identically equal to $1$.} The vector $\xi(y)$ is the market price of risk, i.e., the excess compensation demanded by the investor to bear the risk coming from uncertainty in the stocks' returns. We also define the space of equivalent local martingale measures (henceforth p.m. for short) as
\begin{align}\label{eq:Q}
{\cal Q}&:=\big\{{\rm p.m.}\ {\mathbb{Q}}:\ {\mathbb{Q}}\sim\Px\text{ on }\G_T,{\rm and\ }(B_t^{-1}\tilde{P}_t)_{t\in[0,T]}\ {\rm is\ a}\ {(\Qx,\Gx)}\mbox{-}{\rm local\ martingale}\big\}.
\end{align}

Let $C_b^1$ denote the set of all bounded $C^1$-functions on $D$, which also admit bounded first-order partial derivatives in $y\in D$. Throughout the paper, we make the following assumptions.
\begin{itemize}
  \item[({\bf A1})] There exists a sequence $(D_{\ell})_{\ell\in\N}$ of bounded domains with $C^2$-boundary and closure $\bar{D}_{\ell}\subset D$ such that $\cup_{\ell=1}^\infty D_{\ell}=D$. Moreover, for all $(t,y)\in [0,T]\times D$, $\Px(Y_s^{t,y}\in D,\ \forall\ s\in[t,T])=1$.
  \item[({\bf A2})] The vector function $\mu_0(\cdot) \in C^1$ with bounded first-order partial derivatives in $y\in D$, and $\sigma_0(\cdot)\in C_b^1$. For each $z\in{\cal S}$, the vector function $\lambda(\cdot,z)\in C_b^1$.
  \item[({\bf A3})] For $z\in{\cal S}$, let ${\cal C}:=\{f(\cdot)\in C_b^{0,1};\ N(f;\cdot,z)\in C_b^{0,1},\ {\rm and}\ \inf_{(t,y)}N(f;t,y,z)\in(-1,\infty)^n\}$. Here $N(f;\cdot,z):={{\rm diag}((1-z)\lambda^{-1}(\cdot,z))}\sigma(\cdot)(\xi(\cdot)-f(\cdot))$ for $z\in{\cal S}$, and $C_b^{0,1}$ denotes the set of functions which are continuous in $t\in[0,T]$ and are $C_b^1$ in $y\in D$.  Define
  \begin{align*}
  {\Theta}_z&=\big\{(\theta(\cdot,z),h(\cdot,z))\in{\cal C}\times{\cal B};\ \theta(t,y,z)\in\R^n, \ h(t,y,z)\in(-1,\infty)^n,\ {\rm and}\nonumber\\
  &\qquad \sigma(y)(\xi(y)-\theta(t,y,z))={{\rm diag}((1-z)\lambda(y,z))}h(t,y,z),\ {\rm for}\ (t,y)\in[0,T]\times D\big\}.
  \end{align*}
  The set ${\Theta}_z$ is nonempty for each $z\in{\cal S}$. Above, ${\cal B}$ represents the set of all Borel functions on $[0,T]\times D$. Moreover, we set $(1-z)\lambda(y,z):=((1-z_i)\lambda_i(y,z);\ i=1,\ldots,n)^{\top}$ and $(1-z)\lambda^{-1}(y,z):=((1-z_i)\lambda_i^{-1}(y,z);\ i=1,\ldots,n)^{\top}$.
\end{itemize}
\begin{remark}\label{rem:A1-A3}
If the coefficients $\mu_0$ and $\sigma_0$ are Lipschitz continuous on $\R^m$, then the assumption ({\bf A1}) is satisfied with $D=\R^m$. This setting covers the case of a stochastic factor given by an Ornstein-Uhlenbeck (OU) process, also considered by \cite{NLHH}.
Fix $m=1$ and assume $y\geq0$. Under the parameter choices of $\mu_0(y)=a(b-y)$ and $\sigma_0(y)=\kappa\sqrt{y}$, where $a,b,\kappa$ are positive constants satisfying the Feller's condition $2ab\geq\kappa^2$, the assumption ({\bf A1}) is satisfied choosing $D=(0,\infty)$. This specification corresponds to a stochastic factor model given by a Cox-Ingersson-Ross (CIR) process. Assumptions ({\bf A1}) and ({\bf A2}) guarantee that Eq.~\eqref{eq:Yt} admits a unique strong non-exploding solution. We will use the elements of the set ${\Theta}_z$ in assumption ({\bf A3}) to establish the equivalent martingale measure in terms of the Radon-Nikodym density, when we consider the dual of the optimal investment-consumption problem. In most situations of practical interest, the set ${\Theta}_z$ in the assumption ({\bf A3}) is nonempty, see also the example below.
\end{remark}

The proposed framework is rich enough to include several stochastic volatility models considered in the literature as special cases. In addition, it allows for systemic effects through the dependence of the default intensity on the common factor and the default states. To illustrate this, we next present a two-dimensional factor model featuring stochastic volatility and default contagion.

\begin{example}\label{exam:1} Consider a two-dimensional stochastic factor process $Y_t=(Y_t^1,Y_t^2)$ of the OU type:
\begin{align*}
dY_t^1&=(u_1-\mu_{01} Y_t^1)dt + \sum_{i=1}^2\sigma_{0i}[\rho dW_t^i + \sqrt{1-\rho^2}d\bar{W}_t^i],\nonumber\\
dY_t^2&=(u_2-\mu_{02}Y_t^2)dt + \sum_{i=1}^2\bar{\sigma}_{0i}[\rho dW_t^i + \sqrt{1-\rho^2}d\bar{W}_t^i],
\end{align*}
where $u_i\in\R$ and $\mu_{0i},\sigma_{0i},\bar{\sigma}_{0i}\in\R_+$ for $i=1,2$. The state space of $Y$ is given by $D=\R^2$ with sub-domains $D_{\ell}=(-\ell,\ell)^2$, $\ell\in\N$, hence
satisfying the assumption ${\bf (A1)}$. We can view $Y_t$ as a vector of economic state
variables such as growth of real returns, or deflator/inflation processes for the factors of production, which have been shown to exhibit mean reversion (see \cite{Jensen}). The price dynamics of the two defaultable stocks are given by
\begin{align*}
\frac{d\tilde{P}_t^1}{\tilde{P}_{t-}^1} &=\mu_1dt + \sqrt{\vartheta_1(Y_t^1)}dW_t^1-dM_t^1,\\
\frac{d\tilde{P}_t^2}{\tilde{P}_{t-}^2} &=\mu_2dt + \sqrt{\vartheta_2(Y_t^2)}\big[\bar{\rho} dW_t^1+\sqrt{1-\bar{\rho}^2}dW_t^2\big]-dM_t^2.
\end{align*}
In the above expressions, $\bar{\rho}\in[-1,1]$, and $\vartheta_i$'s are positive and $C^1$. We recall that $M_t^i$, $i=1,2$, are $(\Px,\Gx)$-martingales given in Eq.~\eqref{eq:taui}.
We have the following coefficients:
\begin{align*}
\mu_0(y) &= \left[
             \begin{array}{c}
               u_1-\mu_{01}y_1 \\
               u_2-\mu_{02}y_2 \\
             \end{array}
           \right],\ \ \ \ \ \ \ \sigma_0(y)\equiv\left[
                                                    \begin{array}{cc}
                                                    \sigma_{01}   & \sigma_{02} \\
                                                    \bar{\sigma}_{01} & \bar{\sigma}_{02} \\
                                                    \end{array}
                                                  \right],\ \ \ \ \ \ \lambda(y,z)= \left[
             \begin{array}{c}
               \lambda_1(y,z) \\
               \lambda_2(y,z) \\
             \end{array}
           \right],\nonumber\\
\mu(y) &= \left[
             \begin{array}{c}
               \mu_1 \\
               \mu_2 \\
             \end{array}
           \right],\ \ \ \ \ \ \ \sigma(y)=\left[
                                                    \begin{array}{cc}
                                                    \sqrt{\vartheta_1(y_1)}   & 0 \\
                                                      \bar{\rho}\sqrt{\vartheta_2(y_2)} &  \sqrt{1-\bar{\rho}^2}\sqrt{\vartheta_2(y_2)} \\
                                                    \end{array}
                                                  \right].
\end{align*}
Then the inverse of the volatility matrix $\sigma$ is given by
\begin{align*}
\sigma^{-1}(y) = \left[
                              \begin{array}{cc}
                                                    \frac{1}{\sqrt{\vartheta_1(y_1)}}   & 0 \\
                                                      -\frac{\bar{\rho}}{\sqrt{1-\bar{\rho}^2}\sqrt{\vartheta_1(y_1)}} &  \frac{1}{\sqrt{1-\bar{\rho}^2}\sqrt{\vartheta_2(y_2)}} \\
                                                    \end{array}
                                                  \right],
\end{align*}
and the vector $\xi$ is given by
\begin{align*}
\xi(y)=\left[
              \begin{array}{c}
                \frac{\bar{\mu}_1}{\sqrt{\vartheta_1(y_1)}}  \\
                \frac{\bar{\mu}_2}{\sqrt{1-\bar{\rho}^2}\sqrt{\vartheta_2(y_2)}}-\frac{\bar{\rho}\bar{\mu}_1}{\sqrt{1-\bar{\rho}^2}\sqrt{\vartheta_1(y_1)}} \\
              \end{array}
            \right].
\end{align*}
Moreover, for $f(y)=(f_1(y),f_2(y))$, we have
\begin{align*}
N(f;y,z) = \left[
              \begin{array}{c}
                \frac{\sqrt{\vartheta_1(y_1)}}{\lambda_1(y,z)}(1-z_1)(\xi_1(y)-f_1(y))  \\
                \frac{\sqrt{\vartheta_2(y_2)}}{\lambda_2(y,z)}(1-z_2)\big\{\bar{\rho}(\xi_1(y)-f_1(y))+\sqrt{1-\bar{\rho}^2}(\xi_2(y)-f_2(y))\big\} \\
              \end{array}
            \right].
\end{align*}
Above, for $i=1,2$, $\bar{\mu}_i:=\mu_i-r$. We can further compute
\begin{align*}
\frac{\partial\xi_1}{\partial y_1}&=-\frac{\vartheta_1'(y_1)\bar{\mu}_1}{2\vartheta_1^{\frac{3}{2}}(y_1)},\quad \frac{\partial\xi_1}{\partial y_2}=0,\quad
\frac{\partial\xi_2}{\partial y_1}=-\frac{\bar{\rho}}{\sqrt{1-\bar{\rho}^2}}\frac{\partial\xi_1}{\partial y_1},\quad
\frac{\partial\xi_2}{\partial y_2}=-\frac{1}{\sqrt{1-\bar{\rho}^2}}\frac{\vartheta_2'(y_2)\bar{\mu}_2}{2\vartheta_2^{\frac{3}{2}}(y_2)}.
\end{align*}
We next consider two choices for the volatility function $\vartheta_i$, $i=1,2$, previously considered in the literature. Under both choices, we can see that $\xi(\cdot)\in C_b^1$.
\begin{itemize}
  \item[{(I)}] Uniformly elliptic Scott volatility, i.e. $\vartheta_i(y_i)=\varepsilon_i+e^{\gamma_iy_i}$ for $\gamma_i,\varepsilon_i>0$. Then we have
  $\big|\frac{1}{\sqrt{\vartheta_i(y_i)}}\big|\leq\frac{1}{\sqrt{\varepsilon_i}}$, and $\big|\frac{\vartheta_i'(y_1)}{\vartheta_i^{{3}/{2}}(y_i)}\big|=\frac{\gamma_ie^{\gamma_iy_i}}{(\varepsilon_i+e^{\gamma_iy_i})^{3/2}}
\leq\frac{\gamma_i(\varepsilon_i+e^{\gamma_iy_i})}{(\varepsilon_i+e^{\gamma_iy_i})^{3/2}}=\frac{\gamma_i}{(\varepsilon_i+e^{\gamma_iy_i})^{1/2}}
\leq\frac{\gamma_i}{\sqrt{\varepsilon_i}}$. Hence  $\xi(\cdot)\in C_b^1$.
\item[{(II)}] Uniformly elliptic Stein-Stein volatility, i.e. $\vartheta_i(y_i)=\varepsilon_i+\gamma_i|y_i|^2$ for $\gamma_i,\varepsilon_i>0$. Then $\big|\frac{1}{\sqrt{\vartheta_i(y_i)}}\big|\leq\frac{1}{\sqrt{\varepsilon_i}}$, and $\big|\frac{\vartheta_i'(y_1)}{\vartheta_i^{{3}/{2}}(y_i)}\big|=\frac{2\gamma_i|y_i|}{(\varepsilon_i+\gamma_i|y_i|^2)^{3/2}}
\leq\frac{2\sqrt{\gamma_i}(\varepsilon_i+\gamma_i|y_i|^2)^{1/2}}{(\varepsilon_i+\gamma_i|y_i|^2)^{3/2}}=\frac{2\sqrt{\gamma_i}}{(\varepsilon_i+\gamma_i|y_i|^2)^{1/2}}
\leq2\sqrt{\frac{\gamma_i}{\varepsilon_i}}$. Hence  $\xi(\cdot)\in C_b^1$.
\end{itemize}
Let $\theta(t,y,z)=(\theta_1(t,y,z),\theta_2(t,y,z))$ and $h(t,y,z)=(h_1(t,y,z),h_2(t,y,z))$ for $(t,y,z)\in[0,T]\times D\times{\cal S}$. The defining equation of the set $\Theta_z$ in the assumption ({\bf A3}) may be rewritten as
\begin{align}\label{eq:rem1eq}
\left\{
  \begin{array}{ll}
    \frac{(1-z_1)\lambda_1(y,z)}{\sqrt{\vartheta_1(y_1)}}h_1(t,y,z)=\xi_1(y)-\theta_1(t,y,z)\\
     \\
   \frac{1}{\sqrt{1-\bar{\rho}^2}}\frac{(1-z_2)\lambda_2(y,z)}{\sqrt{\vartheta_2(y_2)}}h_2(t,y,z)
-\frac{\bar{\rho}}{\sqrt{1-\bar{\rho}^2}}\frac{(1-z_1)\lambda_1(y,z)}{\sqrt{\vartheta_1(y_1)}}h_1(t,y,z)=\xi_2(y)-\theta_2(t,y,z).
  \end{array}
\right.
\end{align}
{We next verify that the setup satisfies the assumption ({\bf A3})}. Consider the solution $\theta(t,y,z)$ to Eq.~\eqref{eq:rem1eq} in the different default states $z\in{\cal S}=\{0,1\}^2$. When $z=(1,1)$, we deduce from Eq.~\eqref{eq:rem1eq} that $\theta(t,y,(1,1))=\xi(y)$ for $y\in D$, i.e. $\theta(t,y,(1,1))$ is the market price of risk. Meanwhile $N(\theta;y,(1,1))=0$ and hence $\theta(\cdot,(1,1))\in {\cal C}$. Thus, given for any Borel function $h(\cdot,(1,1))$, we have $(\xi(\cdot),h(\cdot,(1,1)))\in\Theta_{(1,1)}$. When $z=(1,0)$, we deduce from Eq.~\eqref{eq:rem1eq} that, for $y\in D$, $\theta_1(t,y,(1,0))=\xi_1(y)$ and we choose $\theta_2(t,y,(1,0))=-\frac{\bar{\rho}\bar{\mu}_1}{\sqrt{1-\bar{\rho}^2}\sqrt{\vartheta_1(y_1)}}\in C_b^1$. Then $N(\theta;y,(1,0))=(0,\frac{\bar{\mu}_2}{\lambda_2(y,(1,0))})$. Let $\lambda_2(y,(1,0))\in C_b^{1}$ and take values in a interval $[\varepsilon,C_2]$ with $\varepsilon<C_2$ and $\frac{\bar{\mu}_2}{C_2}>-1$. Then $\theta(\cdot,(1,0))\in{\cal C}$. Thus for any Borel function $h_1(t,y,(1,0))$, we have $(\theta(\cdot,(1,0)),(h_1(\cdot,(1,0)),\frac{\bar{\mu}_2}{\lambda_2(\cdot,(1,0))})^{\top})\in\Theta_{(1,0)}$. Similar we can discuss the case $z=(0,1)$. For the last case $z=(0,0)$, a direct solution is to take $\theta(\cdot,(1,1))=\xi(\cdot)\in C_b^1$ and hence $N(\theta;y,(0,0))=0$. Then $\theta(\cdot,(0,0))\in{\cal C}$. Obviously it holds that $(\theta(\cdot,(0,0)),0)\in\Theta_{(0,0)}$. We can also take the same $\theta$ as in the case $z=(1,0)$. Then, it holds that $(\theta(\cdot,(0,0)),(0,\frac{\bar{\mu}_2}{\lambda_2(\cdot,(0,0))})^{\top})\in\Theta_{(0,0)}$. Hence, for each default state $z\in\{0,1\}^2$, $\Theta_z$ is nonempty, i,e. the assumption ({\bf A3}) is satisfied.
\end{example}

\subsection{The Optimal Investment/Consumption Problem}\label{sec:primal}
We consider a power investor who wants to maximize his expected utility from consumption plus wealth at the target horizon $T$. He dynamically allocates his wealth into the risk-free money market account, and $n$ defaultable stocks.

{For $t\in[0,T]$,} denote by $\phi^B_t$ the number of shares of the risk-free bank account held by the investor at time $t$. We use $\phi^i_t$ to denote the number of shares of the $i$-th defaultable stock, $i= 1,\ldots,n$, at time $t$. The wealth process $X_t$ associated with the $\Gx$-predictable portfolio process  $(\phi_t^B,\phi_t)$, $\phi_t=(\phi_t^i;\ i=1,\ldots,n )^{\top}$, and with a nonnegative consumption rate process $c_t$, is given by
\begin{align*}
X_t=\phi_t^B B_t + \phi_t^{\top}\tilde{P}_t-\int_0^t c_sds.
\end{align*}
If the wealth process is positive, we may define the fractions of wealth invested in the stocks and money market account as
\begin{align}\label{eq:fraction}
\pi_t^i:=\frac{\phi_t^i \tilde{P}_{t-}^i}{X_{t-}},\ \ \ {\rm and}\ \ \pi_t^B:=1-\pi_t^{\top}e_n,
\end{align}
where $\pi_t=(\pi_t^i;\ i=1,\ldots,n)^{\top}$, and {recall that} $e_n = \big(\underbrace{1,1,\ldots,1}_{n \; ones}\big)^{\top}$. Since the price of the $i$-th stock jumps to zero when the $i$-th stock defaults, the fraction of wealth held by the investor in this stock is zero after its default. In particular, it holds that $\pi_t^i=(1-H_{t-}^i)\pi_t^i$ for $i=1,\ldots,n$.
Using the self-financing condition, we may rewrite the wealth process as
\begin{align}\label{eq:wealth}
dX_t &= X_{t-}\pi_t^{\top}{\rm diag}(\tilde{P}_t)^{-1}d\tilde{P}_t + X_t(1-\pi_t^{\top}e_n)\frac{dB_t}{B_t}-c_tdt\\\notag
&=X_t[r+\pi_t^{\top}(\mu(Y_t)-re_n)]dt + X_t\pi_t^{\top}\sigma(Y_t)dW_t-X_{t-}\pi_t^{\top}dM_t-c_tdt.
\end{align}
The wealth process dynamics is intuitive. When the $i$-th stock defaults, the investor's wealth gets reduced by the amount of wealth allocated to it. Next, we define the space of admissible strategies:
\begin{definition} \label{def:admissi}
Let $(x,{y},{z})\in\R_+\times D\times{\cal S}$. The class ${\cal U}:={\cal U}(x,{y},{z})$ of admissible strategies is the set of $\Gx$-predictable processes ${\pi}=(\pi_t^i;\ i=1,\ldots,n)_{t\in[0,T]}^{\top}$ such that $\Ex[\int_0^T\|{\pi}_t^{\top}\sigma(Y_t)\|^2dt]<+\infty$, $\sum_{i=1}^n\Ex[\int_0^T|\pi_t^i|^2\lambda_i(Y_t,H_t)dt]<+\infty$, and of nonnegative predictable consumption processes $c=(c_t)_{t\in[0,T]}$ satisfying $\Ex\big[\int_0^Tc_tdt\big]<+\infty$, so that the associated wealth process satisfying SDE~\eqref{eq:wealth} is strictly
positive when $X_0^{\pi,c} = x\in\R_+$, $Y_0=y\in D$ and $H_0 = z\in{\cal S}$ (i.e., $X_t^{\pi,c}:=X_t^{{\pi},c,x,{y},{z}}>0$ for all $t\in[0,T]$). Above, we have used the notation $\|x\|^2:=\sum_{i=1}^mx_i^2$ for $x\in\R^m$.
\end{definition}

We choose both the utility extracted from consumption and terminal wealth to be of the HARA type with risk aversion parameter $p$. More specifically, for $x\in\R_+$ we choose $U_i(x)=\frac{K_i}{p}x^{p}$, where $p\neq0$ and $p<1$ ($p=0$ corresponds to the logarithm utility case). The coefficients $K_1$ and $K_2$ are positive constants capturing the tradeoff between the utility extracted from intermediate consumption and terminal wealth. Our objective is to study the following utility maximization problem: {under the initial conditions $(X_0^{\pi,c},Y_0,H_0)=(x,y,z)\in\R_+\times D\times{\cal S}$,} find the optimal investment/consumption strategy and expected utility, also referred to as investor's value function, given by
\begin{align}\label{eq:primal-pro}
V(x,{y},{z}) &:= \sup_{({\pi},c)\in{\cal U}}\Ex\left[U_1\big(X_T^{{\pi},c}\big)+\int_{0}^T U_2(c_s)d s\right],
\end{align}
where $(x,y,z) \in \R_+ \times D\times{\cal S}$ and $X_T^{{\pi},c}>0$, $({\pi},c)\in{\cal U}={\cal U}(x,{y},{z})$, is the wealth of the investor at terminal time $T$.

\section{Dual Formulation and Verification Result} \label{sec:dualver}
Section \ref{sec:dualform} formulates the dual of the primal problem \eqref{eq:primal-pro} and then provides an equivalence relation between their value functions. We analyze the equivalent problem in Section~\ref{sec:beta}. Section~\ref{sec:verif2} gives the verification result.

\subsection{Dual Formulation} \label{sec:dualform}
We start noticing that, using equations~\eqref{eq:Pt} and~\eqref{eq:wealth}, the discounted stock price and consumption-adjusted wealth process admit the following representations: for $(\theta(\cdot,z),h(\cdot,z))\in\Theta_z$ with $z\in{\cal S}$,
\begin{align}\label{eq:Pab-discount}
&d\left(\frac{\tilde{P}_t}{B_t}\right) = {\rm diag}\left(\frac{\tilde{P}_{t-}}{B_{t-}}\right)\big[\sigma(Y_t)d{W}_t^{\theta}-dM_t^h\big],\ \ {\rm and}\nonumber\\
&d\left(\frac{X_t^{{\pi},c}}{B_t}\right) + \frac{c_t}{B_t}dt = \frac{X_{t-}^{\pi,c}}{B_{t-}}\pi_t^{\top}\big[\sigma(Y_t)d{W}_t^{\theta}-dM_t^h\big],
\end{align}
where, for $t\in[0,T]$,
\begin{align}\label{eq:tildeWi}
{W}_t^{\theta} &:= W_t + \int_0^t \theta(s,Y_s,{H}_s)ds,\ \ {\rm and}\ \ M_t^{h,i}:=M_t^i-\int_0^{t\wedge\tau_i}(h_i\lambda_i)(s,Y_s,H_s)ds,\quad i=1,\ldots,n.
\end{align}

Next, we define a set of probability measures which are equivalent to $\Px$ such that the above discounted consumption-adjusted wealth process is a supermartingale w.r.t. the filtration $\Gx$. For any $\theta\in{\cal C}$, denote by ${\cal M}$ the set of $\Gx$-predictable processes $a=(a_t^i;\ i=1,\ldots,n)_{t\in[0,T]}^{\top}$ taking values on $\R^n$ and $h=(h_i(t,Y_{t-},H_{t-});\ i=1,\ldots,n)_{t\in[0,T]}^{\top}$ satisfying $(\theta(\cdot,z),h(\cdot,z))\in\Theta_z$, $z\in{\cal S}$, such that the process
\begin{align}\label{eq:eta}
\Gam_t^{a,h} &:= {\cal E}\left\{-\int_0^{\cdot}\theta(s,Y_{s-},H_{s-})^{\top}dW_s-\int_0^{\cdot}a_s^{\top}d\bar{W}_s+\int_0^{\cdot}h_s^{\top}dM_s\right\}_t,\quad t\in[0,T]
\end{align}
is a $(\Px,\Gx)$-martingale. Above, we use ${\cal E}\{\cdot\}$ to denote the stochastic exponential. Let $\theta_t:=\theta(t,Y_{t-},H_{t-})$. Then, the density process ${\Gam}^{a,h}=({\Gam}_t^{a,h})_{t\in[0,T]}$ satisfies
\begin{eqnarray}
\frac{d\Gam_t^{a,h}}{\Gam_{t-}^{a,h}}=-\theta_t^{\top}dW_t-a_t^{\top}d\bar{W}_t+h_{t}^{\top}dM_t,\ \ \ \ {\Gam}_0^{a,h}=1,
\label{eq:Gameq}
\end{eqnarray}
and admits the closed-form solution given by
\begin{align}\label{eq:solution-density}
\Gam_t^{a,h} &= \exp\bigg\{-\int_0^t\theta_s^{\top}d{W}_s-\frac{1}{2}\int_0^t{\rm tr}[(\theta_s\theta_s^{\top})]ds-\int_0^t a_s^{\top}d\bar{W}_s-\frac{1}{2}\int_0^t{\rm tr}[a_sa_s^{\top}]ds\\
&\qquad\qquad + \sum_{i=1}^n\int_0^t\log(1+h_{s}^i)dM_s^i
+\sum_{i=1}^n\int_0^{t\wedge\tau_i}\big[\log(1+h_s^i)-h_s^i\big]\lambda_i(Y_s,H_s)ds\bigg\}.\nonumber
\end{align}
Then, for $(a,h)\in{\cal M}$, define the p.m. $\Px^{a,h}\sim\Px$ on ${\mathcal{G}}_T$ as
\begin{align}\label{eq:Pmugamma}
\frac{d\Px^{a,h}}{d\Px} = \Gam_T^{a,h}.
\end{align}
Hence, we have that $\Gam_t^{a,h}=\Ex[\frac{d\Px^{a,h}}{d\Px}|{{\mathcal{G}}_t}]$ for $t\in[0,T]$, and under $(\Px^{a,h},\Gx)$, we have that the processes $W_t^{\theta}$, $t\in[0,T]$, defined in \eqref{eq:tildeWi} and
\begin{align}\label{eq:tildeWi2}
\bar{W}_t^{a}:= \bar{W}_t + \int_0^t a_sds,\ \ \ \ \ t\in[0,T]
\end{align}
form a Brownian motion. Moreover, for each $i=1,\ldots,n$, the process $M^{h,i}=(M_t^{h,i})_{t\in[0,T]}$ defined in \eqref{eq:tildeWi} is a martingale.

Notice that for $(a,h)\in{\cal M}$, the p.m. $\Px^{a,h}\in{\cal Q}$ by virtue of \eqref{eq:Pab-discount}. Moreover, for $(\pi,c)\in{\cal U}$, the discounted process $\frac{X_t^{\pi,c}}{B_t} + \int_0^t\frac{c_s}{B_s}ds$, $t\in[0,T]$, is nonnegative. It then follows from \eqref{eq:Pab-discount} and \eqref{eq:tildeWi} that it is a $(\Px^{a,h},\Gx)$-local martingale and hence a $(\Px^{a,h},\Gx)$-supermartingale. This yields that, for all $(\pi,c)\in{\cal U}$,
\begin{align}\label{eq:constranits1}
\sup_{(a,h)\in{\cal M}}\Ex^{a,h}\left[\frac{X_T^{\pi,c}}{B_T} + \int_0^T\frac{c_t}{B_t}dt\right]\leq x,
\end{align}
where $\Ex^{a,h}[\cdot]$ denotes the expectation w.r.t. $\Px^{a,h}$. Using \eqref{eq:Pmugamma} and Lemma 2.5 in \cite{CoxHuang}, for $(a,h)\in{\cal M}$, it holds that
\begin{eqnarray}\label{eq:bayes}
\Ex^{a,h}\left[\frac{X_T^{\pi,c}}{B_T} + \int_0^T\frac{c_s}{B_s}ds\right]=\Ex\left[\frac{\Gam_T^{a,h}}{B_T}X^{\pi,c}_T + \int_0^T\frac{\Gam_s^{a,h}}{B_s}c_sds\right].
\end{eqnarray}
For $(a,h)\in{\cal M}$ and $\kappa>0$, define the dual functional by
\begin{align}\label{eq:Pi}
{\it\Pi}(a,h,\kappa):={\it\Pi}^{x,y,z}(a,h,\kappa):=\Ex\left[\tilde{U}_1\left(\kappa\frac{\Gam_T^{a,h}}{B_T}\right)
+\int_0^T\tilde{U}_2\left(\kappa\frac{\Gam_s^{a,h}}{B_s}\right)ds\right]
+\kappa x,
\end{align}
under the initial condition $(X_0^{\pi,c},Y_0,H_0)=(x,y,z)\in\R_+\times D\times{\cal S}$. For $i=1,2$, $\tilde{U}_i(y)$, $y\in\R_+$ is the Legendre-Fenchel transform of the HARA utility function $U_i(x)$, $x\in\R_+$, given by
\begin{align}\label{eq:tildeU}
\tilde{U}_i(y) &:= \sup_{x\in\R_+}\left\{U_i(x)-xy\right\}.
\end{align}
{\begin{remark}\label{rem:31}
It is well known that the supremum above is
attained at the point $I_i(y):=(U_i')^{-1}(y)$, $y\in\R_+$, and hence $I_i(y)=K_i^{\frac{1}{1-p}}y^{\frac{1}{p-1}}$, i.e., it holds that $\tilde{U}_i(y) =U_i(I_i(y))-I_i(y)y$.
Using the expression for $I_i(y)$, we obtain that $\tilde{U}_{i}(y) = -\frac{1}{q}K_i^{1-q}y^{q}$ where $q:=\frac{p}{p-1}$.
We may alternatively rewrite $I_i(y)$ in terms of $q$ as
$I_i(y)=K_i^{1-q}y^{q-1}$. It may be easily seen that $q\in(0,1)$ if $p<0$, and $q<0$ if $p\in(0,1)$.
\end{remark}}
Hence, from the inequality~\eqref{eq:constranits1}, we obtain that for all $({\pi},c)\in{\cal U}$, $(a,h)\in{\cal M}$, and $\kappa>0$,
\begin{eqnarray}\label{eq:ine0}
&&\Ex\left[U_1(X^{{\pi},c}_T)+\int_{0}^T U_2(c_s)ds\right]\nonumber\\
&&\qquad \leq \Ex\left[U_1(X^{\pi,c}_T)+\int_{0}^T U_2(c_s)ds\right]-\kappa\left\{\Ex\left[\frac{\Gam^{a,h}_T}{B_T}X^{\pi,c}_T + \int_0^T\frac{\Gam_s^{a,h}}{B_s}c_sds\right]-x\right\}\nonumber\\
&&\qquad =\Ex\left[U_1(X_T^{\pi,c})-\kappa\frac{\Gam_T^{a,h}}{B_T}X_T^{\pi,c}\right] + \Ex\left[\int_{0}^T \Big(U_2(c_s)-\kappa\frac{\Gam_s^{a,h}}{B_s}c_s\Big)ds\right]+\kappa x\nonumber\\
&&\qquad\leq 
{\it\Pi}(a,h,\kappa).
\end{eqnarray}
This immediately yields the following
\begin{lemma}\label{lem:ineL}
Let $(x,y,z)\in\R_+\times D\times{\cal S}$. Recall the value function $V(x,y,z)$ defined by \eqref{eq:primal-pro}. We have
\begin{align}\label{eq:ineL}
V(x,y,z) \leq \inf_{(a,h)\in{\cal M},\kappa>0}{\it\Pi}(a,h,\kappa).
\end{align}
\end{lemma}

{Next, we show that there exist $(\hat{a},\hat{h},\hat{\kappa})\in{\cal M}\times\R_+$ and $(\hat{\pi},\hat{c})\in{\cal U}$ so that both  {inequalities}~\eqref{eq:constranits1} and~\eqref{eq:ineL}
become equalities.}
This in turn yields that the optimal trading and consumption strategy are given by {$(\hat{\pi},\hat{c})\in{\cal U}$.}
The proof is standard and reported in the Appendix for completeness.
\begin{proposition}\label{prop:1}
Let $(X_0,Y_0,H_0)=(x,y,z)\in\R_+\times D\times{\cal S}$. Assume that there exist a triple $(\hat{a},\hat{h},\hat{\kappa})\in{\cal M}\times\R_+$, a strategy
$(\hat{\pi},\hat{c})\in{\cal U}={\cal U}(x,y,z)$ such that $X_T^{\hat{\pi},\hat{c}}=I_1\big(\hat{\kappa}\frac{\Gam_T^{\hat{a},\hat{h}}}{B_T}\big)$,
$\hat{c}_t=I_2\big(\hat{\kappa}\frac{\Gam_t^{\hat{a},\hat{h}}}{B_t}\big)$, $t\in[0,T]$, and that the following equality holds
\begin{align}\label{eq:constranits2}
\Ex^{\hat{a},\hat{h}}\left[\frac{X_T^{\hat{\pi},\hat{c}}}{B_T} + \int_0^T\frac{\hat{c}_s}{B_s}ds\right]=x.
\end{align}
Then the following holds for the value function defined in~\eqref{eq:primal-pro},
\begin{align}\label{eq:ineL2}
V(x,y,z)=\Ex\left[U_1\big(X_T^{\hat{\pi},\hat{c}}\big)+\int_0^TU_2(\hat{c}_s)ds\right] = \inf_{(a,h)\in{\cal M},\kappa>0}{\it\Pi}(a,h,\kappa)={\it\Pi}(\hat{a},\hat{h},\hat{\kappa}),
\end{align}
where, for $(a,h)\in{\cal M}$, and $\kappa>0$, the function ${\it\Pi}(a,h,\kappa)$ is given by \eqref{eq:Pi}. This shows that  $(\hat{\pi},\hat{c})$ given above is the optimal investment-consumption strategy to the primal problem \eqref{eq:primal-pro}.
\end{proposition}
Proposition~\ref{prop:1} establishes a link between the {primal} problem \eqref{eq:primal-pro} and the dual (minimization) problem \eqref{eq:ineL2}. The forthcoming sections discuss the solvability of the dual problem \eqref{eq:ineL2}. Next,
we {provide} an equivalent representation for {the criterion} \eqref{eq:ineL2}, which turns out to be more convenient for the analysis of HJB equations {associated with it}. Recall that $q=\frac{p}{p-1}$ {(see Remark~\ref{rem:31})} and the definition of ${\cal M}$ given in Section \ref{sec:dualform}. Introduce the measure $\Px^{q,a,h}$, $(a,h)\in{\cal M}$, specified by $\frac{d\Px^{q,a,h}}{d\Px}|_{{\mathcal{G}}_t}=\Gam_t^{q,a,h}$ for $t\in[0,T]$, where the density process is given by
\begin{align}\label{eq:Xbq}
\Gam_t^{q,a,h} &:=\exp\bigg\{-\int_0^tq\theta_s^{\top}d{W}_s-\frac{1}{2}\int_0^tq^2{\rm tr}[\theta_s\theta_s^{\top}]ds
-\int_0^t qa_s^{\top}d\bar{W}_s-\frac{1}{2}\int_0^tq^2{\rm tr}[a_sa_s^{\top}]ds\\
&\qquad + \sum_{i=1}^n\int_0^tq\log(1+{h}_{s}^i)dM_s^{i}
+\sum_{i=1}^n\int_0^{t\wedge\tau_i}\big[q\log(1+h_s^i)-(1+h_s^i)^q+1\big]\lambda_i(Y_s,H_s)ds\bigg\}.\nonumber
\end{align}
{We recall that $(\theta(\cdot,z),h(\cdot,z))\in\Theta_z$ for $z\in{\cal S}$, and $h_t=h(t,Y_{t-},H_{t-})$. Further, define
\begin{equation}
\Theta_{\theta}:=\{h(\cdot);\ (\theta(\cdot,z),h(\cdot,z))\in\Theta_z,\ z\in{\cal S}\}.
\label{eq:Ttheta}
\end{equation}
}
Let $\Ex^{q,a,h}$ denote the expectation under $\Px^{q,a,h}$. The following lemma proven in the Appendix gives an equivalent simplified representation of the dual problem \eqref{eq:ineL2}.
\begin{lemma}\label{lem:simplication-value}
It holds that
\begin{align}
&\inf_{(a,h)\in{\cal M},\kappa>0}{\it\Pi}(a,h,\kappa)\nonumber\\
 &\qquad= \esssup_{(a,h)\in{\cal M}}\Ex^{q,a,h}\left[K_1^{1-q}e^{\int_0^T\psi(s,a_s,h_s,Y_s,H_s)ds}+K_2^{1-q}\int_0^Te^{\int_0^t\psi(s,a_s,h_s,Y_s,H_s)ds}dt\right],
\label{eq:problem-p<0}
\end{align}
where $\Xi$ denotes ``$\sup$'' if $p<0$, while $\Xi$ denotes ``$\inf$'' if $p\in(0,1)$. Here, for $a=(a_i;\ i=1,\ldots,n)^{\top}\in\R^n$, $h=(h_i;\ i=1,\ldots,n)^{\top}\in(-1,\infty)^n$, and $(y,z)\in D\times{\cal S}$, the function
\begin{align}\label{eq:psi}
\psi(t,a,h,y,z) 
&:=\frac{q(q-1)}{2}\left\{{\rm tr}[(\theta\theta^{\top})(t,y,z)]+{\rm tr}[aa^{\top}]\right\}-q r\nonumber\\
&\quad+\sum_{i=1}^n(1-z_i)\big[(1+h_i)^q-q(1+h_i)+q-1\big]\lambda_i(y,z).
\end{align}
\end{lemma}

\subsection{HJB Equations of Dual Problem} \label{sec:beta}
This section derives and analyzes the HJB equation of {the stochastic control problem~\eqref{eq:problem-p<0}}. We will establish existence and uniqueness of a positive bounded classical solution for the HJB equation, and further prove the boundedness of its gradient. This will play a key role in establishing the admissibility of the optimal strategy in the verification theorem.

For given initial values $(Y_0,H_0)=(y,z)\in D\times{\cal S}$ of the stochastic factor and default state, and
{$t\in[0,T]$, define the value function corresponding to {the optimization problem}~\eqref{eq:problem-p<0}}:
\begin{equation}
g(t,y,z) := \esssup_{(a,h)\in{\cal M}}\Ex^{q,a,h}\left[K_1^{1-q}e^{\int_0^t\psi(s,a_s,h_s,Y_s,H_s)ds}+K_2^{1-q}\int_0^te^{\int_0^s\psi(u,a_u,h_u,Y_u,H_u)du}ds\right].
\label{eq:problem-p<022}
\end{equation}
We first characterize the dynamics of the state processes $(Y,H)$ under $\Px^{q,a,h}$.  Under $(\Px^{q,a,h},\Gx)$, it holds that
\begin{eqnarray}\label{eq:qtildeWi}
W_t^{q,\theta} := W_t + \int_0^t q\theta_sds,\ \ \ \ \ \bar{W}_t^{q,a} := \bar{W}_t + \int_0^t qa_sds,\ \ \ \ \ \ t\in[0,T]
\end{eqnarray}
is a $2n$-dimensional Brownian motion, and for $i=1,\ldots,n$,
\begin{eqnarray}\label{eq:qtildexii}
M_t^{q,h,i} :=
M_t^i -\int_0^{t\wedge\tau_i}\big[(1+h_s^i)^q-1\big]\lambda_i(Y_s,H_s)ds,\ \ \ \ \ t\in[0,T]
\end{eqnarray}
is a martingale. Therefore, using Eq.~\eqref{eq:Yt}, we may rewrite the dynamics of the stochastic factor process under $\Px^{q,a,h}$ as
\begin{eqnarray}\label{eq:qYi}
dY_t =\eta(a_t;t,Y_t,H_t)dt+\sigma_0(Y_t)\big[\rho dW_t^{q,\theta}+\sqrt{1-\rho^2}d\bar{W}_t^{q,a}\big].
\end{eqnarray}
In the above expression, for $(t,a,y,z)\in[0,T]\times\R^d\times D\times{\cal S}$, the drift vector is given by
\begin{eqnarray}\label{eq:drift-q}
\eta(a;t,y,z) :=\mu_0(y)-q\sigma_0(y)\big[\rho\theta(t,y,z)+\sqrt{1-\rho^2}a\big].
\end{eqnarray}

We derive the HJB equation for {the dual problem} using heuristic arguments first. We will then verify that the solution of the HJB equation satisfies regularity conditions and coincides with the value function of the dual problem. Using Eq.~\eqref{eq:problem-p<022} and the {dynamic} programming principle, if $g(t,y,z)$ is $C^1$ in $t$ and $C^2$ in $y$ for each state $z$, it satisfies the HJB equation given by
\begin{align}\label{eq:hjnp<0}
\frac{\partial g(t,y,z)}{\partial t} &= K_2^{1-q} + \frac{1}{2}{\rm tr}\big[\big(\sigma_0\sigma_0^{\top}D_y^2g\big)(t,y,z)\big]\nonumber\\
&\quad
+\esssup_{(a,h)\in{\cal B}\times\Theta_{\theta}}\Bigg\{g(t,y,z)\left(\frac{q(q-1)}{2}{\rm tr}[(\theta\theta^{\top})(t,y,z)]-q r\right)\nonumber\\
&\quad+\frac{q(q-1)}{2}g(t,y,z){\rm tr}[aa^{\top}]
+\eta(a;t,y,z)^{\top}D_yg(t,y,z)\nonumber\\
&\quad+g(t,y,z)\sum_{i=1}^n\big[(1+h_i)^q-q(1+h_i)+q-1\big](1-z_i)\lambda_i(y,z)\nonumber\\
&\quad+\sum_{i=1}^n\left[g(t,y,\bar{z}^i)-g(t,y,z)\right](1-z_i)(1+h_i)^q\lambda_i(y,z)\bigg\},
\end{align}
with initial condition $g(0,y,z)=K_1^{1-q}$ for all $(y,z)\in D\times{\cal S}$. In the above expression, for $i=1,\ldots,n$, we are using the notation
\begin{eqnarray}
\bar{z}^i:=(z_1,\ldots,z_{i-1},1-z_i,z_{i+1},\ldots,z_n),
\label{eq:zjdef}
\end{eqnarray}
i.e. $\bar{z}^i$ is obtained by flipping the $i$-th component of $z$. 
Using the first-order conditions for optimality, we obtain that $a$ satisfies the {linear equation} given by
\begin{eqnarray*}\label{eq:foc-a}
q(q-1)g(t,y,z)a-q\sqrt{1-\rho^2}\sigma_0^{\top}(y)D_yg(t,y,z)=0,
\end{eqnarray*}
leading to
\begin{equation}\label{eq:hatai}
\hat{a}=\hat{a}(t,y,z)=-\frac{\sqrt{1-\rho^2}}{1-q}\frac{\sigma_0^{\top}(y)D_yg(t,y,z)}{g(t,y,z)}.
\end{equation}
On the other hand, we set
\begin{align}\label{eq:hathi}
&\hat{h}=\hat{h}(t,y,z)\in \argXi_{h\in\Theta_{\theta}}\Bigg\{\frac{q(q-1)}{2}g(t,y,z){\rm tr}[(\theta\theta^{\top})(t,y,z)]+\sum_{i=1}^ng(t,y,\bar{z}^i)(1-z_i)(1+h_i)^q\lambda_i(y,z)\nonumber\\
&\qquad-q\rho(\sigma_0(y)\theta(t,y,z))^{\top}D_yg(t,y,z)+g(t,y,z)\sum_{i=1}^n\big[q-1-q(1+h_i)\big](1-z_i)\lambda_i(y,z)\bigg\},
\end{align}
where ${\rm arg}\Xi$ denotes ``${\rm argmax}$'' if $p<0$, while ${\rm arg}\Xi$ denotes ``${\rm argmin}$'' if $p\in(0,1)$. Notice that for any $\theta\in{\cal C}$, the set $\Theta_{\theta}$ defined in Eq.~\eqref{eq:Ttheta} is a singleton $\{\hat{h}\}$. By the definition of $\Theta_z$ given in Assumption ({\bf A3}), $\hat{h}\in C_b^{0,1}$ and takes values in the interval $(-1+\varepsilon,\infty)^n$ for some $\varepsilon>0$.

Plugging the above expressions for $\hat{a}_i$, $i=1,\ldots,n$, into the master HJB equation~\eqref{eq:hjnp<0}, we obtain the following nonlinear PDE:
\begin{align}\label{eq:hjn2p<0}
\frac{\partial g(t,y,z)}{\partial t} &= K_2^{1-q} + \frac{1}{2}{\rm tr}\big[\big(\sigma_0\sigma_0^{\top}D_y^2g\big)(t,y,z)\big]
+\nu(t,y,z)^{\top}D_yg(t,y,z)+g(t,y,z)\varphi(t,y,z)\nonumber\\
&\quad+\frac{q(1-\rho^2)}{2(1-q)}g^{-1}(t,y,z)\left\|\sigma_0^{\top}(y)D_yg(t,y,z)\right\|^2\\
&\quad+\sum_{i=1}^n g(t,y,\bar{z}^i)(1-z_i)(1+\hat{h}_i(t,y,z))^q\lambda_i(y,z),\nonumber
\end{align}
with initial condition $g(0,y,z)=K_1^{1-q}$. For $(t,y,z)\in [0,T]\times D\times{\cal S}$, the coefficients in the above equation are defined by
\begin{align}\label{eq:drifts}
\nu(t,y,z) &:= \mu_0(y) - q\rho\sigma_0(y)\theta(t,y,z),\\\notag
\varphi(t,y,z)&:=\frac{q(q-1)}{2}{\rm tr}[(\theta\theta^{\top})(t,y,z)]-q r+\sum_{i=1}^n[q-1-q(1+\hat{h}_i(t,y,z))](1-z_i)\lambda_i(y,z).
\end{align}

\begin{remark}
Eq.~\eqref{eq:hjn2p<0} defines a recursive system of nonlinear PDEs, in which the dependence arises because of systemic effects captured by the term
$\sum_{i=1}^n g(t,y,\bar{z}^i)(1-z_i)(1+\hat{h}_i(t,y,z))^q\lambda_i(y,z)$. Consider first the situation when all $n$ stocks are defaulted, i.e.,
the default state is $z=1$. Then, the above term becomes zero, consistently with the intuition that no contagion is present if all
stocks are already defaulted. Next, consider the situation when ${z} = (\underbrace{0,0,\ldots 0}_{k \; terms}, \underbrace{1,1,\ldots,1}_{n-k \; terms})$,
i.e. the first $k$ stocks are alive and the remaining $n-k$ are defaulted. The above term will then become $\sum_{i=1}^k g(t,y,\bar{z}^i)(1+\hat{h}_i(t,y,z))^q\lambda_i(y,z)$.
Hence, the solution to the PDE in Eq.~\eqref{eq:hjn2p<0} depends on the set $\{g(t,y,\bar{z}^i)\}_{i=1}^k$. Such a set consists of solutions to {$k$} PDEs of the form~\eqref{eq:hjn2p<0}, but associated with the default state ${\bar{z}}^i$, $i=1,\ldots,k$, in which the $i$-th stock defaults. This reflects the fact that the investor needs to account for the optimal expected utilities achievable in any state in which an additional stock defaults when determining his strategy. This observation will also guide the analysis of Eq.~\eqref{eq:hjn2p<0} in the following sections.
\end{remark}

Let us analyze the structure of Eq.~\eqref{eq:hjn2p<0}. It is readily seen that it contains a term given by the product of the quadratic gradient and the reciprocal of the solution. In addition, the solution of the PDE associated with the augmented default state $\bar{z}^i$ appears nonlinearly. Our semi-linear PDE exhibits a similar structure to that analyzed by \cite{Ludow} in the context of insurance. In their case, the exponential nonlinearity in the PDE comes from mortality risk.

Our first step is to transform it into a semi-linear PDE, getting rid off the term given by the product of the quadratic gradient and the reciprocal of the solution. This is achieved by adopting the power transform developed by \cite{Thaleia01} which allows us to reduce Eq.~\eqref{eq:hjn2p<0} to a recursive system of nonlinear PDEs without quadratic gradient. Concretely, we look for a solution of Eq.~\eqref{eq:hjn2p<0} of the form $g(t,y,z)=f^{\beta}(t,y,z)$, where $\beta\neq0$ is a free parameter which needs to be determined. Under this ansatz,
 $\frac{\partial g}{\partial t}=\beta f^{\beta-1}\frac{\partial f}{\partial t}$, $D_yg=\beta f^{\beta-1}D_yf$, and $D_y^2g=\beta(\beta-1)f^{\beta-2}D_yf(D_yf)^{\top}+\beta f^{\beta-1}D_y^2f$.
 Plugging the above derivatives expressions into \eqref{eq:hjn2p<0}, we get
\begin{align*}
\frac{\partial f(t,y,z)}{\partial t} &= K_2^{1-q}\beta^{-1}f^{1-\beta}(t,y,z) + \frac{1}{2}{\rm tr}\big[\big(\sigma_0\sigma_0^{\top}D_y^2{f}\big)(t,y,z)\big]
+\nu(t,y,z)^{\top}D_yf(t,y,z)\nonumber\\
&\quad+\beta^{-1}\varphi(t,y,z)f(t,y,z)
+\left[\frac{q(1-\rho^2)}{2(1-q)}\beta+\frac{1}{2}(\beta-1)\right]f^{-1}(t,y,z)\left\|\sigma_0^{\top}(y)D_yf(t,y,z)\right\|^2\\
&\quad+\beta^{-1}f^{1-\beta}(t,y,z)\sum_{i=1}^n f^{\beta}(t,y,\bar{z}^i)(1-z_i)(1+\hat{h}_i(t,y,z))^q\lambda_i(y,z).\nonumber
\end{align*}
Choosing {the parameter}
$$\beta=\frac{1-q}{1-q\rho^2},$$
then $\frac{q(1-\rho^2)}{2(1-q)}\beta+\frac{1}{2}(\beta-1)$=0, yielding the following recursive system of semi-linear PDEs
\begin{align}\label{eq:hjn2pf2}
\frac{\partial f(t,y,z)}{\partial t} &= \frac{1}{2}{\rm tr}\big[\big(\sigma_0\sigma_0^{\top}D_y^2f\big)(t,y,z)\big]
+\nu(t,y,z)^{\top}D_yf(t,y,z)+\beta^{-1}\varphi(t,y,z)f(t,y,z)\nonumber\\
&\quad+\beta^{-1}f^{1-\beta}(t,y,z)\left[K_2^{1-q}+\sum_{i=1}^n f^{\beta}(t,y,\bar{z}^i)(1-z_i)(1+\hat{h}_i(t,y,z))^q\lambda_i(y,z)\right]
\end{align}
with initial condition $f(0,y,z)=K_1^{\frac{1-q}{\beta}}=K_1^{\frac{1}{1-q\rho^2}}$ for all $(y,z)\in D\times{\cal S}$. {Eq.~\eqref{eq:hjn2pf2} still contains nonlinear terms which are non-Lipschitz continuous. This prevents the application of standard existence and uniqueness results of global classic solutions to semi-linear PDEs.
{Our approach is to study the solvability of Eq.~\eqref{eq:hjn2pf2} by first developing a truncation technique which reduces the above system of equations to semi-linear PDEs belonging to a class investigated by \cite{BechererSchweizer}. We then establish the equivalence between the solutions of the transformed system and those of the original system using probabilistic representations of classical solutions.}


Before proceeding to analyze the system \eqref{eq:hjn2pf2} of PDEs, we need an auxiliary result which guarantees existence and uniqueness of the underlying state factor process. This is given in the following lemma proven in the Appendix.


\begin{lemma}\label{lem:sde-hatY}
Let $z\in{\cal S}$. Consider the following $m$-dimensional SDE given by
\begin{align}\label{eq:barY}
d\hat{Y}_t^z = \nu(t,\hat{Y}_t^z,z)dt + \sigma_0(\hat{Y}_t^z)dW_t,
\end{align}
where the $\R^m$-valued drift term $\nu$ has been defined in Eq.~\eqref{eq:drifts}, and the $n$-dimensional Brownian motion $W$ under $\Px$ has been introduced in Section~\ref{sec:model}.
For each $(t,y)\in[0,T]\times D$, let $\hat{Y}^{t,y,z}=(\hat{Y}^{t,y,z}_s)_{s\in[t,T]}$ be a strong solution of Eq.~\eqref{eq:barY} with initial value $y$ at time $t$. Under assumptions $({\bf A1})$-$({\bf A3})$, $\hat{Y}^{t,y,z}$ exists and is unique for each $y\in D$. Further, the solution does not leave $D$ before $T$ for each $z\in{\cal S}$, i.e. $\Px(\hat{Y}_s^{t,y,z}\in D,\ \forall\ s\in[t,T])=1$.
\end{lemma}

Recall that $\hat{h}\in\Theta_{\theta}$ with $\theta\in{\cal C}$, and the assumption {({\bf A3})}.
Define the following default-state dependent constants:
\begin{align*}
\bar{m}^{\theta}(z)&:=\sum_{j=1}^n\left\|\theta_j(\cdot,z)\right\|^2_{\infty},\ \ \bar{m}^{\lambda}_i(z):=(1-z_i)\left\|1+\hat{h}_i(\cdot,z)\right\|_{\infty},\quad \bar{m}^{\hat{h}}_i(z):=(1-z_i)\left\|\hat{h}_i(\cdot,z)\right\|_{\infty},
\end{align*}
where $z\in{\cal S}$ and $\|l\|_{\infty}:=\sup_{(t,y)\in[0,T]\times D}|l(t,y)|$ for any bounded continuous function $l(t,y)$ defined on $(t,y)\in[0,T]\times D$. Using that $q<1$ for $(t,y,z)\in [0,T]\times D\times{\cal S}$, it holds that
\begin{align}\label{eq:verphi-bound}
\varphi(t,y,z)\left\{
  \begin{array}{ll}
    <0, & {\rm if}\ q\in(0,1),\\
    >-[\frac{q(1-q)}{2}\bar{m}^{\theta}(z)+qr+(1-q)\sum_{i=1}^n\bar{m}_i^{\lambda}(z)+q\sum_{i=1}^n\bar{m}_i^{\lambda}(z)\bar{m}_i^{\hat{h}}(z)],   & \\ \\
    <\frac{q(q-1)}{2}\bar{m}^{\theta}(z)-qr-q\sum_{i=1}^n\bar{m}_i^{\lambda}(z)\bar{m}_i^{\hat{h}}(z), & {\rm if}\ q<0.\\
    >-(1-q)\sum_{i=1}^n\bar{m}_i^{\lambda}(z).
  \end{array}
\right.
\end{align}
Next, we rewrite Eq.~\eqref{eq:hjn2pf2} in a more convenient form for the analysis. To this purpose, define the $\Px$-infinitesimal generator ${\cal A}_z$ of the diffusion process $\hat{Y}^z$ satisfying the SDE~\eqref{eq:barY}. It depends on the default state $z\in{\cal S}$ and acts on any {$C^2$-function $\varpi$ defined on $D$} as
\begin{align*}
{\cal A}_{t,z}\varpi(y) :=\frac{1}{2}{\rm tr}\big[(\sigma_0\sigma_0^{\top}D_y^2\varpi)(y)\big]+\nu(t,y,z)^{\top}D_y\varpi(y),\ \ \ \ \ y\in D.
\end{align*}
We may then rewrite Eq.~\eqref{eq:hjn2pf2} as
\begin{eqnarray}\label{eq:hjn2pf3}
\frac{\partial f(t,y,z)}{\partial t} = {\cal A}_{t,z}f(t,y,z)+\beta^{-1}\varphi(t,y,z)f(t,y,z)+{\it\Phi}(t,y,f(t,y,z),z),
\end{eqnarray}
where, for $(t,y,v,z)\in[0,T]\times D\times\R_+\times{\cal S}$, the solution-dependent nonlinear term is
\begin{align}\label{eq:Phi}
{\it\Phi}(t,y,v,z):=\beta^{-1}v^{1-\beta}\left(K_2^{1-q}+\sum_{i=1}^n f^{\beta}(t,y,\bar{z}^i)(1-z_i)(1+\hat{h}_i(t,y,z))^q\lambda_i(y,z)\right).
\end{align}

The following theorem establishes existence and uniqueness of a bounded global classical solution of the interlinked system \eqref{eq:hjn2pf2} of semi-linear PDEs. Notice that the linkage is through the {nonlinear} term ${\it\Phi}(t,y,v,z)$ and more specifically via the dependence on the default variable $\bar{z}^i$.

\begin{theorem}\label{thmhjb-p<0}
Under assumptions $({\bf A1})$-$({\bf A3})$, the recursive system of semi-linear PDEs \eqref{eq:hjn2pf2} admits a unique classical solution $f(t,y,z)$, which is $C^1$ in $t$, and $C^2$ in $y$, for each default state $z\in{\cal S}$. Moreover, there exist two positive default-state-dependent constants $0<\underline{K}(z)<\bar{K}(z)<+\infty$ such that $f(t,y,z)\in[\underline{K}(z),\bar{K}(z)]$ for all $(t,y,z)\in[0,T]\times D\times{\cal S}$.
\end{theorem}

\noindent{\it Proof.}\quad
Our proof strategy is based on a recursive procedure, starting from the default state $z=1$ (i.e., all stocks are defaulted) and proceeding backward to the default state $z=0$ (i.e., all stocks are alive). {We use $0$ to denote the row vector consisting of all zero entries}. We also introduce a general default state representation
$z=0^{j_1,\ldots,j_k}$ for indices $j_1\neq\cdots\neq j_k$ belonging to $\{1,\ldots,n\}$, and $k\in\{0,1,\ldots,n\}$. Such a vector $z$ is obtained by setting the entries
$j_1,j_2,\ldots,j_k$ of the zero vector to one, i.e. $z_{j_1}=\cdots=z_{j_k}=1$, and $z_{j}=0$ for $j\notin\{j_1,\ldots,j_k\}$ (if $k=0$, we set $z=0^{j_1,\ldots,j_k}=0$). Clearly $0^{j_1,\ldots,j_n}=e_n^{\top}$, where we recall that {$e_n$ denotes the $n$-dimensional column vector consisting of all entries equal to $1$.}
\begin{itemize}
  \item[{(I)}] The case $k=n$, i.e., $z=e_n^{\top}$, and all stocks are defaulted. In this case, using the expression \eqref{eq:Phi}, we have that
  ${\it\Phi}(t,y,v,e_n^{\top})=K_2^{1-q}\beta^{-1}v^{1-\beta}$. Specializing Eq.~\eqref{eq:hjn2pf3} to this default state, we obtain
      \begin{eqnarray}\label{eq:hjn2pf4}
\frac{\partial f(t,y,e_n^{\top})}{\partial t} = {\cal A}_{t,e_n^{\top}}f(t,y,e_n^{\top})+\beta^{-1}\varphi(t,y,e_n^{\top})f(t,y,e_n^{\top})+K_2^{1-q}\beta^{-1}f^{1-\beta}(t,y,e_n^{\top})
\end{eqnarray}
with initial condition $f(0,y,e_n^{\top})=K_1^{\frac{1}{1-q\rho^2}}$ for all $y\in D$. Using the estimates~\eqref{eq:verphi-bound} and setting $z=e_n^{\top}$, we can prove that Eq.~\eqref{eq:hjn2pf4} admits a unique bounded classical solution. The detailed proof of this fact will be presented below for the case of a general default state representation.
\item[{(II)}] The case $0\leq k\leq n-1$, i.e., $z=0^{j_1,\ldots,j_k}$ and stocks $j_1,\ldots,j_k$ are defaulted. Set the solution $f_{j_1,\ldots,j_k}(t,y):=f(t,y,0^{j_1,\ldots,j_k})$, the operator ${\cal A}_{t,j_1,\ldots,j_k}:={\cal A}_{t,0^{j_1,\ldots,j_k}}$, the functions $\varphi_{j_1,\ldots,j_k}(t,y):=\varphi(t,y,0^{j_1,\ldots,j_k})$, $\lambda_{j;j_1,\ldots,j_k}(y):=\lambda_j(y,0^{j_1,\ldots,j_k})$ and ${\it\Phi}_{j_1,\ldots,j_k}(t,y,v):={\it\Phi}(t,y,v,0^{j_1,\ldots,j_k})$. Then $f_{j_1,\ldots,j_k}(t,y)$ solves the following semi-linear PDE:
\begin{eqnarray}\label{eq:hjn2pf5}
\frac{\partial V(t,y)}{\partial t} = {\cal A}_{t,j_1,\ldots,j_k}V(t,y)+\beta^{-1}\varphi_{j_1,\ldots,j_k}(t,y)V(t,y)+{\it\Phi}_{j_1,\ldots,j_k}(t,y,V(t,y)),
\end{eqnarray}
where, for $(t,y,v)\in[0,T]\times D\times\R_+$,
\begin{align}\label{eq:Phi6}
{\it\Phi}_{j_1,\ldots,j_k}(t,y,v)&=\beta^{-1}v^{1-\beta}\\
&\quad\times\left(K_2^{1-q}+\sum_{j\notin\{j_1,\ldots,j_k\}} f_{j_1,\ldots,j_k,j}^{\beta}(t,y)(1+\hat{h}_{j;j_1,\ldots,j_k}(t,y))^q\lambda_{j;j_1,\ldots,j_k}(y)\right).\nonumber
\end{align}
From the structural form of Eq.~\eqref{eq:hjn2pf5}, it can be seen that the PDE satisfied by $f_{j_1,\ldots,j_k}(t,y)$ also depends on the sequence of solutions
$(f_{j_1,\ldots,j_k,j}(t,y))_{j\notin\{j_1,\ldots,j_k\}}$ of Eq.~\eqref{eq:hjn2pf2} associated with default states $z=0^{j_1,\ldots,j_k,j}$, for all $j\notin\{j_1,\ldots,j_k\}$.
This establishes a backward recursive relation between the solutions through default states $z\in{\cal S}$ and guides us to solve Eq.~\eqref{eq:hjn2pf2} in an iterative backward fashion.
Concretely, we proceed inductively and first assume that for all $j\notin\{j_1,\ldots,j_k\}$, Eq.~\eqref{eq:hjn2pf2} admits a unique bounded classical solution
$f_{j_1,\ldots,j_k,j}(t,y)\in[\underline{K}_{j},\bar{K}_j]$ when the default state is $z=0^{j_1,\ldots,j_k,j}$,
where $0<\underline{K}_{j}<\bar{K}_{j}<+\infty$. Then, we prove that Eq.~\eqref{eq:hjn2pf5} has a unique bounded classical solution $V(t,y)=f_{j_1,\ldots,j_k}(t,y)$ when
$z=0^{j_1,\ldots,j_k}$. {This proof will be based on a suitably designed truncation technique along with probabilistic Feynman-Kac's representations}.
Since we are able to show that for the base inductive case $z=e_n^{\top}$ there is a unique bounded classical solution, the inductive proof is complete.
\end{itemize}

In the sequel, we rigorously implement the steps just discussed. We start with (II). Let $\underline{m}_{j_1,\ldots,j_k}^{\varphi,q}$ and $\bar{m}_{j_1,\ldots,j_k}^{\varphi,q}$ denote the lower and upper bounds obtained in Eq.~\eqref{eq:verphi-bound} when the default state is $z=0^{j_1,\ldots,j_k}$. Define
\begin{align}\label{eq:underK}
\underline{K}_{j_1,\ldots,j_k}&:= K_1^{\frac{1}{1-q\rho^2}}\exp\left\{\beta^{-1}T\big(\underline{m}_{j_1,\ldots,j_k}^{\varphi,q}\wedge0\big)\right\},\ \ \ \ \ell(t):=\exp\left\{-\beta^{-1}\bar{m}_{j_1,\ldots,j_k}^{\varphi,q}t\right\},\nonumber\\
\bar{K}_{j_1,\ldots,j_k}(t)&:=\ell(T-t)\exp\left\{{\it\Theta}\big(\underline{K}_{j_1,\ldots,j_k}\big)t\right\},\ \ \ \ \ \ \ \ \ t\in[0,T],
\end{align}
where, for $x>0$, the positive function
\begin{align}
{\it\Theta}(x) := \beta^{-1}x^{-\beta}\left[K_2^{1-q}+\sum_{j\notin\{j_1,\ldots,j_k\}} \bar{K}_{j}^{\beta}\left\|(1+\hat{h}_{j;j_1,\ldots,j_k})^q\right\|_{\infty}\left\|\lambda_{j;j_1,\ldots,j_k}\right\|_{\infty}\right].
\label{eq:Thetafun}
\end{align}
Notice that by the assumption ({\bf A3}), we have that $\Theta_{\theta}\subseteq C_b^{0,1}$ and hence $\hat{h}(\cdot,z)\in C_b^{0,1}$ with $z\in{\cal S}$. Then $\left\|(1+\hat{h}_{j;j_1,\ldots,j_k})^q\right\|_{\infty}\leq \left\|1+h_{j;j_1,\ldots,j_k}\right\|^q_{\infty}$ if $q\in(0,1)$ and  $\left\|(1+\hat{h}_{j;j_1,\ldots,j_k})^q\right\|_{\infty}\leq \varepsilon^{q}$ if $q<0$ for some $\varepsilon>0$.
Apply the following truncation to the function ${\it\Phi}_{j_1,\ldots,j_k}$ defined in \eqref{eq:Phi6}: for $(t,y,v)\in[0,T]\times D\times\R$,
\begin{eqnarray}\label{eq:H1K}
{\it\Phi}_{j_1,\ldots,j_k}^{(K)}(t,y,v) := {\it\Phi}_{j_1,\ldots,j_k}\big(t,y,(\underline{K}_{j_1,\ldots,j_k}\vee v)\wedge\bar{K}_{j_1,\ldots,j_k}(t)\big).
\end{eqnarray}
Using~\eqref{eq:Phi6}, and recalling that $\beta=\frac{1-q}{1-q\rho^2}>0$ (since $\rho\in(-1,1)$ and $q<1$), we obtain
\begin{align}\label{eq:bounfPhiK}
0<&{\it\Phi}_{j_1,\ldots,j_k}^{(K)}(t,y,v)\leq K_2^{1-q}\beta^{-1}\big[(\underline{K}_{j_1,\ldots,j_k}\vee v)\wedge\bar{K}_{j_1,\ldots,j_k}(t)\big]^{1-\beta}\nonumber\\
 &+\beta^{-1}\big[(\underline{K}_{j_1,\ldots,j_k}\vee v)\wedge\bar{K}_{j_1,\ldots,j_k}(t)\big]^{1-\beta}\sum_{j\notin\{j_1,\ldots,j_k\}}
 \bar{K}_{j}^{\beta}\left\|(1+\hat{h}_{j;j_1,\ldots,j_k})^q\right\|_{\infty}\left\|\lambda_{j;j_1,\ldots,j_k}\right\|_{\infty}\nonumber\\
&\leq {\it\Theta}\big(\underline{K}_{j_1,\ldots,j_k}\big)\bar{K}_{j_1,\ldots,j_k}(t).
\end{align}

Consider the following PDE with terminal condition $\bar{V}^{(K)}(T,y)= K_1^{\frac{1}{1-q\rho^2}}$ for $y\in D$, and on $(t,y)\in[0,T)\times D$,
\begin{align}\label{eq:fhjn2p1KT-t}
0&=\frac{\partial \bar{V}^{(K)}(t,y)}{\partial t} + {\cal A}_{t,j_1,\ldots,j_k}\bar{V}^{(K)}(t,y)+\beta^{-1}\varphi_{j_1,\ldots,j_k}(t,y)\bar{V}^{(K)}(t,y)\nonumber\\
&\quad+{\it\Phi}_{j_1,\ldots,j_k}^{(K)}\big(T-t,y,\bar{V}^{(K)}(t,y)\big).
\end{align}
{By the inductive hypothesis,} $f_{j_1,\ldots,j_k,j}(t,y)\in[\underline{K}_{j},\bar{K}_j]$ is the classical solution of Eq.~\eqref{eq:hjn2pf2} at the default state
$z=0^{j_1,\ldots,j_k,j}$ with $j\notin\{j_1,\ldots,j_k\}$. Then, under assumptions ({\bf A1})-({\bf A3}) and recalling the bounded sub-domains $(D_{\ell})_{\ell\in\N}$ given in ({\bf A1}), we have that $\varphi_{j_1,\ldots,j_k}$ is $C^1$. Moreover, ${\it\Phi}_{j_1,\ldots,j_k}^{(K)}(t,y,v)$ is Lipschitz continuous on $[0,T]\times D_{\ell}\times\R$, and it is also Lipschitz in $v$ uniformly on $(t,y)\in[0,T]\times D$. Using Proposition 2.1 in \cite{BechererSchweizer}, we can conclude
that Eq.~\eqref{eq:fhjn2p1KT-t} admits a unique classical solution, which admits the following Feynman-Kac representation
\begin{align}\label{eq:lowerbarf}
\bar{V}^{(K)}(t,y) &=\Ex\left[K_1^{\frac{1}{1-q\rho^2}}\exp\left(\beta^{-1}\int_t^T\varphi_{j_1,\ldots,j_k}(s,\hat{Y}_s^{t,y,k})ds\right)\right]\\\notag
&\quad+\Ex\left[\int_t^T{\it\Phi}_{j_1,\ldots,j_k}^{(K)}\big(T-u,\hat{Y}_u^{t,y,k},\bar{V}^{(K)}(u,\hat{Y}_u^{t,y,k})\big)
\exp\left(\beta^{-1}\int_t^u\varphi_{j_1,\ldots,j_k}(s,\hat{Y}_s^{t,y,k})ds\right)du\right].
\end{align}
In the above expression, we are using the simplified notation $Y_s^{t,y,k}:=Y_s^{t,y,0^{j_1,\ldots,j_k}}$, $s\in[t,T]$, where $Y_s^{t,y,z}$ has been defined in Lemma~\ref{lem:sde-hatY}. Since $\beta>0$, using the bounds in Eq.~\eqref{eq:verphi-bound} we have that
\begin{align}\label{eq:lowerBound}
\bar{V}^{(K)}(t,y) &\geq\Ex\left[K_1^{\frac{1}{1-q\rho^2}}\exp\left(\beta^{-1}\int_t^T\varphi_{j_1,\ldots,j_k}(s,\hat{Y}_s^{t,y,k})ds\right)\right]\nonumber\\
&\geq \Ex\left[K_1^{\frac{R1}{1-q\rho^2}}\exp\big(\beta^{-1}\underline{m}_{j_1,\ldots,j_k}^{\varphi,q}(T-t)\big)\right]\geq \underline{K}_{j_1,\ldots,j_k}.
\end{align}
On the other hand, define the stopping time
\begin{eqnarray}\label{eq:stop-time}
\tau := \inf\big\{s\in[t,T];\ \bar{V}^{(K)}\big(s,\hat{Y}_s^{t,y,k}\big)\leq \bar{K}(T-s)\big\}.
\end{eqnarray}
Then, it obviously holds that $\bar{V}^{(K)}\big(\tau,\hat{Y}_{\tau}^{t,y,k}\big)\leq\bar{K}(T-\tau)$, $\Px$-a.s.. Moreover, using Eq.~\eqref{eq:bounfPhiK} and recalling
the quantity ${\it \Theta}(\cdot)$ defined in~\eqref{eq:Thetafun}, it holds that
${\it\Phi}_{j_1,\ldots,j_k}^{(K)}\big(T-s,\hat{Y}_s^{t,y,k},\bar{V}^{(K)}\big(s,\hat{Y}_s^{t,y,k}\big)\big) \leq {\it\Theta}(\underline{K}_{j_1,\ldots,j_k})\bar{K}(T-s)$ for all $s\in[t,T]$.
Using the Feynman-Kac representation, we obtain
\begin{align}\label{eq:upperbarf}
\bar{V}^{(K)}(t,y) &= \Ex\left[\bar{V}^{(K)}\big(\tau,\hat{Y}_{\tau}^{t,y,k}\big)\exp\left(\beta^{-1}\int_t^{\tau}\varphi_{j_1,\ldots,j_k}(s,\hat{Y}_s^{t,y,k})ds\right)\right]\nonumber\\
&\quad+\Ex\left[\int_t^{\tau} {\it\Phi}_{j_1,\ldots,j_k}^{(K)}\big(T-u,\hat{Y}_u^{t,y,k},\bar{V}^{(K)}(u,\hat{Y}_u^{t,y,k})\big)\exp\left(\beta^{-1}\int_t^u\varphi_{j_1,\ldots,j_k}(s,\hat{Y}_s^{s,y,k})ds\right) du\right] \nonumber\\
&\leq  \Ex\left[\bar{K}_{j_1,\ldots,j_k}(T-\tau)e^{\beta^{-1}\bar{m}_{j_1,\ldots,j_k}^{\varphi,q}(\tau-t)}
+{\it\Theta}(\underline{K}_{j_1,\ldots,j_k})\int_t^{\tau}\bar{K}_{j_1,\ldots,j_k}(T-u)e^{\beta^{-1}\bar{m}_{j_1,\ldots,j_k}^{\varphi,q}(u-t)}du\right]\nonumber\\
&=\bar{K}_{j_1,\ldots,j_k}(T-\tau)\ell(t-\tau)+{\it\Theta}(\underline{K}_{j_1,\ldots,j_k})\int_t^{\tau}\bar{K}_{j_1,\ldots,j_k}(T-u)\ell(t-u)du\nonumber\\
&=\ell(t)\exp\left\{{\it\Theta}\big(\underline{K}_{j_1,\ldots,j_k}\big)(T-t)\right\}=\bar{K}_{j_1,\ldots,j_k}(T-t).
\end{align}
Using~\eqref{eq:lowerBound} and \eqref{eq:upperbarf}, we obtain the following bounds for the solution $\bar{V}^{(K)}(t,y)$ given by
\begin{align}\label{eq:boundvarV}
\underline{K}_{j_1,\ldots,j_k}\leq\bar{V}^{(K)}(t,y)\leq \bar{K}_{j_1,\ldots,j_k}(T-t),\ \ \ \ \ \ \ \forall\ (t,y)\in[0,T]\times D.
\end{align}
Recall that $V(t,y)$ is the solution of Eq.~\eqref{eq:hjn2pf5}. It follows from \eqref{eq:H1K} and \eqref{eq:boundvarV} that ${\it\Phi}_{j_1,\ldots,j_k}^{(K)}(T-t,y,\bar{V}^{(K)}(t,y))={\it\Phi}_{j_1,\ldots,j_k}(T-t,y,\bar{V}^{(K)}(t,y))$ for all $(t,y)\in[0,T]\times D$. By the uniqueness of the classical solution to Eq.~\eqref{eq:fhjn2p1KT-t}, this yields $V(t,y)=\bar{V}^{(K)}(T-t,y)$. Thus we obtain existence and uniqueness of a classical solution
$V(t,y)$ of Eq.~\eqref{eq:hjn2pf5}. Moreover, using equations~\eqref{eq:underK} and \eqref{eq:boundvarV}, we also have that, for all $(t,y)\in[0,T]\times D$,
\begin{align}\label{eq:bounds-V}
\underline{K}_{j_1,\ldots,j_k}\leq V(t,y)\leq \bar{K}_{j_1,\ldots,j_k}(t)\leq \exp\big\{\big[{\it\Theta}(\underline{K}_{j_1,\ldots,j_k})-\beta^{-1}\big(\bar{m}_{j_1,\ldots,j_k}^{\varphi,q}\wedge0\big)\big]T\big\}.
\end{align}

We next turn to the proof of the case {(I)}. Using the above argument, it is enough to modify the definition of the constants given in Eq.~\eqref{eq:underK}.
Since $z=0^{j_1,\ldots,j_n}=e_n^{\top}$, we define
\begin{align}\label{eq:underK2}
\underline{K}_{j_1,\ldots,j_n}&:= K_1^{\frac{1}{1-q\rho^2}}\exp\left\{\beta^{-1}T\big(\underline{m}_{j_1,\ldots,j_n}^{\varphi,q}\wedge0\big)\right\},\ \ \ \ \ell(t):=\exp\left\{-\beta^{-1}\bar{m}_{j_1,\ldots,j_n}^{\varphi,q}t\right\},\nonumber\\
\bar{K}_{j_1,\ldots,j_n}(t)&:=\ell(T-t)\exp\big\{\beta^{-1}K_2^{1-q}\underline{K}_{j_1,\ldots,j_n}^{-\beta}t\big\},\ \ \ \ \ \ \ \ \ t\in[0,T].
\end{align}
Then we can conclude that Eq.~\eqref{eq:hjn2pf4} admits a unique classical solution $f(t,y,1)$, and further
\begin{align}\label{eq:bounds-V2}
\underline{K}_{j_1,\ldots,j_n}\leq f(t,y,1)\leq \bar{K}_{j_1,\ldots,j_n}(t)\leq \exp\big\{\beta^{-1}\big[K_2^{1-q}\underline{K}_{j_1,\ldots,j_n}^{-\beta}-\big(\bar{m}_{j_1,\ldots,j_n}^{\varphi,q}\wedge0\big)\big]T\big\}.
\end{align}
This completes the proof of the theorem. \hfill$\Box$


{Notice that the solution $\hat{a}$ of the system of first-order conditions given by~\eqref{eq:hatai} includes the gradient of the solution to the HJB equation. Hence, in order to prove that the strategy is admissible, we need to estimate the gradient $D_yf(t,y,z)$ of the bounded classical solution $f(t,y,z)$ of Eq.~\eqref{eq:hjn2pf2}.}
Such an estimate is given in the following
\begin{proposition}\label{prop:gradient-est}
Under assumptions {\rm ({\bf A1})-({\bf A3})}, there exists a finite constant $C=C(m,n,T)$ (it may depend on $m,n,T$) such that $\sup_{(t,y)\in[0,T]\times D}\|D_yf(t,y,z)\|\leq C$ for each default state $z\in{\cal S}$. We recall here that $\|x\|^2:=\sum_{i=1}^mx_i^2$ for $x\in\R^m$.
\end{proposition}

\noindent{\it Proof.}\quad We adopt a probabilistic method based on a Feynman-Kac representation of the classical solution $f(t,y,z)$ of Eq.~\eqref{eq:hjn2pf2}, for each default state $z\in{\cal S}$. Recall the strong solution $\hat{Y}_s^{t,y,z}$, $s\in[t,T]$, obtained in Lemma~\ref{lem:sde-hatY}. Set $\hat{Y}_t^{y,k}:=\hat{Y}_t^{0,y,z}$ for $t\in[0,T]$, when $z=0^{j_1,\ldots,j_k}$, and use $\hat{Y}_t^{i,y,k}$ to denote the $i$-th component of solution vector $\hat{Y}_t^{y,k}$ for $i=1,\ldots,m$. We next estimate
the {gradient} of $\hat{Y}_t^{i,y,z}$ w.r.t $y\in D$. To this purpose, let $\phi_t^{ij,z}:=\frac{\partial\hat{Y}_t^{i,y,z}}{\partial y_j}$ for
$i,j=1,\ldots,m$. Then, using Theorem 4.6.5 in \cite{Kun97}, pp. 173-174, for each state $z\in{\cal S}$, it satisfies
\begin{align}\label{eq:phiij}
\phi_t^{ij,z}=\delta_{ij} + \sum_{l=1}^m\int_0^t\frac{\partial\nu^i(s,\hat{Y}_s^{y,z},z)}{\partial y_l}\phi_s^{lj,z}ds +\sum_{l=1}^m\sum_{\ell=1}^n\int_0^t\frac{\partial\sigma_0^{i\ell}(\hat{Y}_s^{y,z})}{\partial y_l}\phi_s^{lj,z}dW_s^{\ell}.
\end{align}
In the sequel of the proof, we use $C=C(m,n,T)$ to denote a finite positive constant depending on $m$, $n$ and $T$ which may be different from line to line. Under assumptions ({\bf A2})-({\bf A3}), it holds that
\begin{align*}
\big|\phi_t^{ij,z}\big|&\leq \delta_{ij}+ C\int_0^t\sum_{l=1}^m\big|\phi_s^{lj,z}\big|ds+C\sum_{\ell=1}^n\int_0^t\sum_{l=1}^m\big|\phi_s^{lj,z}\big|dW_s^{\ell}.
\end{align*}
It follows from H\"older's inequality that $\Ex\big[\big|\phi_t^{ij,z}\big|^2\big] \leq C+ C\int_0^t\sum_{l=1}^m\Ex\big[\big|\phi_s^{lj,z}\big|^2\big]ds$. Then, from Gronwall's lemma, one gets
\begin{align}\label{eq:estimate-phi}
\sup_{t\in[0,T]}\sum_{i,j=1}^m\Ex\big[\big|\phi_t^{ij,z}\big|^2\big]\leq Ce^{CT}.
\end{align}

We next estimate the gradient $D_yf(t,y,z)$ iteratively in terms of the default state variable $z\in{\cal S}$. {As in the proof of Theorem~\ref{thmhjb-p<0}, we consider two cases}:
\begin{itemize}
  \item[{(I)}] $k=n$, i.e. all stocks are defaulted and $z=e_n^{\top}$. {In this case, using the Feynman-Kac's representations given by Eq.~\eqref{eq:lowerbarf} in the proof of
  Theorem~\ref{thmhjb-p<0}}, we have
  \begin{align*}
  f(t,y,e_n^{\top}) &=\Ex\bigg[K_1^{\frac{1}{1-q\rho^2}}e^{\int_0^t\beta^{-1}\varphi(s,\hat{Y}_s^{y,n},e_n^{\top})ds}\nonumber\\
              &\quad+\int_0^tK_2^{1-q}\beta^{-1}f^{1-\beta}(s,\hat{Y}_s^{y,n},e_n^{\top})e^{\int_0^s\beta^{-1}\varphi(u,\hat{Y}_u^{y,n},e_n^{\top})du}ds\bigg].
  \end{align*}
\end{itemize}
Then, for $j=1,\ldots,m$, by virtue of Eq.~\eqref{eq:verphi-bound} and the dominated convergence theorem, we have
\begin{align*}
&D_{y_j}f(t,y,1) =\Ex\left[K_1^{\frac{1}{1-q\rho^2}}e^{\int_0^t\beta^{-1}\varphi(s,\hat{Y}_s^{y,n},e_n^{\top})ds}\int_0^t\beta^{-1}
\sum_{i=1}^mD_{y_i}\varphi(s,\hat{Y}_s^{y,n},e_n^{\top})\phi_s^{ij,n}ds\right]\nonumber\\
        &\quad+\Ex\left[\int_0^tK_2^{1-q}\beta^{-1}(1-\beta)f^{-\beta}(s,\hat{Y}_s^{y,n},e_n^{\top})\sum_{i=1}^mD_{y_i}f(s,\hat{Y}_s^{y,n},e_n^{\top})
\phi_s^{ij,n}e^{\int_0^s\beta^{-1}\varphi(u,\hat{Y}_u^{y,n},e_n^{\top})du}ds\right]\nonumber\\
&\quad+\Ex\left[\int_0^tK_2^{1-q}\beta^{-2}f^{1-\beta}(s,\hat{Y}_s^{y,n},e_n^{\top})e^{\int_0^s\beta^{-1}\varphi(u,\hat{Y}_u^{y,n},e_n^{\top})du}\beta^{-1}\left(\int_0^s
\sum_{i=1}^mD_{y_i}\varphi(u,\hat{Y}_u^{y,n},e_n^{\top})\phi_u^{ij,n}du\right)ds\right].
\end{align*}
Above, $\phi_t^{ij,k}:=\phi_t^{ij,z}$ when $z=0^{j_1,\ldots,j_k}$ and $k\in\{0,1,\ldots,n\}$. Under assumptions ({\bf A2})-({\bf A3}), $D_y\varphi(\cdot,e_n^{\top})$ is bounded on $D$, {where we recall that $\varphi(y,z)$ has been defined in \eqref{eq:drifts}}. Using the estimate~\eqref{eq:verphi-bound}, under assumptions ({\bf A1})-({\bf A3}), using the boundedness of the classical solution $f$ proved in Theorem~\ref{thmhjb-p<0}, we deduce the existence of a constant $C>0$ so that
\begin{align*}
\left|D_{y_j}f(t,y,e_n^{\top})\right|&\leq C\int_0^t\sum_{i=1}^m\Ex\big[\big|\phi_s^{ij,n}\big|\big]ds+C\int_0^t\int_0^s\sum_{i=1}^m\Ex\big[\big|\phi_u^{ij,n}\big|\big]duds\nonumber\\
&\quad+C\int_0^t\sum_{i=1}^m\Ex\big[\big|D_{y_i}f(s,\hat{Y}_s^{y,n},e_n^{\top})\phi_s^{ij,n}\big|\big]ds.
\end{align*}
In the estimates below, we keep using $C=C(m,n,T)$ to denote a finite positive constant depending on $m$, $n$ and $T$ possibly different from line to line. Using H\"older's inequality and Jensen's inequality,
\begin{align*}
&\sum_{j=1}^m\left|D_{y_j}f(t,y,e_n^{\top})\right|^2\leq C\int_0^t\sum_{i,j=1}^m\Ex\big[\big|\phi_s^{ij,n}\big|^2\big]ds+C\int_0^t\sum_{i=1}^m\Ex\big[\big|D_{y_i}f(s,\hat{Y}_s^{y,n},e_n^{\top})\big|^2\big]
\sum_{j=1}^n\Ex\big[\big|\phi_s^{ij,n}\big|^2\big]ds\nonumber\\
&\quad\leq C\int_0^t\sum_{i,j=1}^m\Ex\big[\big|\phi_s^{ij,n}\big|^2\big]ds+C\int_0^t\sup_{(u,y)\in[0,s]\times D}\sum_{j=1}^m\big|D_{y_j}f(u,y,e_n^{\top})\big|^2
\sum_{i,j=1}^n\Ex\big[\big|\phi_s^{ij,n}\big|^2\big]ds\nonumber\\
&\quad\leq CT\sup_{t\in[0,T]}\sum_{i,j=1}^m\Ex\big[\big|\phi_t^{ij,n}\big|^2\big]+C\sup_{t\in[0,T]}\sum_{i,j=1}^m\Ex\big[\big|\phi_t^{ij,n}\big|^2\big]\int_0^t\sup_{(u,y)\in[0,s]\times D}\sum_{j=1}^m\big|D_{y_j}f(u,y,e_n^{\top})\big|^2ds.
\end{align*}
Then, from Gronwall's lemma, we obtain using \eqref{eq:estimate-phi} that
\begin{align}\label{eq:bound-derivative1}
\sup_{(t,y)\in[0,T]\times D}\big\|D_yf(t,y,e_n^{\top})\big\|^2\leq CT\sup_{t\in[0,T]}\sum_{i,j=1}^m\Ex\big[\big|\phi_t^{ij,n}\big|^2\big]\exp\bigg(CT\sup_{t\in[0,T]}\sum_{i,j=1}^m\Ex\big[\big|\phi_t^{ij,n}\big|^2\big]\bigg).
\end{align}
\begin{itemize}
\item[{(II)}] $0\leq k\leq n-1$, i.e., $z=0^{j_1,\ldots,j_k}$, and the stocks $j_1,\ldots,j_k$ are defaulted. In this case, {using the PDE representation in Eq.~\eqref{eq:hjn2pf5}, we obtain
the following Feynman-Kac's representation}
  \begin{align*}
  f_{j_1,\ldots,j_k}(t,y) &=\Ex\left[K_1^{\frac{1}{1-q\rho^2}}e^{\int_0^t\beta^{-1}\varphi_{j_1,\ldots,j_k}(s,\hat{Y}_s^{y,k})ds}\right]\nonumber\\
              &\quad+\Ex\left[\int_0^t{\it\Phi}_{j_1,\ldots,j_k}(t,\hat{Y}_s^{y,k},f_{j_1,\ldots,j_k}(s,\hat{Y}_s^{y,k}))
e^{\int_0^s\beta^{-1}\varphi_{j_1,\ldots,j_k}(u,\hat{Y}_u^{y,k})du}ds\right],
  \end{align*}
\end{itemize}
where the function ${\it\Phi}_{j_1,\ldots,j_k}$ has been defined in Eq.~\eqref{eq:Phi6}. Then, for $j=1,\ldots,m$, we obtain using the estimate~\eqref{eq:verphi-bound} and the dominated convergence theorem that
\begin{align}\label{eq:dyjk}
&D_{y_j}f_{j_1,\ldots,j_k}(t,y)=\Ex\left[K_1^{\frac{1}{1-q\rho^2}}e^{\int_0^t\beta^{-1}\varphi_{j_1,\ldots,j_k}(s,\hat{Y}_s^{y,k})ds}\int_0^t\beta^{-1}
\sum_{i=1}^mD_{y_i}\varphi_{j_1,\ldots,j_k}(s,\hat{Y}_s^{y,k})\phi_s^{ij,k}ds\right]\nonumber\\
&\quad+\Ex\Bigg[\int_0^t{\it\Phi}_{j_1,\ldots,j_k}(t,\hat{Y}_s^{y,k},f_{j_1,\ldots,j_k}(s,\hat{Y}_s^{y,k}))\nonumber\\
&\qquad\qquad\times\beta^{-1}\left(\int_0^s
\sum_{i=1}^mD_{y_i}\varphi_{j_1,\ldots,j_k}(u,\hat{Y}_u^{y,k})\phi_u^{ij,k}du\right)
e^{\int_0^s\beta^{-1}\varphi_{j_1,\ldots,j_k}(u,\hat{Y}_u^{y,k})du}ds\Bigg]\nonumber\\
&\quad+\Ex\left[\int_0^t\sum_{i=1}^mD_{y_i}{\it\Phi}_{j_1,\ldots,j_k}(t,\hat{Y}_s^{y,k},f_{j_1,\ldots,j_k}(s,\hat{Y}_s^{y,k}))\phi_s^{ij,k}
e^{\int_0^s\beta^{-1}\varphi_{j_1,\ldots,j_k}(u,\hat{Y}_u^{y,k})du}ds\right]\\
&\quad+\Ex\left[\int_0^tD_v{\it\Phi}_{j_1,\ldots,j_k}(t,\hat{Y}_s^{y,k},f_{j_1,\ldots,j_k}(s,\hat{Y}_s^{y,k}))\sum_{i=1}^mD_{y_i}f_{j_1,\ldots,j_k}(s,\hat{Y}_s^{y,k})
\phi_s^{ij,k}
e^{\int_0^s\beta^{-1}\varphi_{j_1,\ldots,j_k}(u,\hat{Y}_u^{y,k})du}ds\right].\nonumber
\end{align}
Using the expression for ${\it\Phi}$ given in~\eqref{eq:Phi6}, we can compute
\begin{align}\label{eq:derivPhi7}
D_{y_i}{\it\Phi}_{j_1,\ldots,j_k}(t,y,v)&=\beta^{-1}v^{1-\beta}\Bigg[\sum_{j\notin\{j_1,\ldots,j_k\}} \beta f_{j_1,\ldots,j_k,j}^{\beta-1}(t,y)D_{y_i}f_{j_1,\ldots,j_k,j}(t,y)\nonumber\\
&\qquad\qquad\qquad\times\big(1+\hat{h}_{j;j_1,\ldots,j_k}(t,y)\big)^q\lambda_{j;j_1,\ldots,j_k}(y)\nonumber\\
&\quad+\sum_{j\notin\{j_1,\ldots,j_k\}}f_{j_1,\ldots,j_k,j}^{\beta}(t,y)D_{y_i}[\big(1+\hat{h}_{j;j_1,\ldots,j_k}(t,y)\big)^q\lambda_{j;j_1,\ldots,j_k}(y)]\Bigg],\ \ {\rm and}\nonumber\\
D_v{\it\Phi}_{j_1,\ldots,j_k}(t,y,v)&=\beta^{-1}(1-\beta)v^{-\beta}\\
&\quad\left(K_2^{1-q}+\sum_{j\notin\{j_1,\ldots,j_k\}} f_{j_1,\ldots,j_k,j}^{\beta}(t,y)\big(1+\hat{h}_{j;j_1,\ldots,j_k}(t,y)\big)^q\lambda_{j;j_1,\ldots,j_k}(y)\right).\nonumber
\end{align}
Eq.~\eqref{eq:dyjk} along with the expressions for the derivatives given in Eq.~\eqref{eq:derivPhi7} indicate that there is also a recursive dependence between the derivatives, i.e. the term $D_yf_{j_1,\ldots,j_k}$ depends on $D_yf_{j_1,\ldots,j_k,j}$ for $j\notin\{j_1,\ldots,j_k\}$.
The analysis in {(I)} indicates that the gradient $D_yf$ is bounded when all stocks are defaulted. Next, we proceed by induction and assume that
$\sup_{(t,y)\in[0,T]\times D}\|D_yf_{j_1,\ldots,j_k,j}(t,y)\|<+\infty$ for $j\notin\{j_1,\ldots,j_k\}$. Then we want to prove that $\sup_{(t,y)\in[0,T]\times D}\|D_yf_{j_1,\ldots,j_k}(t,y)\|<+\infty$.
First, notice that $\hat{h}_{j_1,\ldots,j_k}(\cdot)\in{\Theta}_{\theta}\subseteq C_b^{0,1}$. By the assumption ({\bf A3}), this implies that $\left\|(1+\hat{h}_{j;j_1,\ldots,j_k})^q\right\|_{\infty}\leq \left\|1+\hat{h}_{j;j_1,\ldots,j_k}\right\|^q_{\infty}$ if $q\in(0,1)$ and $\left\|(1+\hat{h}_{j;j_1,\ldots,j_k})^q\right\|_{\infty}\leq \varepsilon^{q}$ if $q<0$ for some $\varepsilon>0$. Also using the assumption ({\bf A3}), the estimates given in Eq.~\eqref{eq:verphi-bound} under assumptions ({\bf A2})-({\bf A3}) and the boundedness of the classic solution $f$ proved in Theorem~\ref{thmhjb-p<0}, it holds that ${\it\Phi}_{j_1,\ldots,j_k}(t,y,f_{j_1,\ldots,j_k}(t,y))$, $\|D_{y}{\it\Phi}_{j_1,\ldots,j_k}(t,y,f_{j_1,\ldots,j_k}(t,y))\|$ and $D_{v}{\it\Phi}_{j_1,\ldots,j_k}(t,y,f_{j_1,\ldots,j_k}(t,y))$ are all bounded. Thus for $j=1,\ldots,m$, we have
\begin{align*}
\big|D_{y_j}f_{j_1,\ldots,j_k}(t,y)\big|&\leq C\int_0^t\sum_{i=1}^m\Ex\big[\big|\phi_s^{ij,k}\big|\big]ds+C\int_0^t\int_0^s\sum_{i=1}^m\Ex\big[\big|\phi_u^{ij,k}\big|\big]duds\nonumber\\
&\quad+C\int_0^t\sum_{i=1}^m\Ex\big[\big|D_{y_i}f_{j_1,\ldots,j_k}(s,\hat{Y}_s^{y,k})\phi_s^{ij,k}\big|\big]ds.
\end{align*}
Using H\"older's inequality and Jensen's inequality, it follows that
\begin{align*}
&\big\|D_{y}f_{j_1,\ldots,j_k}(t,y)\big\|^2\leq CT\sup_{t\in[0,T]}\sum_{i,j=1}^m\Ex\big[\big|\phi_t^{ij,n}\big|^2\big]
+C\int_0^t\Ex\big[\big\|D_{y}f_{j_1,\ldots,j_k}(s,\hat{Y}_s^{y,k})\big\|^2\big]\sum_{i,j=1}^m\Ex\big[\big|\phi_s^{ij,k}\big|^2\big]ds\nonumber\\
&\quad\leq CT\sup_{t\in[0,T]}\sum_{i,j=1}^m\Ex\big[\big|\phi_t^{ij,k}\big|^2\big]
+C\sup_{t\in[0,T]}\sum_{i,j=1}^m\Ex\big[\big|\phi_t^{ij,k}\big|^2\big]\int_0^t\sup_{(u,y)\in[0,s]\times D}\big\|D_{y}f_{j_1,\ldots,j_k}(u,y)\big\|^2ds.
\end{align*}
Then by Gronwall's lemma and the estimate \eqref{eq:estimate-phi}, we have that $\sup_{(t,y)\in[0,T]\times D}\|D_yf_{j_1,\ldots,j_k}(t,y)\|<+\infty$. This completes the proof.\hfill$\Box$

\subsection{The Verification Result}\label{sec:verif2}
This section gives the verification theorem for the dual problem \eqref{eq:problem-p<0}. Before proving this theorem, we recall that $f(t,y,z)$
is the unique positive bounded classical solution to the recursive system~\eqref{eq:hjn2pf2} of PDEs and that $\hat{a}(t,y,z)$ given by \eqref{eq:hatai}
is the solution to the corresponding system of first-order conditions. Also recall that $\hat{h}(t,y,z)$ is given by \eqref{eq:hathi}.
We define the value function {of the stochastic control problem \eqref{eq:problem-p<0}} as follows
\begin{align}\label{eq:FahTyz}
F(T,y,z):=\esssup_{(a,h)\in{\cal M}}{\Ex^{q,a,h}}\left[K_1^{1-q}e^{\int_0^T\psi(s,a_s,h_s,Y_s,H_s)ds}+K_2^{1-q}\int_0^Te^{\int_0^t\psi(s,a_s,h_s,Y_s,H_s)ds}dt\right],
\end{align}}
for a given initial condition $(Y_0,H_0)=(y,z)\in D\times{\cal S}$. Then, we have the following
\begin{proposition}\label{prop:verificationp<0}
Under assumptions $({\bf A1})$-$({\bf A3})$, the following statements hold
\begin{itemize}
  \item[{\rm(I)}] Let $\hat{a}_t=\hat{a}(T-t,Y_{t-},H_{t-})$ and $\hat{h}_t=\hat{h}(T-t,Y_{t-},H_{t-})$ for $t\in[0,T]$. Then $(\hat{a},\hat{h})\in{\cal M}$.
  \item[{\rm(II)}] The value function $F(T,y,z) = f^{\beta}(T,y,z)$ with $\beta=\frac{1-q}{1-q\rho^2}$. In particular, the Markov feedback controls $(\hat{a},\hat{h})=(\hat{a}_t,\hat{h}_t)_{t\in[0,T]}$ given in {(I)} {are the optimal controls in \eqref{eq:FahTyz}.}
\end{itemize}
\end{proposition}

\noindent{\it Proof.}\quad We first prove {(I)}. {Recall that $g(t,y,z)$, $(t,y,z)\in[0,T]\times D\times{\cal S}$, is the solution of the HJB equation~\eqref{eq:hjn2p<0}. Then $g(t,y,z)=f^{\beta}(t,y,z)$, with $\beta=\frac{1-q}{1-q\rho^2}$.}
Thus, using \eqref{eq:hatai}, it follows that
\begin{eqnarray*}\label{eq:hatai2}
\hat{a}(t,y,z)=-\frac{\sqrt{1-\rho^2}}{1-q}\frac{\sigma_0^{\top}(y)D_yg(t,y,z)}{g(t,y,z)}=-\frac{\sqrt{1-\rho^2}}{1-q}
\frac{\beta\sigma_0^{\top}(y)D_yf(t,y,z)}{f(t,y,z)}.
\end{eqnarray*}
Since both $f$ and its gradient are bounded from Theorem~\ref{thmhjb-p<0} and Proposition~\ref{prop:gradient-est} respectively, we get that for each state $z\in{\cal S}$,
\begin{align}\label{eq:bounda}
\sup_{(t,y)\in[0,T]\times D}\left\|\hat{a}(t,y,z)\right\|\leq C\frac{\sup_{(t,y)\in[0,T]\times D}\|D_yf(t,y,z)\|}{\underline{K}(z)}<\infty.
\end{align}
On the other hand, by \eqref{eq:hathi}, we have $\hat{h}\in\Theta_{\theta}$ while $\Theta_{\theta}\subseteq C_b^{0,1}$ using the assumption ({\bf A3}).
Thus, together with the estimates \eqref{eq:bounda}, it implies that $(\hat{a},\hat{h})$ given in {(I)} belongs to the space ${\cal M}$.

Next, we turn to the proof of {(II)}. From Eq.~\eqref{eq:hjnp<0}, we can define the Hamiltonian ${\it\Psi}(a,h;t,y,z)$ for $(a,h)\in\R^n\times(-1,\infty)^n$ and $(t,y,z)\in[0,T]\times D\times{\cal S}$, which is given by the r.h.s. of Eq.~\eqref{eq:hjnp<0}.
It can be easily verified that ${\it\Psi}(\hat{a},\hat{h};t,y,z)=\esssup_{(a,h)\in{\cal B}\times\Theta_{\theta}}{\it\Psi}(a,h;t,y,z)$ for $(t,y,z)\in[0,T]\times D\times{\cal S}$ a.s. Then, applying It\^o's formula to $g(T-t,Y_t,H_t)$ and noticing that $g(t,y,z)$ satisfies the HJB equation~\eqref{eq:hjn2p<0}, under the p.m. $\Px^{q,\hat{a},\hat{h}}$, we obtain
\begin{eqnarray}\label{eq:gver}
\Big(\frac{\partial}{\partial t} +\hat{\cal A}_t^{\eta} + \psi(t,\hat{a}_t,\hat{h}_t,Y_t,H_t)\Big)g(T-t,Y_t,H_t) = -K_2^{1-q},\ \ \ \ t\in[0,T),
\end{eqnarray}
where the coefficient $\psi$ is given by \eqref{eq:psi}, and the operator $\hat{\cal A}_t^{\eta}$ is defined as
\begin{align}\label{eq:hatAeta}
\hat{\cal A}_t^{\eta}l(t,y,z):=\frac{1}{2}{\rm tr}\big[(\sigma_0\sigma_0^{\top}D_y^2l)(t,y,z)\big]+\eta(\hat{a}_t;t,y,z)^{\top}D_yl(t,y,z)
\end{align}
with $l(t,y,z)$ being $C^2$ in $y$ for fixed $(t,z)\in[0,T]\times{\cal S}$. Notice that $g(0,y,z)=K_1^{1-q}$ for all $(y,z)\in D\times{\cal S}$. Then, the equality $F(T,y,z)=g(T,y,z)$ follows from Feynman-Kac's formula and the equality~\eqref{eq:gver}.\hfill$\Box$
\begin{remark}\label{rem:martGamqah}
For some $(a,h)\in{\cal M}$, the process $\Gam^{q,a,h}=(\Gam_t^{q,a,h})_{t\in[0,T]}$ defined by \eqref{eq:Xbq} may fail to be a $(\Px,\Gx)$-martingale. However, because $\theta\in{\cal C}$ and $(a,h)\in{\cal M}$, we have that $(\theta(\cdot,z),h(\cdot,z))$ are  $C_{b}^{0,1}$ and $h(t,y,z)\in(-1+\varepsilon,\infty)^n$ for some $\varepsilon\in(0,1)$. Define a subset ${\cal M}_0$ of ${\cal M}$ by ${\cal M}_0=\{(a,h)\in{\cal M};\ a\text{ is bounded}\}$. Then, the Novikov's condition is satisfied, and thus the process $\Gam^{q,a,h}$ is a $(\Px,\Gx)$-martingale for all $(a,h)\in{\cal M}_0$. Thanks to Proposition 3.9, it holds that
$\inf_{(a,h)\in{\cal M}_0,\kappa>0}{\it\Pi}(a,h,\kappa)=\inf_{(a,h)\in{\cal M},\kappa>0}{\it\Pi}(a,h,\kappa)$ because the optimal control $\hat{a}$ given in Proposition 3.9 is indeed bounded (i.e., $(\hat{a},\hat{h})\in{\cal M}_0$). This implies that in the constrained optimization problem \eqref{eq:problem-p<022}, we can replace the set of admissible policies ${\cal M}$ with ${\cal M}_0$. Hence, we can solve the problem \eqref{eq:problem-p<022} under the assumption that $\Gam^{q,a,h}$ is a $(\Px,\Gx)$-martingale.
\end{remark}

\section{Optimal Investment/Consumption Strategies}\label{sec:optimal-ic}
In this section, we derive the admissible optimal trading strategy {denoted by }$\hat{\pi}_t=(\hat{\pi}_t^i)_{i=1,\ldots,n}^{\top}$, $t\in[0,T]$, and the admissible optimal consumption process
{denoted by} $\hat{c}_t$, $t\in[0,T]$. This is achieved by exploiting the primal-dual relation provided in Proposition~\ref{prop:1}. 

{We start recalling the quantity $(\hat{a},\hat{h})\in{\cal M}$ given by Proposition~\ref{prop:verificationp<0}. Let the
density process $\Gam_t^{\hat{a},\hat{h}}$, $t\in[0,T]$, be given by~\eqref{eq:eta} with $(a,h)\in{\cal M}$ replaced by $(\hat{a},\hat{h})$, and $F^{\hat{a},\hat{h}}(T,y,z)$ be defined by \eqref{eq:FahTyz} with $(a,h)$ replaced by $(\hat{a},\hat{h})$. Then $F^{\hat{a},\hat{h}}(T,y,z)=g(T,y,z)=f^{\beta}(T,y,z)$ using the verification result given in Proposition~\ref{prop:verificationp<0}. The functions $g(t,y,z)$ and $f(t,y,z)$ are, respectively, the unique positive bounded classic solutions of Eq.~\eqref{eq:hjn2p<0} and Eq.~\eqref{eq:hjn2pf2}, where $\beta=\frac{1-q}{1-q\rho^2}$. } We also recall the quantity $I_i(y)=K_i^{1-q}y^{q-1}$, $i=1,2$, $y\in\R_+$, given in Section \ref{sec:dualform}. Using Proposition~\ref{prop:1}, {under the initial condition $(X^{\hat{\pi},\hat{c}}_0,Y_0,H_0)=(x,y,z)\in\R_+\times D\times{\cal S}$,} the optimal terminal wealth is given by
\begin{eqnarray}\label{eq:optimal-wealth-T}
X^{\hat{\pi},\hat{c}}_T = I_1\left(\hat{\kappa}\frac{{\Gam}_T^{\hat{a},\hat{h}}}{B_T}\right)=\frac{K_1^{1-q}x}{F^{\hat{a},\hat{h}}(T,y,z)}\left(\frac{\Gam_T^{\hat{a},\hat{h}}}{B_T}\right)^{q-1},
\end{eqnarray}
and the optimal consumption process is given by
\begin{eqnarray}\label{eq:optimal-consumption-t}
\hat{c}_t = I_2\left(\hat{\kappa}\frac{{\Gam}_t^{\hat{a},\hat{h}}}{B_t}\right)=\frac{K_2^{1-q}x}{F^{\hat{a},\hat{h}}(T,y,z)}\left(\frac{{\Gam}_t^{\hat{a},\hat{h}}}{B_t}\right)^{q-1},\ \ \ \ \ \ \ \ \ t\in[0,T],
\end{eqnarray}
{where $\hat{\kappa}=(\frac{F^{\hat{a},\hat{h}}(T,y,z)}{x})^{\frac{1}{1-q}}$.

Next, we want to establish the optimal admissible trading/consumption process $(\hat{\pi},\hat{c})$ of the investor. Our objective is to provide a representation for the optimal wealth process $\hat{X}_t$ which satisfies the dynamics given in Eq.~\eqref{eq:Pab-discount}. This will be achieved using the expressions for the optimal terminal wealth given in
Eq.~\eqref{eq:optimal-wealth-T} as well as of the consumption process~\eqref{eq:optimal-consumption-t}. We start defining
\begin{eqnarray*}
{\hat{X}_t} := B_t\Ex^{\hat{a},\hat{h}}\left[\frac{X^{\hat{\pi},\hat{c}}_T }{B_T}+\int_t^T\frac{\hat{c}_s}{B_s}ds\bigg|\G_t\right],\ \ \ \ \ t\in[0,T].
\end{eqnarray*}
Using that $g(0,y,z)=K_1^{1-q}$, and Lemma 2.5 in \cite{CoxHuang}, we have, for $t\in[0,T]$,
\begin{align}\label{eq:hatXB0}
\frac{\hat{X}_t}{B_t}  &=\frac{x}{g(T,y,z)\Gam_t^{\hat{a},\hat{h}}}\Ex\left[K_1^{1-q}\left(\frac{\Gam_T^{\hat{a},\hat{h}}}{B_T}\right)^{q}
+K_2^{1-q}\int_t^T\left(\frac{\Gam_s^{\hat{a},\hat{h}}}{B_s}\right)^{q}ds\bigg|\G_t\right]\nonumber\\
&= \frac{x}{g(T,y,z)\Gam_t^{\hat{a},\hat{h}}}\Ex\left[{G}_T-K_2^{1-q}\int_0^t\left(\frac{\Gam_s^{\hat{a},\hat{h}}}{B_s}\right)^{q}ds\bigg|\G_t\right].
\end{align}
In the above expression, for $t\in[0,T]$, we have defined the process
\begin{eqnarray}\label{eq:mart-Gt}
{G}_t :=K_2^{1-q}\int_0^t \left(\frac{\Gam_s^{\hat{a},\hat{h}}}{B_s}\right)^qds + \left(\frac{\Gam_t^{\hat{a},\hat{h}}}{B_t}\right)^qg(T-t,Y_t,H_t).
\end{eqnarray}
We then have the following lemma proven in the Appendix.
\begin{lemma}\label{lem:G-mart}
The process $G=(G_t)_{t\in[0,T]}$ defined in Eq.~\eqref{eq:mart-Gt} is a positive $(\Px,\Gx)$-martingale.
\end{lemma}
Using Lemma~\ref{lem:G-mart} and Eq.~\eqref{eq:mart-Gt}, we may rewrite Eq.~\eqref{eq:hatXB0} as
\begin{align}\label{eq:hatXB}
\frac{\hat{X}_t}{B_t} 
&=\frac{x}{g(T,y,z)\Gam_t^{\hat{a},\hat{h}}}{\left[{G}_t-K_2^{1-q}\int_0^t\left(\frac{\Gam_s^{\hat{a},\hat{h}}}{B_s}\right)^{q}ds\right]}\nonumber\\
&= \frac{x}{g(T,y,z)\Gam_t^{\hat{a},\hat{h}}}\left(\frac{\Gam_t^{\hat{a},\hat{h}}}{B_t}\right)^qg(T-t,Y_t,H_t)\nonumber\\
&= \frac{x}{g(T,y,z)}\frac{\left(\Gam_t^{\hat{a},\hat{h}}\right)^{q-1}}{B_t^q}g(T-t,Y_t,H_t).
\end{align}
Then, using Eq.~\eqref{eq:optimal-consumption-t} along with the above expression, the optimal consumption process may be rewritten in terms of the time-$t$ optimal wealth as
\begin{eqnarray}\label{eq:optimal-consumption-t2}
\hat{c}_t= \frac{K_2^{1-q}x}{g(T,y,z)}\frac{\hat{X}_t}{\frac{x}{g(T,y,z)}g(T-t,Y_t,H_t)}
=\frac{K_2^{1-q}\hat{X}_t}{g(T-t,Y_t,H_t)},\ \ \ \ \ \ \ \ \ t\in[0,T].
\end{eqnarray}
In order to achieve our goal of expressing the dynamics $d\left(\frac{\hat{X}_t}{B_t}\right)+\frac{\hat{c}_t}{B_t}dt$ in the form~\eqref{eq:Pab-discount}, we first derive the dynamics of $\frac{\hat{X}_t}{B_t}$. This is done in the following lemma proven in the Appendix.
\begin{lemma}\label{lem:P-dyn-hatXB}
It holds that
\begin{align}\label{eq:P-dyn-hatXB}
&d\left(\frac{\hat{X}_t}{B_t}\right) = \frac{x}{g(T,y,z)}\frac{(\Gam_t^{\hat{a},\hat{h}})^{q-1}}{B_t^q}\bigg\{-K_2^{1-q}dt+g(T-t,Y_t,H_t)\nonumber\\
&\quad\qquad\qquad\times\bigg[(1-q)\theta_t^{\top}+\rho\frac{D_yg(T-t,Y_t,H_t)^{\top}\sigma_0(Y_t)}{g(T-t,Y_t,H_t)}\bigg]dW_t^{\theta}\bigg\}\\
&\quad\qquad\qquad+\frac{x}{g(T,y,z)}\frac{(\Gam_{t-}^{\hat{a},\hat{h}})^{q-1}}{B_{t-}^q}g(T-t,Y_t,H_{t-})\sum_{i=1}^n
\left[\big(1+\hat{h}_{t-}^i\big)^{q-1}\frac{g(T-t,Y_t,\bar{H}_{t-}^i)}{g(T-t,Y_t,H_{t-})}-1\right]dM_t^{\hat{h},i}.\nonumber
\end{align}
Above, for $t\in[0,T]$, ${W}^{\theta}_t$ and $M_t^{\hat{h},i}$ for $i=1,\ldots,n$ are defined in \eqref{eq:tildeWi} with $h$ replaced by $\hat{h}$.
\end{lemma}
Using the above dynamics along with equations~\eqref{eq:Pab-discount} and~\eqref{eq:optimal-consumption-t}, we obtain the following characterization of the optimal feedback strategy. {For $(t,y,z)\in [0,T]\times D\times{\cal S}$, define
\begin{align}\label{eq:LamJ}
\Lambda^{\theta,\hat{h}}(t,y,z)&:=(1-q)\theta(t,y,z)^{\top}+\rho\frac{D_yg(T-t,y,z)^{\top}\sigma_0(y)}{g(T-t,y,z)},\nonumber\\
J_i^{\theta,\hat{h}}(t,y,z)&:=1-(1+\hat{h}_i(t,y,z))^{q-1}\frac{g(T-t,y,\bar{z}^i)}{g(T-t,y,z)},\quad i=1,\ldots,n,
\end{align}
and $J^{\theta,\hat{h}}(t,y,z)=(J_i^{\theta,\hat{h}}(t,y,z);\ i=1,\ldots,n)^{\top}$.}
\begin{proposition}\label{prop:optium}
{Let assumptions~({\bf A1})-({\bf A3}) hold. Suppose that there exists a $\hat{h}\in\Phi_{\theta}$ such that $J^{\theta,\hat{h}}(t,y,z)^{\top}\sigma(y)=\Lambda^{\theta,\hat{h}}(t,y,z)$.} Then, the $\Gx$-predictable optimal feedback strategy $\hat{\pi}_t=\hat{\pi}(t,Y_{t-},H_{t-})$, $t\in[0,T]$, is given by
\begin{align}\label{eq:optimal}
\hat{\pi}_t^{\top}\sigma(Y_{t-}) &=\left[(1-q)\theta(t,Y_{t-},H_{t-})^{\top}+\rho\beta\frac{D_yf(T-t,Y_{t-},H_{t-})^{\top}\sigma_0(Y_{t-}) }{f(T-t,Y_{t-},H_{t-})}\right]{\rm diag}(1-H_{t-}^i;\ i=1,\ldots,n),\nonumber\\
\hat{\pi}_t^i&=\left[1-(1+\hat{h}_i(t,Y_{t-},H_{t-}))^{q-1}\left(\frac{f(T-t,Y_{t-},\bar{H}^i_{t-})}{f(T-t,Y_{t-},H_{t-})}\right)^{\beta}\right](1-H_{t-}^i),\quad i=1,\ldots,n,
\end{align}
where $\beta=\frac{1-q}{1-q\rho^2}$, and $f(t,y,z)$ is the unique classical solution of the recursive system {of PDEs} \eqref{eq:hjn2pf2}.
\end{proposition}
{Let us analyze the structure of the optimal investment strategy given by the first equation in~\eqref{eq:optimal}. On the right hand side, the first component is the so-called myopic portion and admits the same functional form as in the classical Merton's model, but adjusted to also reflect the contribution coming credit risk. The second component is the excess hedging demand generated by the correlated movements (the correlation coefficient $\rho \neq 0$) of the stochastic factors and stock prices.}

\noindent{\it Proof.}\quad Using Lemma~\ref{lem:P-dyn-hatXB} along with the equality \eqref{eq:hatXB}, we obtain the dynamics given by
\begin{align}\label{eq:stoch-inte}
d \left(\frac{\hat{X}_t}{B_t}\right)+\frac{\hat{c}_t}{B_t}dt &= \left(\frac{\hat{X}_{t-}}{B_{t-}}\right)\big[\Lambda^{\theta,\hat{h}}(t,Y_{t-},H_{t-})dW_t^{\theta}-J^{\theta,\hat{h}}(t,Y_{t-},H_{t-})^{\top}dM_t^{\hat{h}}\big],
\end{align}
where $\Lambda^{\theta,\hat{h}}(t,y,z)$ and $J^{\theta,\hat{h}}(t,y,z)$ are defined in \eqref{eq:LamJ}. For $\hat{h}\in\Phi_{\theta}$ given above, we let a predictable process $\pi^{\hat{h}}$ satisfy the system~\eqref{eq:optimal}. This further gives that
\begin{align}\label{eq:wealthsdehat}
d\left(\frac{\hat{X}_t}{B_t}\right) + \frac{\hat{c}_t}{B_t}dt = \frac{\hat{X}_{t-}}{B_{t-}}(\pi^{\hat{h}}_{t})^{\top}\big[\sigma(Y_t)d{W}_t^{\theta}-dM_t^{\hat{h}}\big].
\end{align}
Then a direct comparison of Eq.~\eqref{eq:wealthsdehat} and Eq.~\eqref{eq:Pab-discount} indicates that the optimal feedback strategy is given by $\pi^{\hat{h}}$, where we also used the fact that $g(T-t,y,z)=f^{\beta}(T-t,y,z)$ and that the fraction of wealth held by the investor in a stock is zero after it defaults.
On the other hand, using Lemma~\ref{lem:ineL} and the assumption ({\bf A3}), it follows that the value function of the primal problem $V(x,y,z)\leq \inf_{(a,h)\in{\cal M},\kappa>0}{\it\Pi}(a,h,\kappa)\leq{\it\Pi}(\hat{a},\hat{h},\hat{\kappa})$ where $\hat{a}_t=\hat{a}(t,Y_{t-},H_{t-})$ with $\hat{a}$ given by \eqref{eq:hatai}, $\hat{\kappa}$ is given in \eqref{eq:optimal-consumption-t}, and $\hat{h}_t=\hat{h}(t,Y_{t-},H_{t-})$ with $\hat{h}\in\Phi_{\theta}$ given above. By virtue of \eqref{eq:Pi}, \eqref{eq:solution-density}, \eqref{eq:optimal-consumption-t} and \eqref{eq:wealthsdehat}, a direct computation yields that ${\it\Pi}(\hat{a},\hat{h},\hat{\kappa})=V(x,y,z)$. Therefore, it holds that $\inf_{(a,h)\in{\cal M},\kappa>0}{\it\Pi}(a,h,\kappa)={\it\Pi}(\hat{a},\hat{h},\hat{\kappa})$. This shows that $\hat{h}\in\Phi_{\theta}$ given above satisfies \eqref{eq:hathi}. We next verify that the above strategy ${\pi}^{\hat{h}}=({\pi}^{\hat{h}}_t)_{t\in[0,T]}$ is admissible. Using the first equality in \eqref{eq:optimal} together with Theorem~\ref{thmhjb-p<0} and Proposition~\ref{prop:gradient-est}, we deduce that $\Ex[\int_0^T\|({\pi}^{\hat{h}}_t)^{\top}\sigma(Y_t)\|^2dt]<+\infty$ under assumptions~({\bf A1})-({\bf A3}). Using the second equality in \eqref{eq:optimal} together with Theorem~\ref{thmhjb-p<0}, we obtain that $\sum_{i=1}^n\Ex[\int_0^T|{\pi}_t^{\hat{h},i}|^2\lambda_i(Y_t,H_t)dt]<+\infty$. Further, under assumptions~({\bf A1})-({\bf A3}), it holds that ${\pi}_t^{\hat{h},i}\in(-\infty,1)$ for each $i=1,\ldots,n$. Hence, the strategy ${\pi}^{\hat{h}}$ is admissible as specified in Definition~\ref{def:admissi}. Thus $\hat{\pi}:=\pi^{\hat{h}}$ is the optimal optimal feedback strategy. \hfill$\Box$
\begin{remark}\label{rem:optimum}
The $\Gx$-predictable optimal feedback strategy $\hat{\pi}_t=\hat{\pi}(t,Y_{t-},H_{t-})$, $t\in[0,T]$, is given in \eqref{eq:optimal}. Observe that the solution $g(t,y,z)=f^{\beta}(t,y,z)$ of the HJB equation~\eqref{eq:hjn2p<0} also depends on $\hat{h}\in\Phi_{\theta}$. \cite{Mile1271} consider an optimal portfolio problem with jump-diffusion dynamics, but without stochastic factors, and solve it using the martingale approach. In the absence of stochastic factors and default contagion, the functions defined in \eqref{eq:LamJ} reduce to $\Lambda^{\theta,\hat{h}}(t,z)=(1-q)\theta(t,z)^{\top}$ and $J_i^{\theta,\hat{h}}(t,z)=1-(1+\hat{h}_i(t,z))^{q-1}$, respectively. Thus, the equation
$J^{\theta,\hat{h}}(t,z)^{\top}\sigma=\Lambda^{\theta,\hat{h}}(t,z)$ takes a similar form to that in Corollary 1 of \cite{Mile1271}
(choosing $\xi=\sigma$ and $\gamma_i=1$). In the absence of default risk, $f(t,y,\bar{z}^i)=f(t,y,z)=f(t,y)$ and hence we have $\hat{h}(t,y,z)=0$. Thus $J_i^{\theta,0}(t,y,z)=0$ for $i=1,\ldots,n$, and the default martingale part in \eqref{eq:stoch-inte} vanishes. Comparing the dynamics \eqref{eq:stoch-inte} and~\eqref{eq:Pab-discount} when $dM^{\hat{h}}=0$ (no default risk), we deduce that the optimal strategy is given by
\[\hat{\pi}_t^{\top}=\left[(1-q)\theta(t,Y_{t})^{\top}+\rho\frac{D_yg(T-t,Y_{t})^{\top}\sigma_0(Y_t)}{g(T-t,Y_{t})}\right]\sigma^{-1}(Y_t),\]
where $g(t,y)$ is the solution of the HJB equation corresponding to the same optimal control problem but ignoring default risk. When $n=1$, we recover the optimal strategy by \cite{NLHH}
(see the last section of their paper, and choose the volatility function of the one-dimensional stochastic factor to be constant).
\end{remark}
\begin{remark} We present a specialization of our framework, in which the optimal strategies and value functions can be explicitly computed. The portfolio model consists of one risky stock ($n=1$) and a constant one-dimensional factor $Y$. In this setup, we obtain a closed-form
solution $f$ for Eq.~\eqref{eq:hjn2pf2}, with an explicit dependence on $h=\hat{h}$. The function $f$ satisfies the following Bernoulli equation
given by
\begin{align*}
\left\{
  \begin{array}{ll}
    f'(t,1) = \frac{\varphi(t,1)}{\beta}f(t,1)+\frac{K_2^{\beta}}{\beta}f^{1-\beta}(t,1),\\ \\
    f'(t,0) = \frac{\varphi(t,0)}{\beta}f(t,0)+\frac{K_2^{\beta}+ f^{\beta}(t,1)(1+h(t,0))^q\lambda(0)}{\beta}f^{1-\beta}(t,0)
  \end{array}
\right.
\end{align*}
with initial conditions given, respectively, by $f(0,1)=f(0,0)=K_1^{(1-q)/\beta}=K_1$ (since $\beta=1-q$). The coefficients of the equation are given by
\begin{align*}
\varphi(t,1)&:=\frac{q(q-1)}{2}\theta^2(t,1)-q r=\frac{q(q-1)}{2}\xi^2-q r,\\
\varphi(t,0)&:=\frac{q(q-1)}{2}\theta^2(t,0)-q r+[q-1-q(1+h(t,0))]\lambda(0).
\end{align*}
If the stock has defaulted ($z=1$) we have $\theta(t,1)\equiv\xi=\sigma^{-1}(\mu-r)$, i.e. it equals the market price of risk. Hence, we can take $h(t,1)=0$ given that the default intensity does not play any role if the stock's default has already occurred. In the state $z=0$, i.e. when the stock is alive, we have that the pair $(\theta(t,0),h(t,0))$ satisfies the equation $\xi-\theta(t,0)=\sigma^{-1}\lambda(0)h(t,0)$. Thus the above coefficients may be rewritten as
\begin{align}\label{eq:varphit0}
\varphi(t,1)&\equiv\varphi(1):=\frac{q(q-1)}{2}\xi^2-q r,\\
\varphi(t,0)&:=\frac{q(q-1)\lambda_1^2(0)}{2\sigma^2}h^2(t,0)-q\lambda(0)\left((q-1)\frac{\xi}{\sigma}+1\right)h(t,0)-qr-\lambda(0)+\frac{q(q-1)}{2}\xi^2.\nonumber
\end{align}
Define $f(t,z)=\tilde{f}(t,z)^{1/\beta}$ for $z=0,1$. Then, for $z=0,1$, $\tilde{f}(t,z)$ satisfies the linear ODE with initial data $\tilde{f}(0,1)=\tilde{f}(0,0)=K_1^{\beta}$, and
\begin{align}\label{eq:bernoulliODE2}
\left\{
  \begin{array}{ll}
    \tilde{f}'(t,1) = \varphi(1)\tilde{f}(t,1)+K_2^{\beta},\\ \\
    \tilde{f}'(t,0) = \varphi(t,0)\tilde{f}(t,0)+K_2^{\beta}+ \tilde{f}(t,1)(1+h(t,0))^q\lambda(0).
  \end{array}
\right.
\end{align}
For $t\in[0,T]$, let $x(t):=1+\tilde{h}(t,0)$, where we set $\tilde{h}(T-t,0)=h(t,0)$. Then we can rewrite the coefficient $\varphi(t,0)$ given in \eqref{eq:varphit0} as
\begin{align}\label{eq:varphix}
\varphi(t,0)=\varphi(x(T-t)) &:= a x(T-t)^2 +b x(T-t) +c,
\end{align}
where the constants $a:=\frac{q(q-1)\lambda^2(0)}{2\sigma^2}$, $b:=\frac{q(1-q)\lambda^2(0)}{\sigma^2}+q\lambda(0)((1-q)\frac{\xi}{\sigma}-1)$ and $c:=q\lambda(0)((q-1)\frac{\xi}{\sigma}+1)-qr-\lambda(0)+\frac{q(q-1)}{2}\xi^2+\frac{q(q-1)\lambda^2(0)}{2\sigma^2}$. The closed-form solution of the second equation in \eqref{eq:bernoulliODE2} is given by
\begin{align*}
 \tilde{f}^x(t,0) &= e^{\int_0^{t}\varphi(x(T-s))ds}\left[\int_0^{t}e^{-\int_0^s\varphi(x(T-v))dv}\big(K_2^{\beta}+ \ell(s)x(T-s)^q\big)ds
+K_1^{\beta}\right].
\end{align*}
In the above expression, $\ell(t):=\lambda(0)\tilde{f}(t,1)=\lambda(0)e^{\varphi(1)t}\big[K_2^{\beta}\int_0^te^{-\varphi(1)v}dv+K_1^{\beta}\big]$ for $t\in[0,T]$, which is independent of $x(t)$.
Fix $t\in[0,T]$ and let $u=T-t$ be the time to maturity. Then the nonlinear equation $J^{\theta,h}(t,0)\sigma=\Lambda^{\theta,h}(t,0)$ in Proposition~\ref{prop:optium} reduces to
\begin{align}\label{eq:eqncons6}
x(u)=\frac{\tilde{b}}{\tilde{a}}+\frac{\ell(u)}{\tilde{a} \tilde{f}^x(u,0)x(u)^{\beta}},
\end{align}
where $\tilde{a}:=\frac{\beta\lambda^2(0)}{\sigma^2}$ and $\tilde{b}:=\lambda(0)\frac{\beta(\xi\sigma+\lambda(0))-\sigma^2}{\sigma^2}$. We consider the case $q=0$ and $\tilde{b}>0$, which corresponds to an investor with logarithmic utility. Let $\varepsilon:=\tilde{b}/\tilde{a}$ and $I:=[\varepsilon,\infty)$. For all $y\in C_I:=C_{I}([0,T])$, define the continuous mapping on $C_I$:
\begin{align*}
F(y)(u) &:=\varepsilon+\frac{\tilde{a}^{-1}y(u)^{-\beta}\ell(u)e^{-cu}}{\int_0^{u}e^{-cs}\big(K_2^{\beta}+ \ell(s)\big)ds
+K_1^{\beta}}.
\end{align*}
Our objective is to show the existence of a unique fixed point of $F(x)$. For any $y_1,y_2\in C_I$, it holds that
\begin{align*}
&\left|F(y_1)(u)-F(y_2)(u)\right| = \frac{\ell(u)}{\tilde{a}}e^{-cu}\frac{\left|y_1(u)^{-\beta}-y_2(u)^{-\beta}\right|}{\int_0^{u}e^{-cs}\big(K_2^{\beta}+ \ell(s)\big)ds+K_1^{\beta}}\nonumber\\
&\qquad\qquad\leq \frac{\ell(u)}{\tilde{a}}\beta\varepsilon^{-\beta-1}\frac{\left|y_1(u)-y_2(u)\right|}{\int_0^{u}e^{c(u-s)}\big(K_2^{\beta}+ \ell(s)\big)ds+e^{cu}K_1^{\beta}}
:=G(u)\left|y_1(u)-y_2(u)\right|.
\end{align*}
Notice that $\ell(t)$ is bounded on $t\in[0,T]$. Then, there exists $\varepsilon>0$ such that $\sup_{u\in[0,T]}G(u)\in(0,1)$. Hence, standard techniques based on Picard iterations yield the unique fixed point of $F(x)$.
\end{remark}

Using Eq.~\eqref{eq:optimal-consumption-t}, we also obtain
\begin{eqnarray}\label{eq:optimal-consum}
\hat{c}_t=\frac{K_2^{1-q}x}{f^{\beta}(T,y,z)}\left(\frac{\Gam_t^{\hat{a},\hat{h}}}{B_t}\right)^{q-1}.
\end{eqnarray}
By virtue of Theorem~\ref{thmhjb-p<0} and Lemma~\ref{lem:density-moment} in the Appendix, we have that $\Ex\big[\int_0^T\hat{c}_tdt\big]<+\infty$. This {shows} that $(\hat{\pi},\hat{c})\in{\cal U}={\cal U}(x,y,z)$. Using Eq.~\eqref{eq:hatXB}, the optimal wealth process is 
\begin{eqnarray}\label{eq:optimal-wealth}
X_t^{\hat{\pi},\hat{c}}
=x\left(\frac{f(T-t,Y_t,H_t)}{f(T,y,z)}\right)^{\beta}\left(\frac{\Gam_t^{\hat{a},\hat{h}}}{B_t}\right)^{q-1},\ \ \ \ \ \ \ \ t\in[0,T].
\end{eqnarray}

Next, we continue the Example~\ref{exam:1} given in Section~\ref{sec:modelinvestor}. 

\noindent {\bf Example~\ref{exam:1} continued:}
We apply the theoretical analysis developed above to derive the optimal strategies and value functions associated with the setup given in the Example~\ref{exam:1}. Both
the optimal strategies and the consumption process depend on the solution $f$ of the recursive system \eqref{eq:hjn2pf2} of semi-linear PDEs.
In the context of our example, the system reduces to

\begin{eqnarray}\label{exam:hjn2pf2-01}
\left\{
  \begin{array}{ll}
    \frac{\partial f_{11}(t,y)}{\partial t} = \frac{1}{2}\sum_{i=1}^2(\sigma_{0i}^2+\bar{\sigma}_{0i}^2)\frac{\partial^2 f_{11}(t,y)}{\partial y_i^2}
+\nu_{11}(t,y)^{\top}D_yf_{11}(t,y)+\beta^{-1}\varphi_{11}(t,y)f_{11}(t,y)\\
      \qquad\qquad\quad+K_2^{1-q}\beta^{-1}f_{11}^{1-\beta}(t,y),\\ \\
    \frac{\partial f_{01}(t,y)}{\partial t} = \frac{1}{2}\sum_{i=1}^2(\sigma_{0i}^2+\bar{\sigma}_{0i}^2)\frac{\partial^2 f_{01}(t,y)}{\partial y_i^2}
+\nu_{01}(t,y)^{\top}D_yf_{01}(t,y)+\beta^{-1}\varphi_{01}(t,y)f_{01}(t,y)\\
         \qquad\qquad\quad+\beta^{-1}f_{01}^{1-\beta}(t,y)\big[K_2^{1-q}+f_{11}^{\beta}(t,y)(1+\hat{h}_{1,01}(t,y))^q\lambda_{1,01}(y)\big],\\ \\
     \frac{\partial f_{10}(t,y)}{\partial t} = \frac{1}{2}\sum_{i=1}^2(\sigma_{0i}^2+\bar{\sigma}_{0i}^2)\frac{\partial^2 f_{10}(t,y)}{\partial y_i^2}
+\nu_{10}(t,y)^{\top}D_yf_{10}(t,y)+\beta^{-1}\varphi_{10}(t,y)f_{10}(t,y)\\
         \qquad\qquad\quad+\beta^{-1}f_{10}^{1-\beta}(t,y)\big[K_2^{1-q}+f_{11}^{\beta}(t,y)(1+\hat{h}_{2,10}(t,y))^q\lambda_{2,10}(y)\big],\\ \\
\frac{\partial f_{00}(t,y)}{\partial t}= \frac{1}{2}\sum_{i=1}^2(\sigma_{0i}^2+\bar{\sigma}_{0i}^2)\frac{\partial^2 f_{00}(t,y)}{\partial y_i^2}
+\nu_{00}(t,y)^{\top}D_yf_{00}(t,y)+\beta^{-1}\varphi_{00}(t,y)f_{00}(t,y)\\
\qquad\qquad\quad+\beta^{-1}f_{00}^{1-\beta}(t,y)\big[f_{10}^{\beta}(t,y)(1+\hat{h}_{1,00}(t,y))^q\lambda_{1,00}(y)
+f_{01}^{\beta}(t,y)(1+\hat{h}_{2,00}(t,y))^q\lambda_{2,00}(y)\big]\\
\qquad\qquad\quad+K_2^{1-q}\beta^{-1}f_{00}^{1-\beta}(t,y)\\
  \end{array}
\right.
\end{eqnarray}
with initial condition $f_{00}(0,y)=f_{01}(0,y)=f_{10}(0,y)=f_{11}(0,y)=K_1^{\frac{1}{1-q\rho^2}}$ for all $y\in D$. Above, notice how default and financial contagion are reflected into
the PDE structure. Consider first the situation in which both stocks are defaulted, i.e., the default state is ${z}=(1,1)$. Then the investor
can only invest in the bank account. Next, consider the situation when ${z} = (0,1)$ or ${z} = (1,0)$, i.e. only one stock is defaulted. In this case, the investor needs to consider the optimal expected utility achievable in the state reached when both stocks are defaulted. This is reflected through the dependence of the PDEs satisfied by $f_{01}$ and $f_{01}$ on $f_{11}$. Similarly, when both stocks are alive, the optimal utility of the investor depends directly on that achieved when
either stock defaults ($f_{00}$ depends on $f_{01}$ and $f_{10}$) and indirectly on the utility achieved when both stocks default ($f_{00}$ depends on $f_{11}$ through
$f_{01}$ and $f_{10}$). Let $\hat{h}\in\Theta_{\theta}$ satisfy $J^{\theta,\hat{h}}(t,y,z)^{\top}\sigma(y)=\Lambda^{\theta,\hat{h}}(t,y,z)$. Using Eq.~\eqref{eq:optimal}, the optimal feedback strategies are given by
\begin{align*}
\left\{
  \begin{array}{ll}
   \hat{\pi}_{1,01}(t,y)=1-(1+\hat{h}_{1,01}(t,y))^{q-1}\left(\frac{f_{11}(T-t,y)}{f_{01}(T-t,y)}\right)^{\beta}, \\
  \hat{\pi}_{2,01}(t,y)=0; \\ 
   \hat{\pi}_{1,10}(t,y)=0, \\
   \hat{\pi}_{2,10}(t,y)=1-(1+\hat{h}_{2,10}(t,y))^{q-1}\left(\frac{f_{11}(T-t,y)}{f_{10}(T-t,y)}\right)^{\beta};\\ 
  \hat{\pi}_{1,00}(t,y)=1-(1+\hat{h}_{1,00}(t,y))^{q-1}\left(\frac{f_{10}(T-t,y)}{f_{00}(T-t,y)}\right)^{\beta},\\
\hat{\pi}_{2,00}(t,y)=1-(1+\hat{h}_{2,00}(t,y))^{q-1}\left(\frac{f_{01}(T-t,y)}{f_{00}(T-t,y)}\right)^{\beta}.
  \end{array}
\right.
\end{align*}

Notice that the optimal strategy adopted by the investor in each default state depends on the gradient of the PDE solution in the same default state. Such a solution depends in turn on the solution of PDEs associated with augmented states in which additional defaults occur. 

\section{Numerical Analysis}\label{sec:numerics}
We develop a numerical analysis to analyze the sensitivity of the optimal investment strategies to the model parameters. To highlight the main economic forces, we consider a portfolio model consisting of two defaultable stocks whose price processes are driven by a single O-U type stochastic factor (i.e., $m=1$ and $n=2$).

\subsection{Model Specification}
The stochastic factor $Y=(Y_t)_{t\in[0,T]}$ and the pre-default price dynamics $P^i=(P^i_t)_{t\in[0,T]}$ of the $i$-th stock are given by
\begin{align}\label{eq:Y}
\left\{
  \begin{array}{ll}
    dY_t=(u_0-\mu_{0} Y_t)dt + \sum_{j=1}^2\sigma_{0j}d\bar{W}_t^j;\\ \\
   \frac{d{P}_t^i}{d{P}_{t}^i}=(\mu_i + \lambda_i(Y_t,H_t))dt + \sigma_idW_t^i,\ i=1,2.
  \end{array}
\right.
\end{align}
For $i=1,2$, $\lambda_i(y,z)$, $(y,z)\in\R\times\{0,1\}^2$, is the default intensity function of the $i$-th stock. We let the stochastic factor be any constant $y$ in interval $(-l,l)$ where $l>0$ is a fixed positive constant. For $i=1,2$, we choose the default intensity function to be of the form $\lambda_i(y,z)=a_{iz}+b_{iz}e^{c_{iz}y}$. The coefficients $a_{iz},b_{iz},c_{iz}$ are positive constants depending on the default state $z\in\{0,1\}^2$. 
{The optimal strategies and  value functions can be recovered by numerically solving the following interacting system of PDEs:}
\begin{eqnarray}\label{exam:hjn2pf2-012}
\left\{
  \begin{array}{ll}
    \frac{\partial f_{11}(t,y)}{\partial t} = \frac{1}{2}\sum_{i=1}^2\sigma_{0i}^2\frac{\partial^2 f_{11}(t,y)}{\partial y^2}
+\nu_{11}(y)\frac{\partial f_{11}(t,y)}{\partial y}+\beta^{-1}\varphi_{11}(y)f_{11}(t,y)+K_2^{1-q}\beta^{-1}f_{11}^{1-\beta}(t,y),\\ \\
    \frac{\partial f_{01}(t,y)}{\partial t} = \frac{1}{2}\sum_{i=1}^2\sigma_{0i}^2\frac{\partial^2 f_{01}(t,y)}{\partial y^2}
+\nu_{01}(y)\frac{\partial f_{01}(t,y)}{\partial y}+\beta^{-1}\varphi_{01}(y)f_{01}(t,y)\\
         \qquad\qquad\quad+\beta^{-1}f_{01}^{1-\beta}(t,y)\big[K_2^{1-q}+f_{11}^{\beta}(t,y)(1+\hat{h}_{1,01}(t,y))^q\lambda_{1,01}(y)\big],\\ \\
     \frac{\partial f_{10}(t,y)}{\partial t} = \frac{1}{2}\sum_{i=1}^2\sigma_{0i}^2\frac{\partial^2 f_{10}(t,y)}{\partial y^2}
+\nu_{10}(y)^{\top}\frac{\partial f_{10}(t,y)}{\partial y}+\beta^{-1}\varphi_{10}(y)f_{10}(t,y)\\
         \qquad\qquad\quad+\beta^{-1}f_{10}^{1-\beta}(t,y)\big[K_2^{1-q}+f_{11}^{\beta}(t,y)(1+\hat{h}_{2,10}(t,y))^q\lambda_{2,10}(y)\big],\\ \\
\frac{\partial f_{00}(t,y)}{\partial t}= \frac{1}{2}\sum_{i=1}^2\sigma_{0i}^2\frac{\partial^2 f_{00}(t,y)}{\partial y^2}
+\nu_{00}(y)\frac{\partial f_{00}(t,y)}{\partial y}+\beta^{-1}\varphi_{00}(y)f_{00}(t,y)\\
\qquad\qquad\quad+\beta^{-1}f_{00}^{1-\beta}(t,y)\big[f_{10}^{\beta}(t,y)(1+\hat{h}_{1,00}(t,y))^q\lambda_{1,00}(y)
+f_{01}^{\beta}(t,y)(1+\hat{h}_{2,00}(t,y))^q\lambda_{2,00}(y)\big]\\
\qquad\qquad\quad+K_2^{1-q}\beta^{-1}f_{00}^{1-\beta}(t,y)\\
  \end{array}
\right.
\end{eqnarray}
with initial condition $f_{00}(0,y)=f_{01}(0,y)=f_{10}(0,y)=f_{11}(0,y)=K_1$ for all $y\in D$. We use the {\sf General Form PDE} interface with {\sf Time-Dependent Study} built-in {\sf COMSOL Multiphysics} to numerically solve the system \eqref{exam:hjn2pf2-012}, assuming $y\in(-l,l)$, and under the conditions of Proposition~\ref{prop:optium} as global constraint equations. We first rewrite the above system in the form (required by
the COMSOL interface) given by
\begin{align}\label{eq:comsol}
e_a\frac{\partial^2 u}{\partial t^2} + d_a\frac{\partial u}{\partial t} + \nabla\Gamma = F.
\end{align}
Eq.~\eqref{exam:hjn2pf2-012} is a system of PDEs with solution vector $u(t,y)=[u_1,u_2,u_3,u_4](t,y)$ for $(t,y)\in[0,T]\times(-l,l)$.
Notice that $\nu(y)u_y=\nabla(\nu(y)u)+\mu_0u$. Let $a:=\frac{1}{2}\sum_{i=1}^2\sigma_{0i}^2$. Then it holds that
\begin{align*}
-au_{yy}-\nu(y)u_y = \nabla\left(-au_y-\nu(y)u\right)-\mu_0u.
\end{align*}
Thus the system \eqref{exam:hjn2pf2-012} of PDEs is obtained by a direct specification of Eq.~\eqref{eq:comsol}, in which we set $e_a=0$, $d_a=I$ (here $I$ denotes the identity matrix),
\begin{align}\label{eq:system}
\Gamma = \left[
           \begin{array}{c}
             -au_{1y}-\nu(y)u_1 \\
             -au_{2y}-\nu(y)u_2 \\
             -au_{3y}-\nu(y)u_3 \\
             -au_{4y}-\nu(y)u_4 \\
           \end{array}
         \right],
\end{align}
and the source term
\begin{align}\label{eq:F}
F=\left[
           \begin{array}{c}
             (\beta^{-1}\varphi_{11}(y)+\mu_0)u_1+\beta^{-1}K_2^{\beta}u_1^{q}\\
             (\beta^{-1}\varphi_{01}(y)+\mu_0)u_2+\beta^{-1}K_2^{\beta}u_2^q+\beta^{-1}u_2^{q}u_1^{\beta}(1+\hat{h}_{101}(y))^q\lambda_{101}(y)\\
             (\beta^{-1}\varphi_{10}(y)+\mu_0)u_3+\beta^{-1}K_2^{\beta}u_3^q+\beta^{-1}u_3^{q}u_1^{\beta}(1+\hat{h}_{210}(y))^q\lambda_{210}(y)\\
             (\beta^{-1}\varphi_{00}(y)+\mu_0)u_4+\beta^{-1}K_2^{\beta}u_4^q+\beta^{-1}u_4^{q}u_3^{\beta}(1+\hat{h}_{100}(y))^q\lambda_{100}(y)\\
             +\beta^{-1}u_4^{q}u_2^{\beta}(1+\hat{h}_{200}(y))^q\lambda_{200}(y)
           \end{array}
         \right].
\end{align}
The initial value $u(0)=[u_1,u_2,u_3,u_4](0)=[K_1,K_1,K_1,K_1]$. We use the following benchmark specification of the parameters: $a_{100}=0.6$, $a_{200}=0.5$, $a_{101}=a_{210}=0.8$, $b_{100}=0.4$, {$b_{200}=0.3$}, $b_{101}=b_{210}=0.6$, $c_{100}=c_{200}=c_{101}=c_{210}=0.1$, $u_0=0.5$, $\mu_0=1.2$, $\sigma_{01}=0.6$, $\sigma_{02}=0.4$, $\mu_1=\mu_2=r=0.2$, $\sigma_1=\sigma_2=0.8$ and $K_1=K_2=1$. We set the investment horizon to $T=1$.}

\subsection{Comparative Statics Analysis}
\begin{figure}
\centering
\begin{tabular}{cc}
\epsfig{file={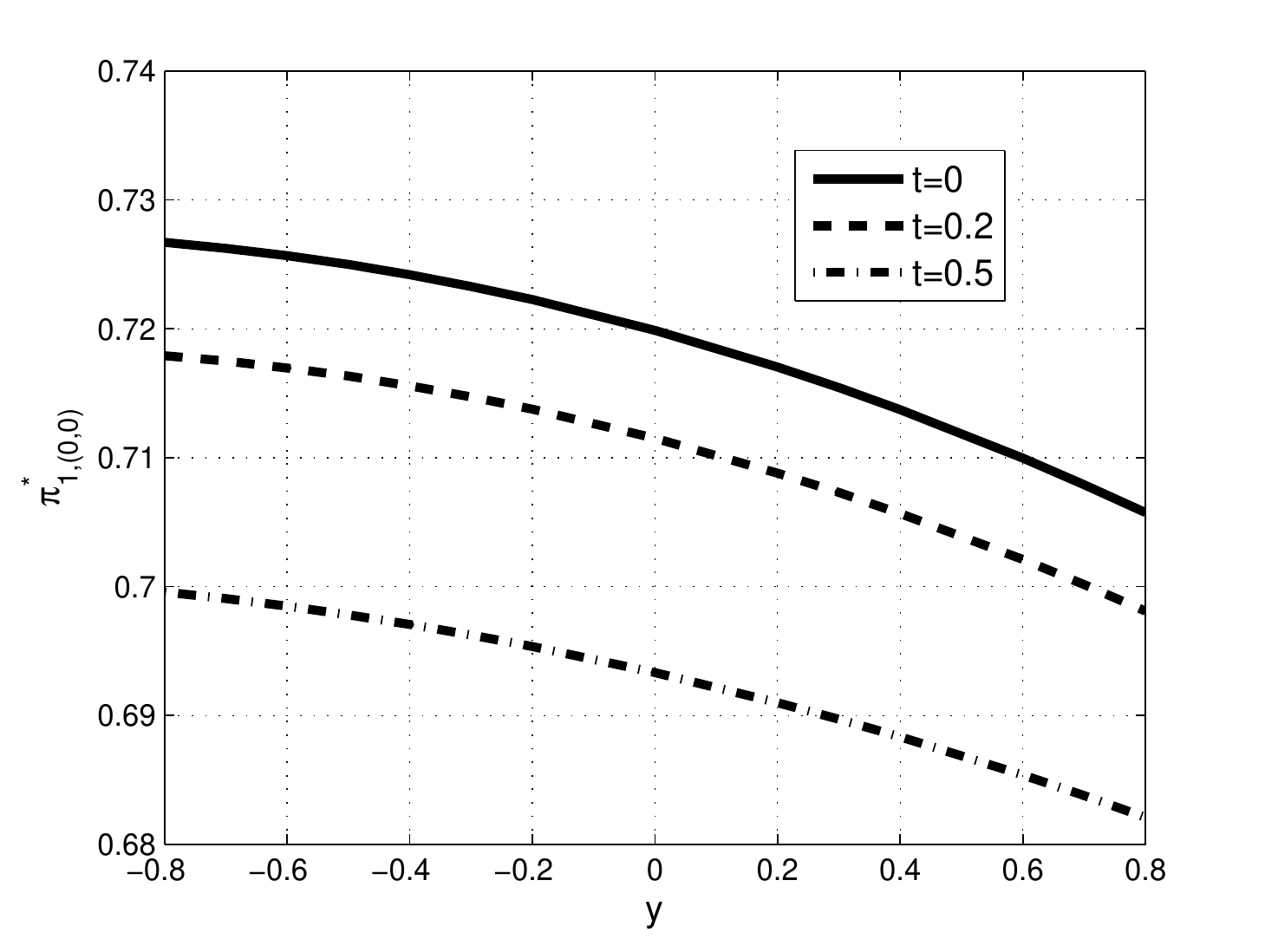},width=0.4\linewidth,clip=}
\epsfig{file={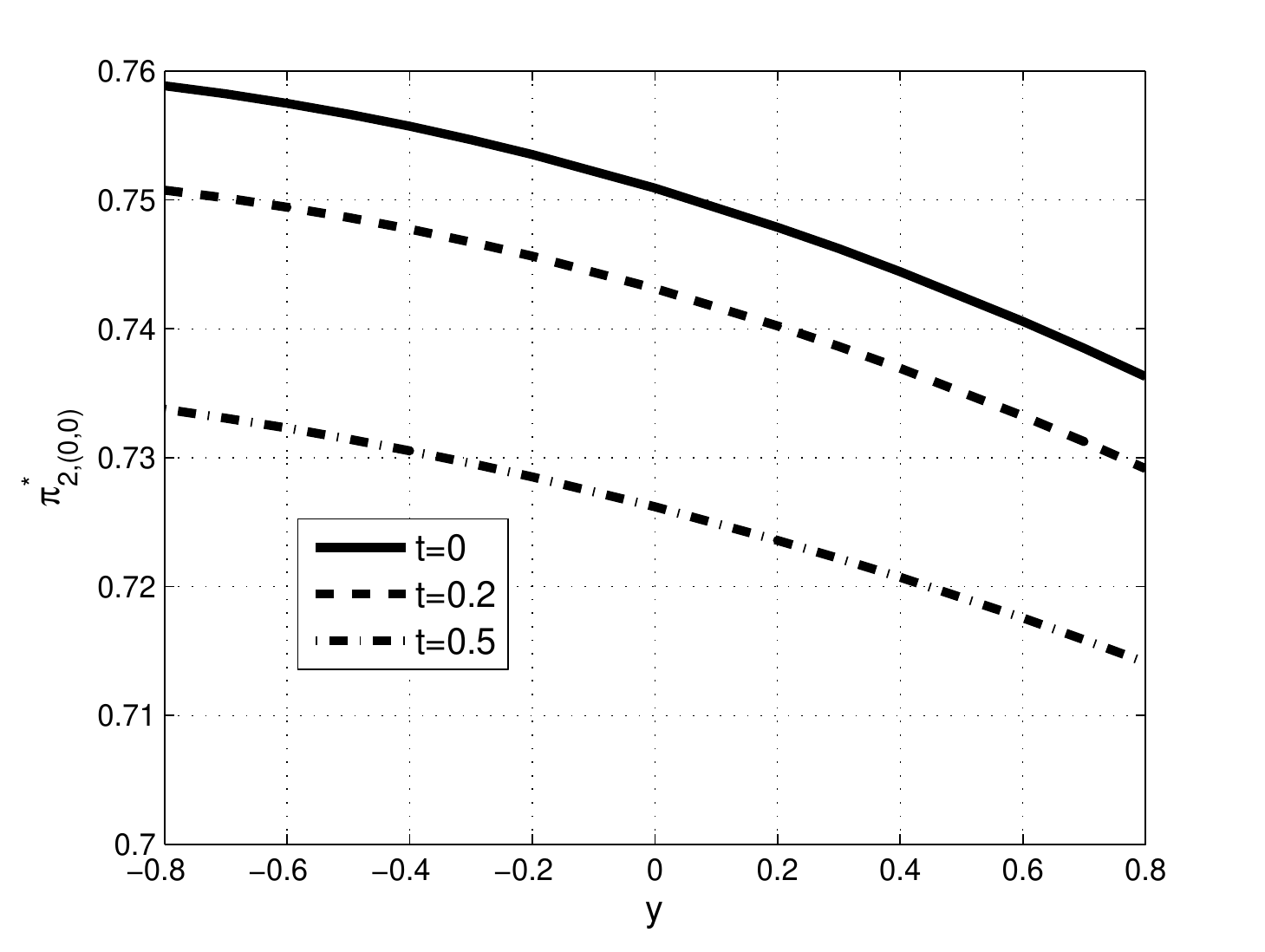},width=0.4\linewidth,clip=} \\
\epsfig{file={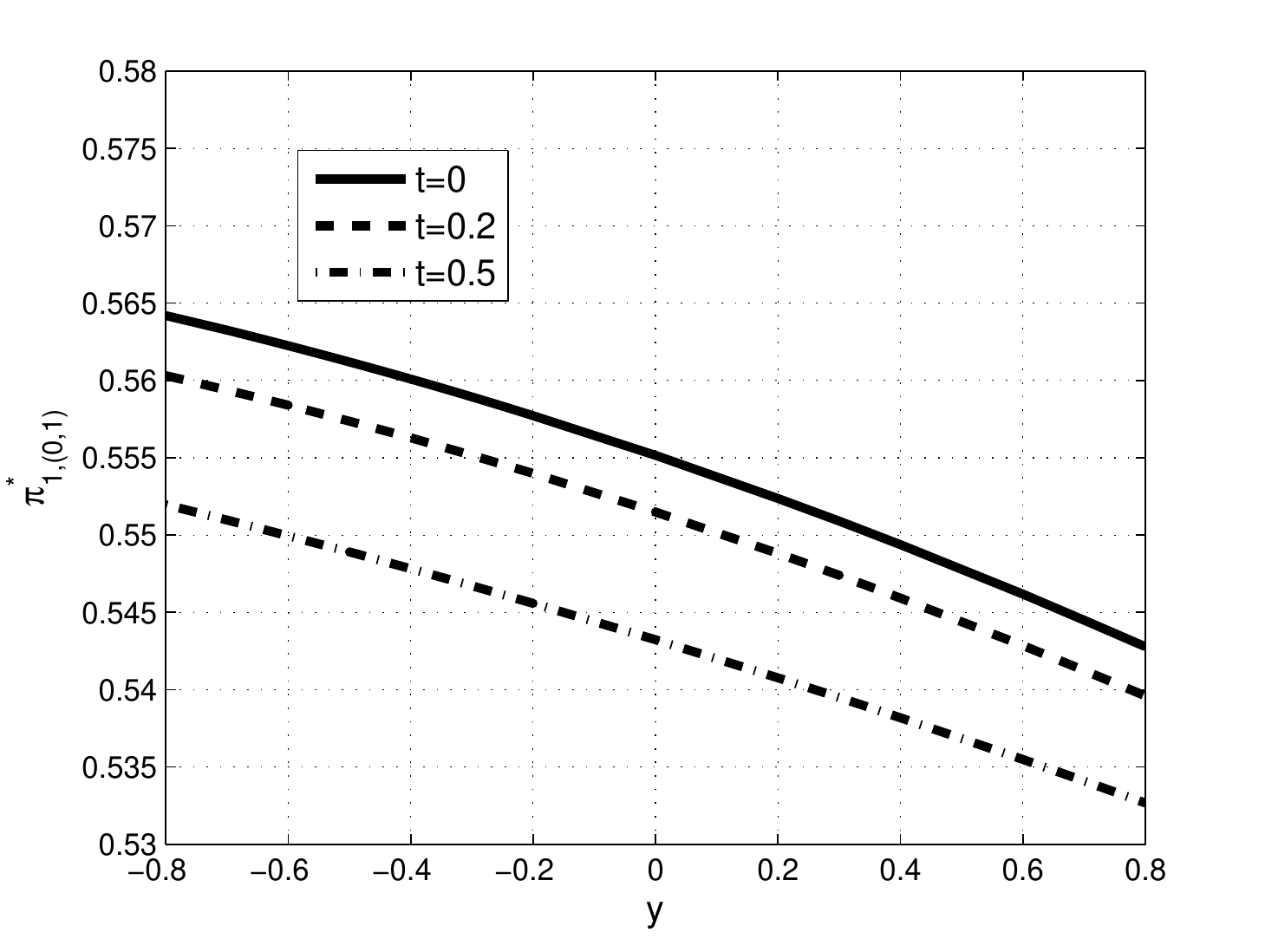},width=0.4\linewidth,clip=}
\epsfig{file={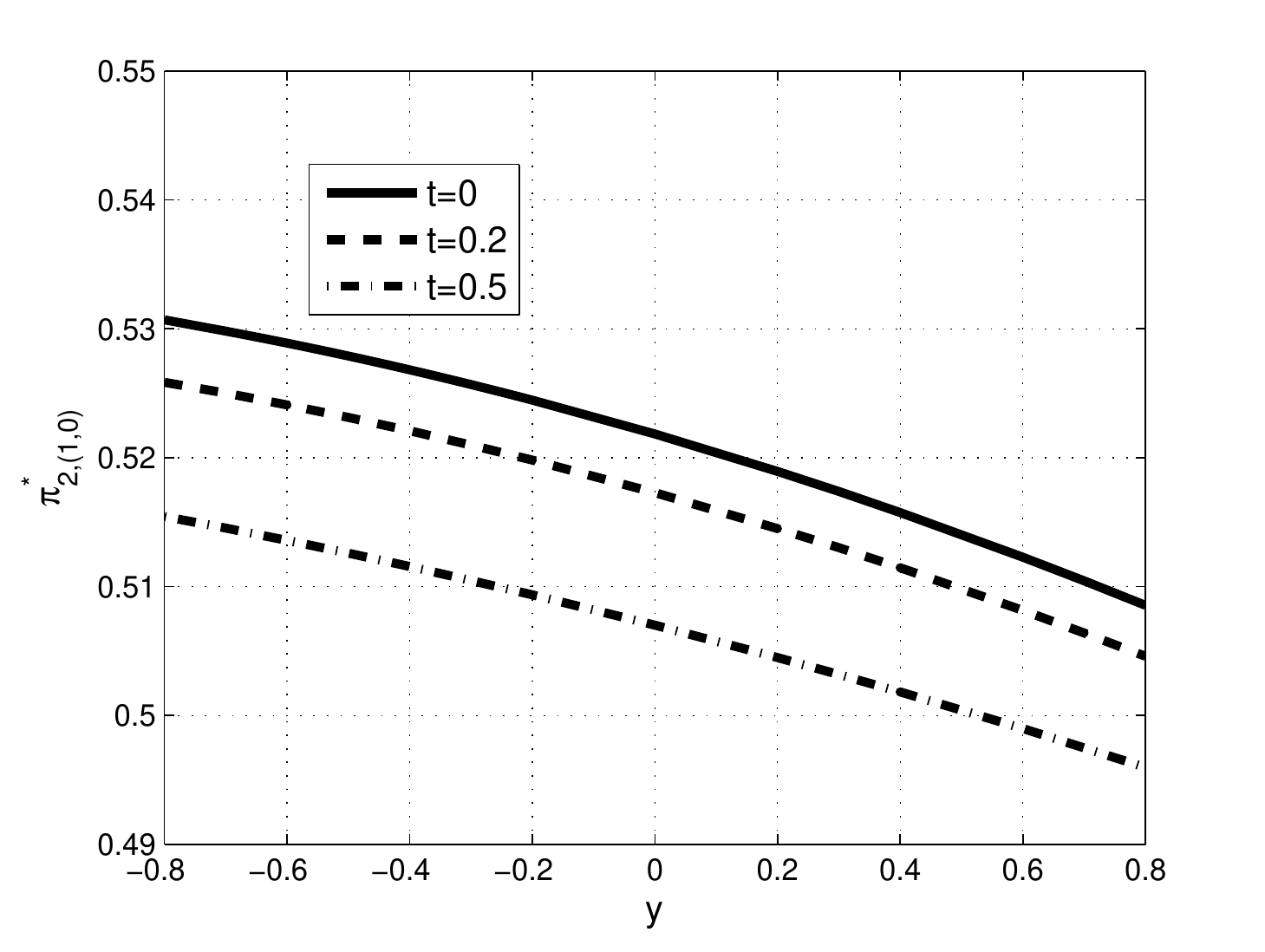},width=0.4\linewidth,clip=} \\
\end{tabular}
\caption{The optimal fractions of wealth invested in stocks versus the value $y$ of the factor. Different lines correspond to  different investment times $t$. Top left panel: investment in stock 1 when both stocks are alive. Top right panel: investment in stock 2 when both stocks are alive. Bottom left panel: investment in stock 1 when stock 2 is defaulted. Bottom right panel: investment in stock 2 when stock 1 is defaulted. The risk aversion parameter is $p=0.8$.}
\label{fig:prestrategy1}
\end{figure}

\begin{figure}
\centering
\begin{tabular}{cc}
\epsfig{file={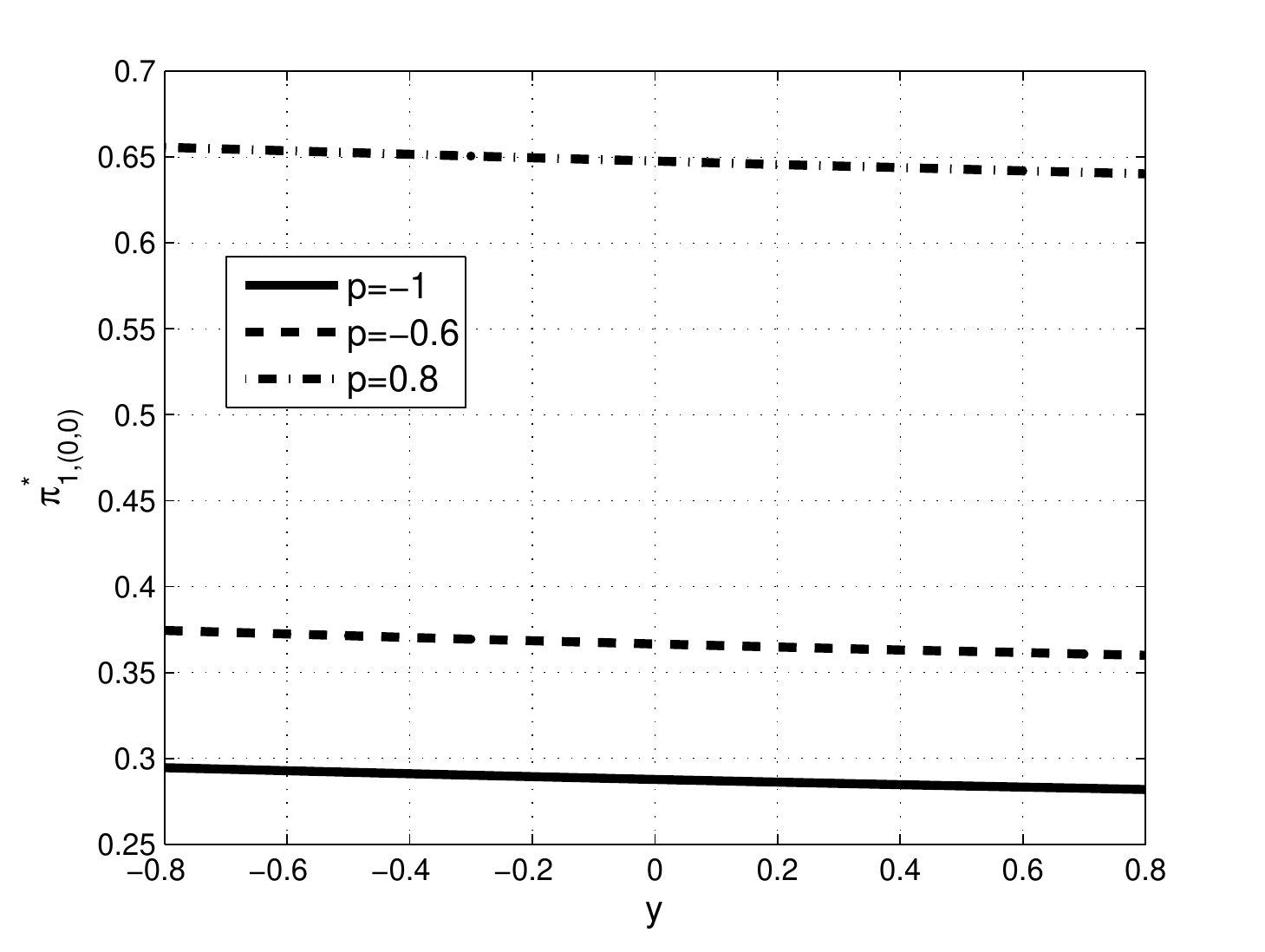},width=0.4\linewidth,clip=}
\epsfig{file={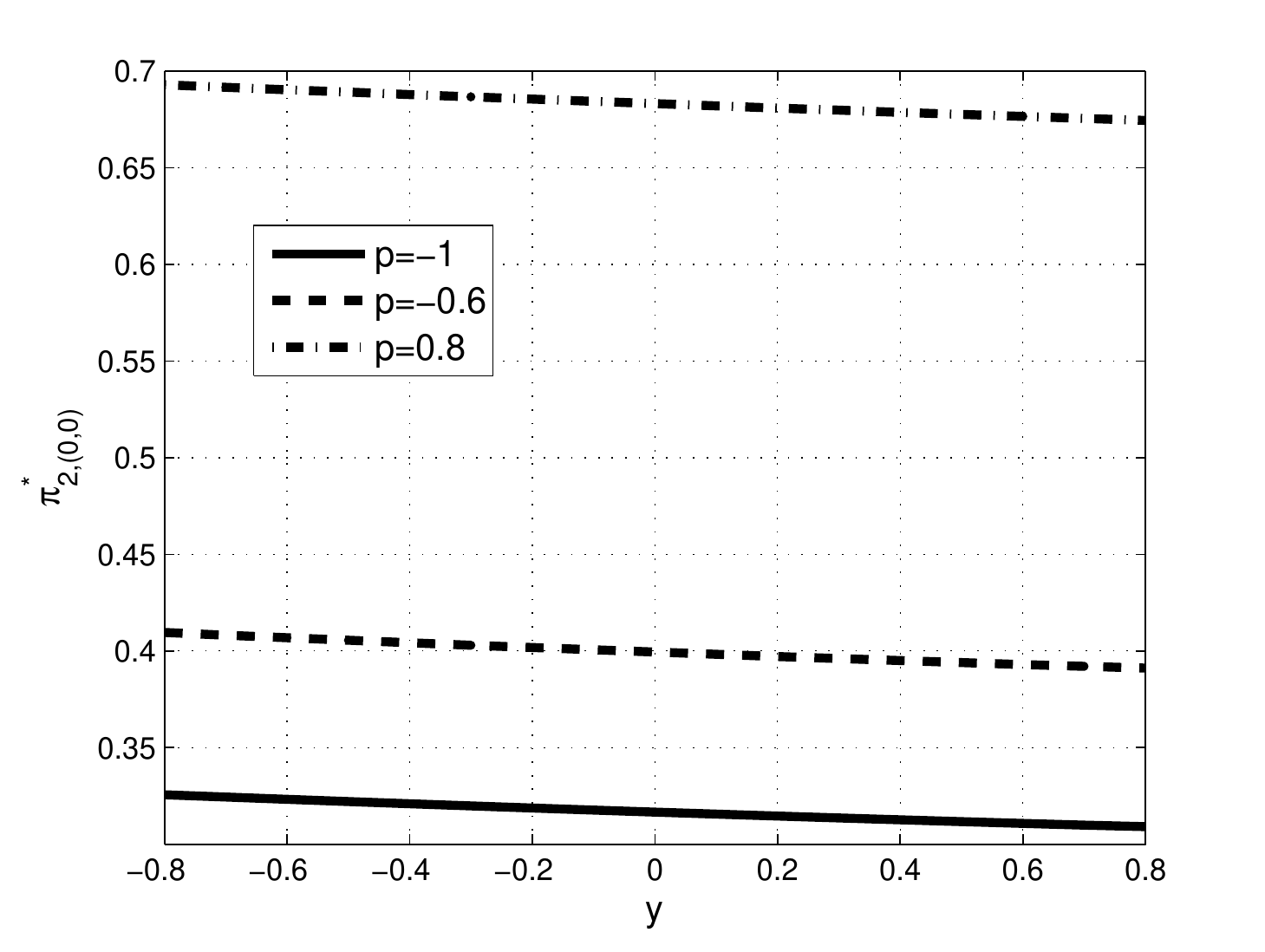},width=0.4\linewidth,clip=} \\
\epsfig{file={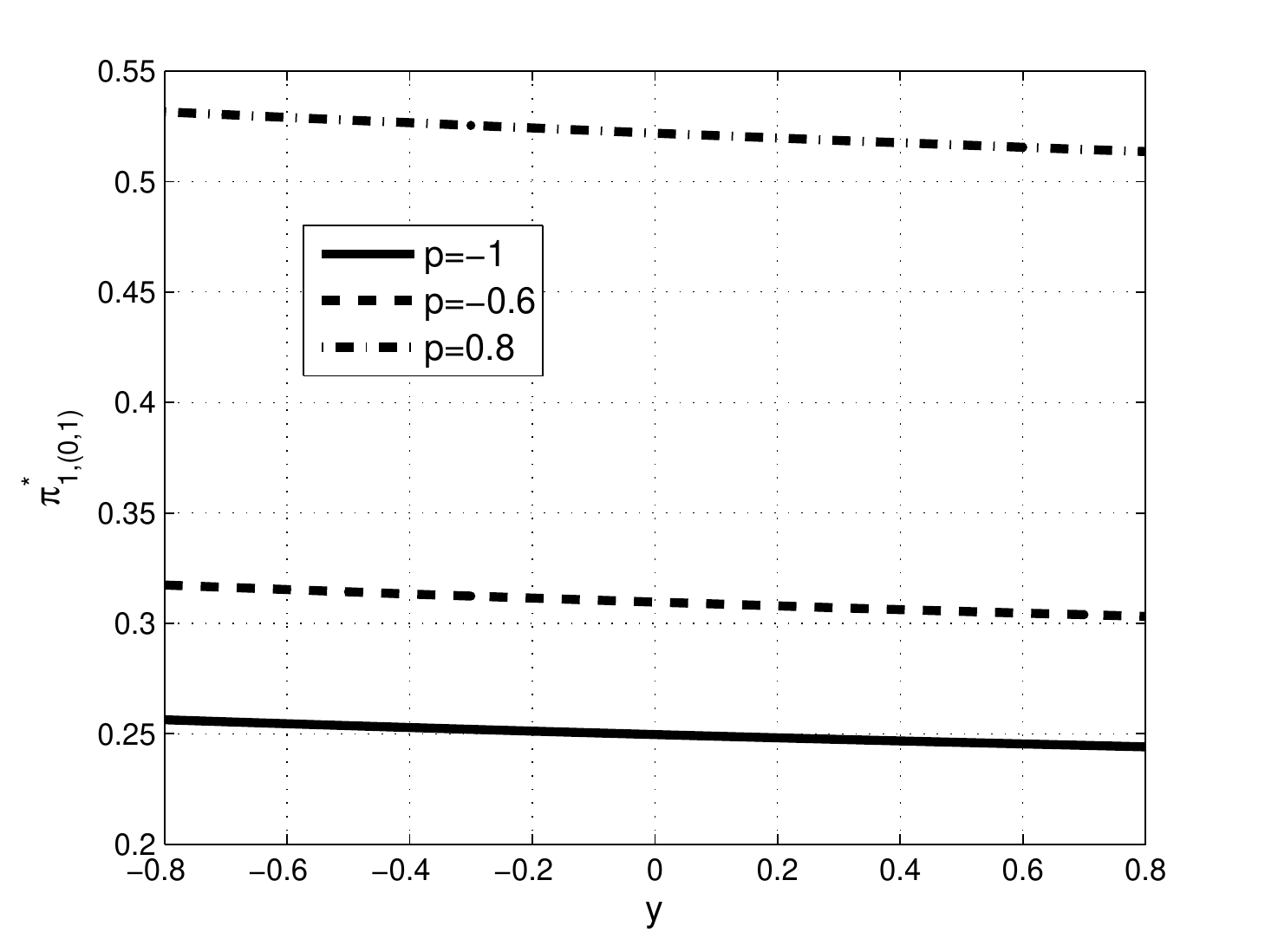},width=0.4\linewidth,clip=}
\epsfig{file={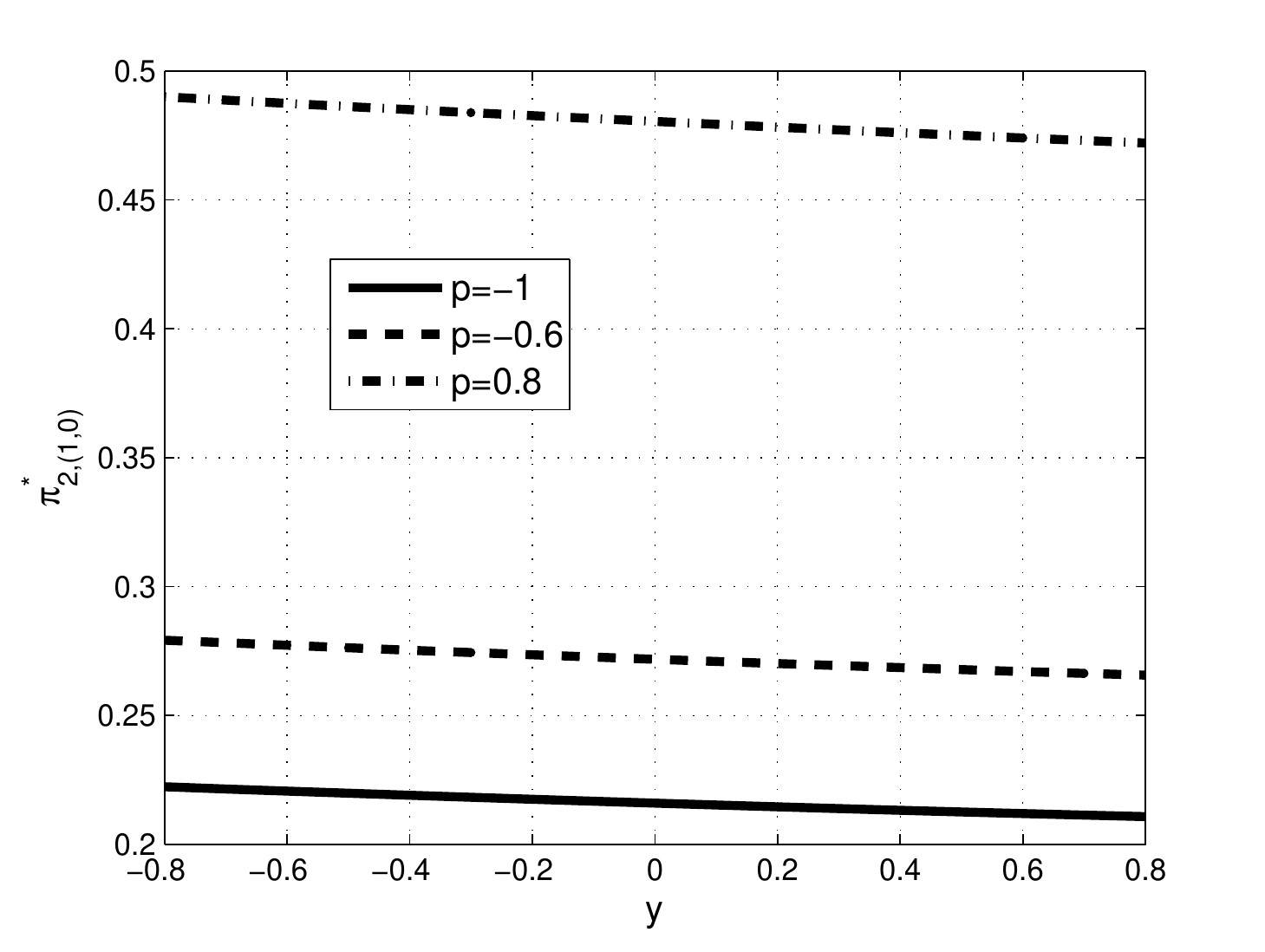},width=0.4\linewidth,clip=} \\
\end{tabular}
\caption{The optimal fractions of wealth invested in stocks versus the value $y$ of the factor. Different lines correspond to  different levels $p$ of the investor's risk aversion. Top left panel: investment in stock 1 when both are alive. Top right panel: investment in stock 2 when both are alive. Bottom left panel: investment in stock 1 when stock 2 is defaulted. Bottom right panel: investment in stock 2 when stock 1 is defaulted. The investment time is $t=0.6$.}
\label{fig:prestrategy2}
\end{figure}

\begin{figure}
\centering
\begin{tabular}{cc}
\epsfig{file={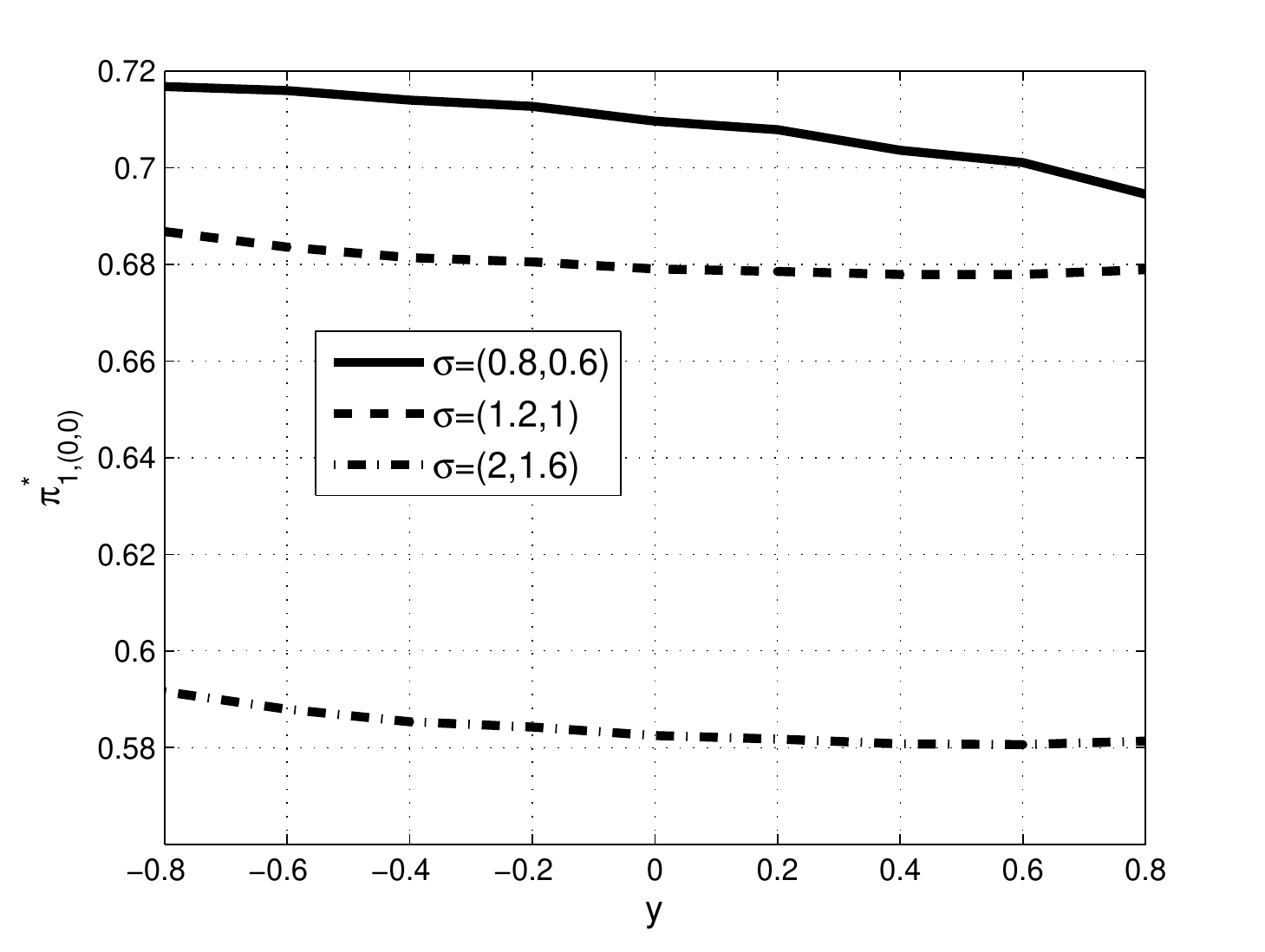},width=0.4\linewidth,clip=}
\epsfig{file={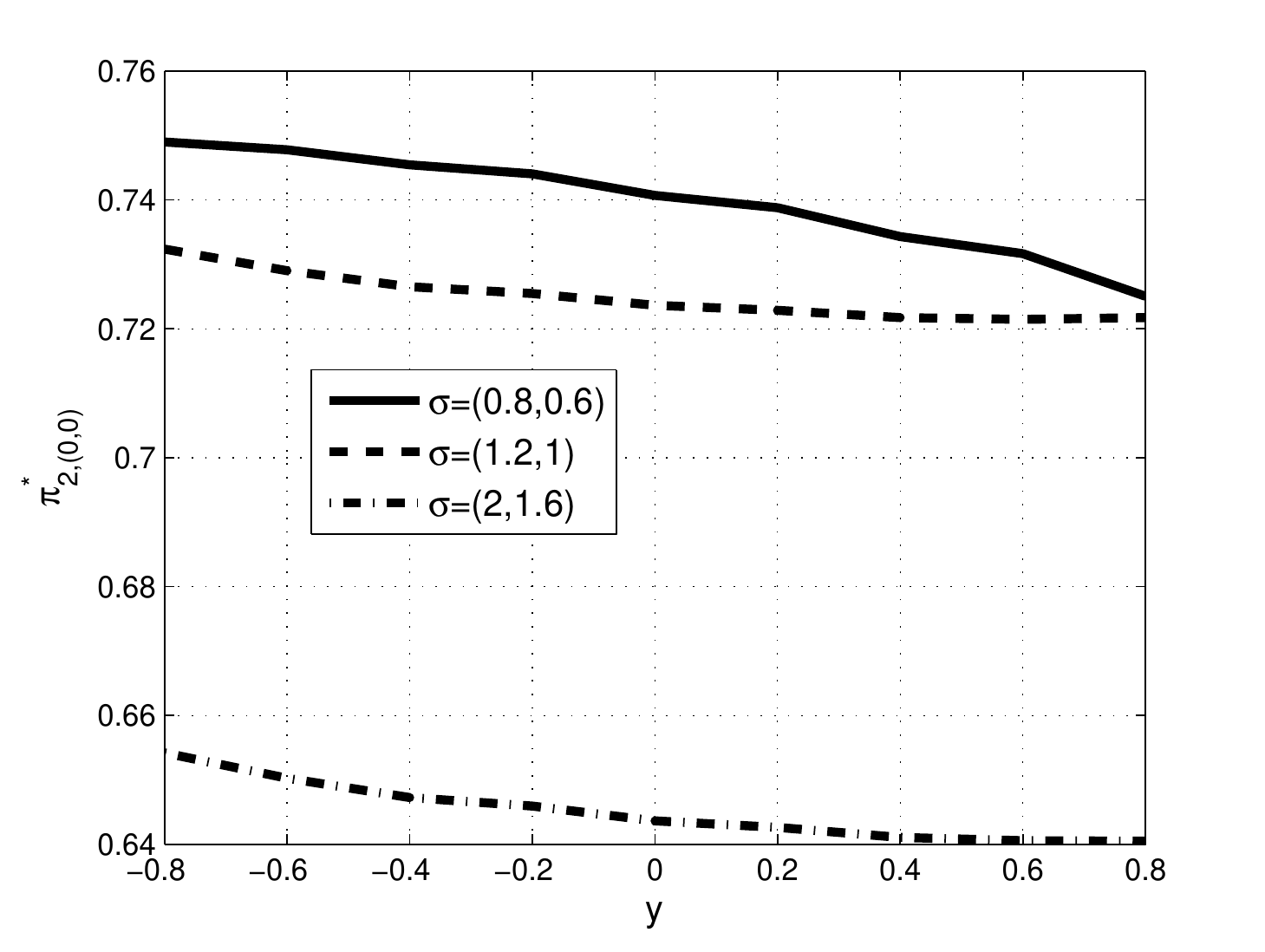},width=0.4\linewidth,clip=} \\
\epsfig{file={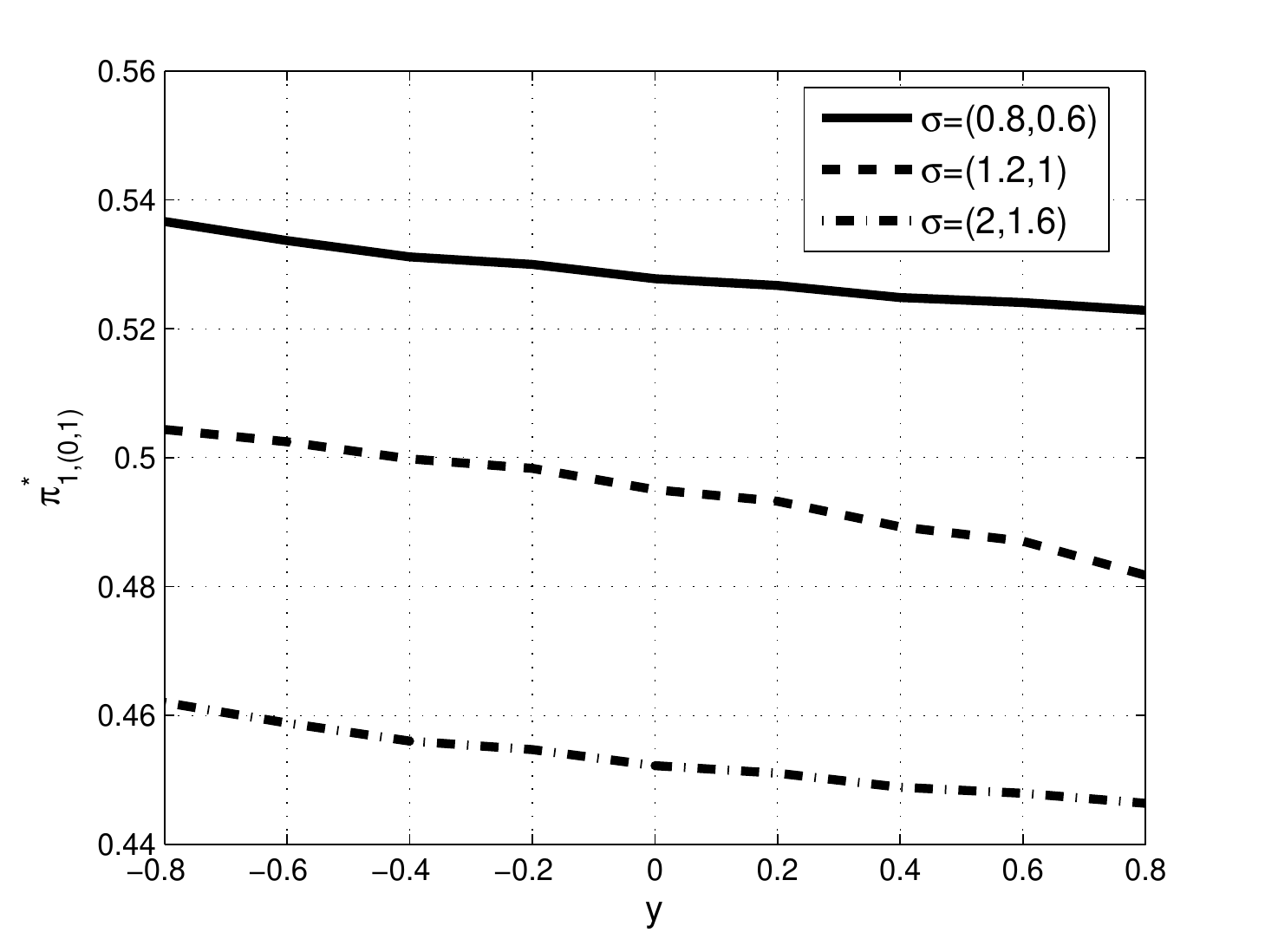},width=0.4\linewidth,clip=}
\epsfig{file={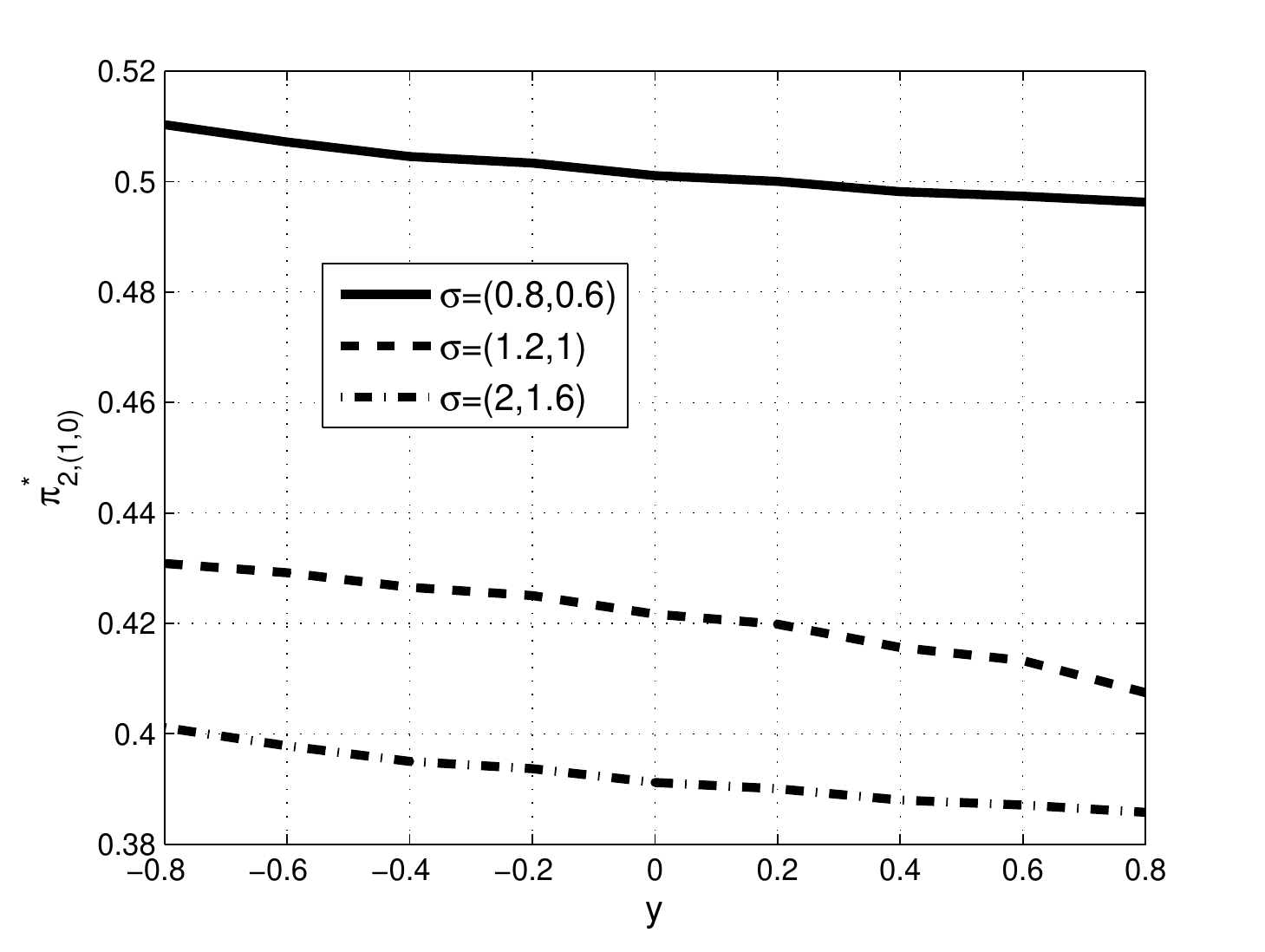},width=0.4\linewidth,clip=} \\
\end{tabular}
\caption{The optimal fractions of wealth invested in stocks versus the value $y$ of the factor. Different lines correspond to  different volatility levels of stocks' price processes. Top left panel: investment in stock 1 when both are alive. Top right panel: investment in stock 2 when both are alive. Bottom left panel: investment in stock 1 when stock 2 is defaulted. Bottom right panel: investment in stock 2 when stock 1 is defaulted. The investment time is $t=0$, and the risk aversion parameter is $p=0.1$.}
\label{fig:prestrategy3}
\end{figure}

Figure \ref{fig:prestrategy1} suggests that, as the factor $y$ increases, the investor reduces the fraction of wealth held in the risky stock. This may be explained by the following two considerations: (i) Default intensities are increasing functions of the factor $y$, and (ii) the risk-averse investor decreases his holdings in the stock as the default probability of the stock increases. A direct comparison of the top graphs in Figure  \ref{fig:prestrategy1} indicates that the investor consistently allocates a smaller fraction of wealth to stock 1, relative to stock 2, as $y$ increases. This follows directly from the fact that $\lambda_1(y,z) > \lambda_2(y,z)$ when the default state $z=(0,0)$.
After a stock defaults, the default intensity of the surviving stock jumps upward. As a result, the investor allocates a smaller fraction of wealth to it, relative to the pre-default case in which both stocks are alive. As the planning horizon $T-t$ increases, the investor is less constrained and therefore willing to take higher risk. Hence, he allocates a higher fraction of his wealth to the risky stock. Consistently with intuition, Figure \ref{fig:prestrategy2} shows that the higher the risk aversion of the investor, and the lower is the fraction of wealth held in the risky stock.
A more volatility path for the stocks' price processes leads the investor to reduce his holdings in risky stocks. This is illustrated in Figure \ref{fig:prestrategy3}, where we can clearly see that a risk averse investor reduces his stocks' exposure as the volatility increases. Being the volatility coefficient of stock ``1'' higher than the corresponding coefficient of stock ``2'', the investor always allocates a higher fraction of wealth to stock ``2'' in all default states of the portfolio; compare top graphs and bottom graphs of Figure \ref{fig:prestrategy3}.

\section{Conclusion}\label{sec:conclusions}

We have studied an optimal investment{/consumption} problem in a market consisting of securities carrying both market and credit risk. The dependence structure of these risks has been modeled via the common dependence of stock volatility and default intensity on stochastic factors. 
Because defaults occur sequentially in our model, the HJB-PDEs of the control problem (and of its dual) are recursively linked. {We have transformed the original recursive system of HJB-PDEs into an equivalent one in which the {default-state dependent} PDEs are semi-linear. The coefficients of these PDEs are, however, {nonlinear} and non-Lipschitz continuous.}
To deal with this nonstandard characteristics of the semi-linear PDEs, we have developed a truncation technique based on a stopping time argument, and then exploited the probabilistic representation of their solutions. We have proven the boundedness of the solution gradient, and used it to establish admissibility of the optimal investment strategy as well as of consumption path via a verification result. These strategies admit explicit representations in terms of the value function and of its gradient. We have complemented our theoretical analysis with a numerical study, illustrating the sensitivities of the investment strategies to the model parameters. The numerical results confirm the economic intuition: the investor reduces his holdings in the risky stocks as the volatility coefficient, default intensity, and risk aversion coefficient increase.


\appendix
\section{Technical Proofs}

\noindent{\it Proof of Proposition~\ref{prop:1}.}\quad Using the definition \eqref{eq:Pi} and the equality \eqref{eq:constranits2}, we obtain
\begin{align*}\label{eq:Pi22}
\inf_{(a,h)\in{\cal M},\kappa>0}{\it\Pi}({a},h,{\kappa})&=\inf_{(a,h)\in{\cal M},\kappa>0}\Ex\left[\tilde{U}_1\left(\kappa\frac{\Gam_T^{a,h}}{B_T}\right)+\int_0^T\tilde{U}_2\left(\kappa\frac{\Gam_s^{a,h}}{B_s}\right)ds+\kappa x\right]\nonumber\\
&\leq\Ex\left[\tilde{U}_1\left(\hat{\kappa}\frac{\Gam_T^{\hat{a},\hat{h}}}{B_T}\right)+\int_0^T\tilde{U}_2\left(\hat{\kappa}\frac{\Gam_s^{\hat{a},\hat{h}}}
{B_s}\right)ds\right]+\hat{\kappa} x\nonumber\\
&=\Ex\Bigg[{U}_1\left(I_1\left(\hat{\kappa}\frac{\Gam_T^{\hat{a},\hat{h}}}{B_T}\right)\right)-I_1\left(\hat{\kappa}\frac{\Gam_T^{\hat{a},\hat{h}}}{B_T}\right)
\hat{\kappa}\frac{\Gam_T^{\hat{a},\hat{h}}}{B_T}\nonumber\\
&\qquad+\int_0^T{U}_2\left(I_2\left(\hat{\kappa}\frac{\Gam_s^{\hat{a},\hat{h}}}{B_s}\right)\right)-I_2\left(\hat{\kappa}\frac{\Gam_s^{\hat{a},\hat{h}}}{B_s}\right)
\hat{\kappa}\frac{\Gam_s^{\hat{a},\hat{h}}}{B_s}ds\Bigg]+\hat{\kappa} x\nonumber\\
&=\Ex\left[U_1\big(X_T^{\hat{\pi},\hat{c}}\big)+\int_0^TU_2(\hat{c}_s)ds\right]-\hat{\kappa}\Ex\left[\Gam_T^{\hat{a},\hat{h}}\frac{X^{\hat{\pi},\hat{c}}_T}{B_T}
+\int_0^T\Gam_s^{\hat{a},\hat{h}}\frac{\hat{c}_s}{B_s}ds\right]
+\hat{\kappa} x\nonumber\\
&=\Ex\left[U_1\big(X_T^{\hat{\pi},\hat{c}}\big)+\int_0^TU_2(\hat{c}_s)ds\right]-\hat{\kappa}\Ex^{\hat{a},\hat{h}}\left[\frac{X_T^{\hat{\pi},\hat{c}}}{B_T}
+\int_0^T\frac{\hat{c}_s}{B_s}ds\right]+\hat{\kappa} x\nonumber\\
&=\Ex\left[U_1\big(X_T^{\hat{\pi},\hat{c}}\big)+\int_0^TU_2(\hat{c}_s)ds\right]\leq V(x,y,z).
\end{align*}
This implies that
\[
\inf_{(a,h)\in{\cal M},\kappa>0}{\it\Pi}({a},h,{\kappa})\leq{\it\Pi}(\hat{a},\hat{h},\hat{\kappa})\leq\Ex\left[U_1\big(X_T^{\hat{\pi},\hat{c}}\big)+\int_0^TU_2(\hat{c}_s)ds\right]
\leq V(x,y,z).
\]
Then, together with the inequality \eqref{eq:ineL}, we obtain
 \[
\inf_{(a,h)\in{\cal M},\kappa>0}{\it\Pi}({a},h,{\kappa})={\it\Pi}(\hat{a},\hat{h},\hat{\kappa})=\Ex\left[U_1\big(X_T^{\hat{\pi},\hat{c}}\big)+\int_0^TU_2(\hat{c}_s)ds\right]
= V(x,y,z),
\]
i.e. the equality \eqref{eq:ineL2} holds. \hfill$\Box$\\

\noindent{\it Proof of Lemma \ref{lem:simplication-value}.}\quad
Using the expression for the dual functional ${\it\Pi}(a,h,\kappa)$, $((a,h),\kappa)\in{\cal M}\times\R_+$, given in~\eqref{eq:Pi} and recalling the Legendre-Fenchel transform $\tilde{U}_i(y)= -\frac{1}{q}K_i^{1-q}y^{q}$, where $q=\frac{p}{p-1}$ and $i=1,2$, we have
\begin{align}\label{eq:Pi2}
{\it\Pi}(a,h,\kappa) &= \kappa x-\frac{1}{q}\kappa^{q}F^{a,h}(T),
\end{align}
where, for $(a,h)\in{\cal M}$, we define
\begin{align}\label{eq:Gamma}
F^{a,h}(T) &:=\Ex\left[K_1^{1-q}\left(\frac{\Gam_T^{a,h}}{B_T}\right)^q+K_2^{1-q}\int_0^T\left(\frac{\Gam_s^{a,h}}{B_s}\right)^qds\right].
\end{align}
Fix $(a,h)\in{\cal M}$. Using the first-order condition for optimality in $\kappa>0$, the value of $\kappa\in\R_+$ maximizing ${\it\Pi}(a,h,\kappa)$ is given by $\hat{\kappa}=(\frac{F^{a,h}(T)}{x})^{\frac{1}{1-q}}$. Hence, we obtain
\begin{align}\label{eq:lambda-pi}
\hat{\it\Pi}(a,h) &:= {\it\Pi}(a,h,\hat{\kappa})=\frac{1}{p}x^{p}\big(F^{a,h}(T)\big)^{1-p},\ \ \ \ \ x>0.
\end{align}

We next develop a more explicit expression for $F^{a,h}(T)$. For $t\in[0,T]$, it follows from \eqref{eq:solution-density} that
\begin{align*}\label{eq:solution-density2}
\big(\Gam_t^{a,h}\big)^q &=\exp\Bigg\{-\int_0^tq\theta_s^{\top}d{W}_s
-\frac{1}{2}\int_0^tq{\rm tr}[\theta_s\theta_s^{\top}]ds-\int_0^t q a_s^{\top}d\bar{W}_s-\frac{1}{2}\int_0^tq{\rm tr}[a_sa_s^{\top}]ds\\\notag
&\qquad\quad + \sum_{i=1}^n\int_0^tq\log\big(1+h_{s}^i\big)dM_s^{i}
+\sum_{i=1}^n\int_0^{t\wedge\tau_i}q\big[\log(1+h_{s}^i)-h_s^i\big]\lambda_i(Y_s,H_s)ds\Bigg\}\\\notag
&=\Gam_t^{q,a,h}\exp\Bigg\{\frac{q(q-1)}{2}\int_0^t{\rm tr}[\theta_s\theta_s^{\top}]ds
+\frac{q(q-1)}{2}\int_0^t{\rm tr}[a_sa_s^{\top}]ds\\\notag
&\qquad\qquad\qquad+\sum_{i=1}^n\int_0^{t\wedge\tau_i}\big[q\log(1+h_s^i)-(1+h_s^i)^q+{q-1}\big]\lambda_i(Y_s,H_s)ds\Bigg\},
\end{align*}
where $\Gam_t^{q,a}$ is given by \eqref{eq:Xbq}.  It then holds that
\begin{align*}
F^{a,h}(T) &=\Ex\left[K_1^{1-q}\left(\frac{\Gam_T^{a,h}}{B_T}\right)^q+K_2^{1-q}\int_0^T\left(\frac{\Gam_s^{a,h}}{B_s}\right)^qds\right]\\\notag
&= \Ex^{q,a,h}\left[K_1^{1-q}e^{\int_0^T\psi(s,a_s,h_s,Y_s,H_s)ds}+K_2^{1-q}\int_0^Te^{\int_0^t\psi(s,a_s,h_s,Y_s,H_s)ds}dt\right].
\end{align*}
Using \eqref{eq:Pi2}, it may be easily seen that the value function of the dual problem given in \eqref{eq:ineL2} coincides with $\esssup_{(a,h)\in{\cal M}}F^{a,h}(T)$,
which concludes the proof. \hfill$\Box$\\

\noindent{\it Proof of Lemma \ref{lem:sde-hatY}.}\quad Under assumptions ({\bf A2})-({\bf A3}), it may be easily seen that the $\R^m$-valued drift function $y\to\nu(y,z)$ is Lipschitz continuous.
Then, by Theorem V.38 in \cite{Protter}), Eq.~\eqref{eq:barY} admits a unique strong solution. 
Next, we define a probability measure $\hat{\Px}\sim\Px$ via the {Radon-Nikodym} derivative $\frac{d\hat{\Px}}{d\Px}|_{\G_t}={\cal E}\big(\int_0^{\cdot}q\rho\theta(s,\hat{Y}_s,z)^{\top}dW_s\big)_t$, $t\in[0,T]$, where ${\cal E}(\cdot)$ denotes the stochastic exponential. Notice that this change of measure is well-defined in light of the assumption ({\bf A3}). Then by Girsanov's theorem, we have $\hat{W}_t^z:=W_t-q\rho\int_0^t\theta(s,\hat{Y}_s,z)ds$, $t\in[0,T]$, is a $n$-dimensional Brownian motion under $\hat{\Px}$. Then we may rewrite Eq.~\eqref{eq:barY} under the probability measure $\hat{\Px}$:
\begin{equation}\label{eq:barY2}
d\hat{Y}_t^z = \mu_0(\hat{Y}_t^z)dt + \sigma_0(\hat{Y}_t^z)d\hat{W}_t^z,
\end{equation}
where we recall the relation between $\mu_0$ and $\nu$ given in Eq.~\eqref{eq:drifts}.
Assumptions ({\bf A1})-({\bf A3}) imply that $\hat{Y}^{t,y,z}$ does not explode to infinity under $\hat{\Px}$ for each state $z\in{\cal S}$ if $y\in D$, i.e. $\hat{\Px}(\hat{Y}_s^{t,y,z}\in D,\ \forall\ s\in[t,T])=1$. The same statement is true under $\Px$ given that $\hat{\Px}\sim\Px$. \hfill$\Box$

\begin{lemma}\label{lem:density-moment}
Let assumptions $({\bf A1})$-$({\bf A3})$ hold. Recall $(\hat{a},\hat{h})$ given by \eqref{eq:hatai} and \eqref{eq:hathi}. Let $\Gam_t^{\hat{a},\hat{h}}$, $t\in[0,T]$, be given by \eqref{eq:eta} with $(a,h)\in{\cal M}$ replaced by $(\hat{a},\hat{h})$. Then $\Ex\big[\int_0^T \big(\Gam_t^{\hat{a},\hat{h}}\big)^{q-1}dt\big]<+\infty$.
\end{lemma}

\noindent{\it Proof.}\quad We first have the decomposition given by $\big(\Gam_t^{\hat{a},\hat{h}}\big)^{q-1}=\Gam_t^{q-1,\hat{a},\hat{h}}Z_t^{q,\hat{a},\hat{h}}$,
where $\Gam_t^{q-1,\hat{a},\hat{h}}$, $t\in[0,T]$, is the density process defined by \eqref{eq:Xbq} with $(q,a,h)$ replaced with $(q-1,\hat{a},\hat{h})$, and  the positive process $Z_t^{q,\hat{a},\hat{h}}$, $t\in[0,T]$, is defined as
\begin{align*}
Z_t^{q,\hat{a},\hat{h}} &:=\exp\Bigg\{\frac{(q-1)(q-2)}{2}\int_0^t{\rm tr}[(\theta_s\theta_s^{\top})]ds
+\frac{(q-1)(q-2)}{2}\int_0^t{\rm tr}[\hat{a}_s\hat{a}_s^{\top}]ds\\\notag
&\qquad\quad+\sum_{i=1}^n\int_0^{t\wedge\tau_i}\big[(1+\hat{h}_{s}^i)^{q-1}-(q-1)(1+\hat{h}_s^i)+{q-2}\big]\lambda_i(Y_s,H_s)ds\Bigg\}.
\end{align*}
Using the conclusion {(I)} of the verification result in Proposition~\ref{prop:verificationp<0} along with the estimates \eqref{eq:bounda}, we obtain that $\Gam_t^{q-1,\hat{a}}$, $t\in[0,T]$, is a $(\Px,\Gx)$-martingale. Moreover, under assumptions $({\bf A2})$-$({\bf A3})$, and the estimates \eqref{eq:bounda}, $Z_t^{q,\hat{a},\hat{h}}$, $t\in[0,T]$, is a bounded positive process. Then, for some $C>0$,
\[
\Ex\left[\int_0^T \big(\Gam_t^{\hat{a},\hat{h}}\big)^{q-1}dt\right]\leq C\int_0^T\Ex\big[\Gam_t^{q-1,\hat{a},\hat{h}}\big]dt=CT<+\infty.
\]
This completes the proof of the lemma. \hfill$\Box$\\

\noindent{\it Proof of Lemma~\ref{lem:G-mart}.}
\quad Using Eq.~\eqref{eq:eta} and applying It\^o's formula, we obtain
the $\Px$-dynamics of the discounted process $\frac{\Gam_t^{\hat{a},\hat{h}}}{B_t}$ given by
\begin{eqnarray*}
\D\left(\frac{\Gam_t^{\hat{a},\hat{h}}}{B_t}\right)=-\left(\frac{\Gam_{t-}^{\hat{a},\hat{h}}}{B_{t-}}\right)\big[rdt+\theta_t^{\top}dW_t
+\hat{a}_t^{\top}d\bar{W}_t-\hat{h}_{t}^{\top}dM_t\big],\ \ \ \ \frac{\Gam_{0}^{\hat{a},\hat{h}}}{B_{0}}=1.
\end{eqnarray*}
Recall the $\Px$-dynamics of stochastic factor process $Y$ given by \eqref{eq:Yt}. Define the operator ${\cal L}$ acting on $l(t,x,y,z)\in C^{1,2,2}([0,T]\times\R_+\times D)$
for each state $z\in{\cal S}$, by
\begin{align}\label{eq:Lab}
{\cal L} l(t,x,y,z) &:=\frac{\partial l}{\partial t} - x\frac{\partial l}{\partial x}\big[r+\hat{h}(t,y,z)^{\top}\lambda(y,z)\big]+\frac{1}{2}x^2\frac{\partial^2 l}{\partial x^2}\big\{{\rm tr}[(\theta\theta^{\top})(t,y,z)]+{\rm tr}[(\hat{a}_t\hat{a}_t^{\top})]\big\}\nonumber\\
&\quad+\hat{\cal A}_t^{\mu_0}l-x\sum_{i=1}^m\frac{\partial^2 l}{\partial y_i\partial x}\sum_{j=1}^n\big[\rho\theta_{j}(t,y,z)+\sqrt{1-\rho^2}\hat{a}_t^{j}\big]\sigma_0^{ij}(y)\nonumber\\
&\quad+\sum_{i=1}^n\big[l\big(t,x(1+\hat{h}^i(t,y,z)),y,\bar{z}^i\big)- l(t,x,y,z)\big](1-z_i)\lambda_i(y,z).
\end{align}
Above, the operator $\hat{\cal A}_t^{\mu_0}$ is defined by \eqref{eq:hatAeta} with $\eta$ replaced by $\mu_0$ therein. Further, recall the vector of functions $\eta$ given by~\eqref{eq:drift-q} and the operator $\hat{\cal A}_t^{\eta}$ defined in \eqref{eq:hatAeta}. Choosing $l(t,x,y,z)=x^qg(T-t,y,z)$, we obtain
\begin{align}\label{eq:hatL=-xq}
&{\cal L} l(t,x,y,z) =x^q\left\{-\frac{\partial g}{\partial t}+\hat{\cal A}_t^{\eta}g+\left(\frac{q(q-1)}{2}{\rm tr}[\theta\theta^{\top}(t,y,z)]+\frac{q(q-1)}{2}{\rm tr}[\hat{a}_t\hat{a}_t^{\top}]-q r\right)g\right\}\nonumber\\
&\quad +x^q\left\{-g\sum_{i=1}^n(q\hat{h}_i(t,y,z)+1)(1-z_i)\lambda_i(y,z)+\sum_{i=1}^n g\big(T-t,y,\bar{z}^i\big)(1-z_i)(1+\hat{h}_i(t,y,z))^q\lambda_i(y,z)\right\}\nonumber\\
&\quad=-K_2^{1-q}x^q,
\end{align}
where we are using the abbreviation $g=g(T-t,y,z)$ above. Using Dynkin's formula, it follows that
\begin{eqnarray*}
{G}_t=  l\left(t,\frac{\Gam_t^{\hat{a},\hat{h}}}{B_t},Y_t,H_t\right) - \int_0^t {\cal L}l\left(s,\frac{\Gam_s^{\hat{a},\hat{h}}}{B_s},Y_s,H_s\right)ds, \ \ \ t\in[0,T]
\end{eqnarray*}
is a $(\Px,\Gx)$-local martingale. Since by {(II)} of Proposition~\ref{prop:verificationp<0}, $\Ex[{G}_T]=\Ex[{G}_0]$, we obtain that ${G}$ is indeed a $(\Px,\Gx)$-martingale. \hfill$\Box$\\

\noindent{\it Proof of Lemma~\ref{lem:P-dyn-hatXB}.}
\quad Recall the $\Px$-dynamics of the density process $\Gam_t^{\hat{a},\hat{h}}$, $t\in[0,T]$, given in~\eqref{eq:Gameq} by
\begin{eqnarray*}
\frac{d\Gam_t^{\hat{a},\hat{h}}}{\Gam_{t-}^{\hat{a},\hat{h}}}=-\sum_{i=1}^n\hat{h}_t^i(1-H^i_t)\lambda_i(Y_t,H_t)dt
-\theta_t^{\top}dW_t-\hat{a}_t^{\top}d\bar{W}_t+\hat{h}_{t}^{\top}dH_t.
\end{eqnarray*}
Application of It\^o's formula gives
\begin{align*}
d(\Gam_t^{\hat{a},\hat{h}})^{q-1} &= (q-1)(\Gam_t^{\hat{a},\hat{h}})^{q-1}\left[-\sum_{i=1}^n\hat{h}_t^i(1-H^i_t)\lambda_i(Y_t,H_t)dt
-\theta_t^{\top}dW_t-\hat{a}_t^{\top}d\bar{W}_t\right]\nonumber\\
&\quad+\frac{(q-1)(q-2)}{2}(\Gam_t^{\hat{a},\hat{h}})^{q-1}\big\{{\rm tr}[(\theta_t\theta_t^{\top})]+{\rm tr}[\hat{a}_t\hat{a}_t^{\top}]\big\}dt\nonumber\\
&\quad+(\Gam_t^{\hat{a},\hat{h}})^{q-1}\sum_{i=1}^n\big[\big(1+\hat{h}_{t}^i\big)^{q-1}-1\big]dH_t^i.
\end{align*}
It can be easily seen that $dB_t^{-q}=-qrB_t^{-q}dt$, where $B_t$ denotes the bank account. Then integration by parts gives

\begin{align*}
& d\left(\frac{(\Gam_t^{\hat{a},\hat{h}})^{q-1}}{B_t^q}\right)\nonumber\\
&\quad= \frac{(\Gam_t^{\hat{a},\hat{h}})^{q-1}}{B_t^q}\left[-qr+(1-q)\sum_{i=1}^n\hat{h}_t^i(1-H^i_t)\lambda_i(Y_t,H_t)
+\frac{(q-1)(q-2)}{2}\big\{{\rm tr}[(\theta_t\theta_t^{\top})]+{\rm tr}[\hat{a}_t\hat{a}_t^{\top}]\big\}\right]dt\nonumber\\
&\qquad+\frac{(\Gam_t^{\hat{a},\hat{h}})^{q-1}}{B_t^q}(1-q)\big[\theta_t^{\top}dW_t+\hat{a}_t^{\top}d\bar{W}_t\big]
+\frac{(\Gam_{t-}^{\hat{a},\hat{h}})^{q-1}}{B_{t-}^q}\sum_{i=1}^n\big[\big(1+\hat{h}_{t}^i\big)^{q-1}-1\big]dH_t^i.
\end{align*}
Moreover, application of It\^o's formula yields that
\begin{align*}
&dg(T-t,Y_t,H_t) = \left[-\frac{\partial g(T-t,Y_t,H_t)}{\partial t} + \hat{\cal A}_t^{\mu_0}g(T-t,Y_t,H_t)\right]dt
\nonumber\\
&\quad+D_yg(T-t,Y_t,H_t)^{\top}\sigma_0(Y_t)[\rho dW_t+\sqrt{1-\rho^2}d\bar{W}_t]
+\sum_{i=1}^n\left[g(T-t,Y_t,\bar{H}_{t-}^i)-g(T-t,Y_t,H_{t-})\right]dH_t^i.
\end{align*}
Using the expression for $\frac{\hat{X}_t}{B_t}$ given in Eq.~\eqref{eq:hatXB} and applying integration by parts, we obtain
\begin{eqnarray}\label{eq:ito-product}
&& \frac{g(T,y,z)}{x}{d \left(\frac{\hat{X}_t}{B_t} \right) }= d\left(\frac{(\Gam_t^{\hat{a},\hat{h}})^{q-1}}{B_t^q}g(T-t,Y_t,H_t)\right)  \nonumber\\
&&\qquad
=\frac{(\Gam_t^{\hat{a},\hat{h}})^{q-1}}{B_t^q}\left[-\frac{\partial g(T-t,Y_t,H_t)}{\partial t} + \hat{\cal A}_t^{\mu_0}g(T-t,Y_t,H_t)\right]dt\nonumber\\
&&\qquad+\frac{(\Gam_t^{\hat{a},\hat{h}})^{q-1}}{B_t^q}g(T-t,Y_t,H_t)\Bigg[-qr+(1-q)\sum_{i=1}^n\hat{h}_t^i(1-H^i_t)\lambda_i(Y_t,H_t)\nonumber\\
&&\qquad\qquad\qquad+\frac{(q-1)(q-2)}{2}\big\{{\rm tr}[(\theta_t\theta_t^{\top})]+{\rm tr}[\hat{a}_t\hat{a}_t^{\top}]\big\}\Bigg]dt\nonumber\\
&&\qquad+\frac{(\Gam_t^{\hat{a},\hat{h}})^{q-1}}{B_t^q}(1-q)\big<\rho\theta_t+\sqrt{1-\rho^2}\hat{a}_t,D_yg(T-t,Y_t,H_t)^{\top}\sigma_0(Y_t)\big>dt\nonumber\\
&&\qquad+\frac{(\Gam_t^{\hat{a},\hat{h}})^{q-1}}{B_t^q}D_yg(T-t,Y_t,H_t)^{\top}\sigma_0(Y_t)[\rho dW_t+\sqrt{1-\rho^2}d\bar{W}_t]\nonumber\\
&&\qquad+\frac{(\Gam_t^{\hat{a},\hat{h}})^{q-1}}{B_t^q}g(T-t,Y_t,H_t)(1-q)\big[\theta_t^{\top}dW_t+\hat{a}_t^{\top}d\bar{W}_t\big]\nonumber\\
&&\qquad+\frac{(\Gam_t^{\hat{a},\hat{h}})^{q-1}}{B_t^q}\sum_{i=1}^n\big[\big(1+\hat{h}_{t}^i\big)^{q-1}g(T-t,Y_t,\bar{H}_{t-}^i)-g(T-t,Y_t,H_{t-})\big]dH_t^i.
\end{eqnarray}
Recall the expression for $(\hat{a},\hat{h})\in{\cal M}$ given in~\eqref{eq:hatai} and \eqref{eq:hathi}, and proven to be the optimal controls in the verification proposition~\ref{prop:verificationp<0}. Then, it holds that, $\Px$-a.s.
\begin{align}\label{eq:eq00}
(1-q)g(T-t,Y_t,H_t)\hat{a}_t^{\top}+\sqrt{1-\rho^2}D_yg(T-t,Y_t,H_t)^{\top}\sigma_0(Y_t)=0.
\end{align}
Plugging the optimum $(\hat{a},\hat{h})\in{\cal M}$ given in~\eqref{eq:hatai} and \eqref{eq:hathi} to have that
\begin{align}\label{eq:ito-product2}
&\frac{g(T,y,z)}{x}{d \left(\frac{\hat{X}_t}{B_t} \right) }
=\frac{(\Gam_t^{\hat{a},\hat{h}})^{q-1}}{B_t^q}\left[-\frac{\partial g(T-t,Y_t,H_t)}{\partial t} + \hat{\cal A}_t^{\mu_0}g(T-t,Y_t,H_t)\right]dt\nonumber\\
&\qquad+\frac{(\Gam_t^{\hat{a},\hat{h}})^{q-1}}{B_t^q}g(T-t,Y_t,H_t)\left(-qr+(1-q)\sum_{i=1}^n\hat{h}_t^i(1-H^i_t)\lambda_i(Y_t,H_t)+\frac{q(q-1)}{2}{\rm tr}[(\theta_t\theta_t^{\top})]\right)dt\nonumber\\
&\qquad+\frac{(\Gam_t^{\hat{a},\hat{h}})^{q-1}}{B_t^q}\frac{q(1-\rho^2)}{2(1-q)}g^{-1}(T-t,Y_t,H_t)\left\|\sigma_0^{\top}(Y_t)D_yg(T-t,Y_t,H_t)\right\|^2\nonumber\\
&\qquad+\frac{(\Gam_t^{\hat{a},\hat{h}})^{q-1}}{B_t^q}D_yg(T-t,Y_t,H_t)^{\top}\sigma_0(Y_t)\rho dW_t^{\theta}+\frac{(\Gam_t^{\hat{a},\hat{h}})^{q-1}}{B_t^q}\big(\nu(t,Y_t,H_t)-\mu_0(Y_t)\big)\nonumber\\
&\qquad+\frac{(\Gam_t^{\hat{a},\hat{h}})^{q-1}}{B_t^q}g(T-t,Y_t,H_t)(1-q)\theta_t^{\top}dW_t+\frac{(\Gam_t^{\hat{a},\hat{h}})^{q-1}}{B_t^q}g(T-t,Y_t,H_t)(1-q){\rm tr}[(\theta_t\theta_t^{\top})]\nonumber\\
&\qquad+\frac{(\Gam_t^{\hat{a},\hat{h}})^{q-1}}{B_t^q}g(T-t,Y_t,H_{t-})\sum_{i=1}^n\left[\big(1+\hat{h}_{t}^i\big)^{q-1}\frac{g(T-t,Y_t,\bar{H}_{t-}^i)}{g(T-t,Y_t,H_{t-})}-1\right]dM_t^{\hat{h},i}\\
&\qquad+\frac{(\Gam_t^{\hat{a},\hat{h}})^{q-1}}{B_t^q}g(T-t,Y_t,H_{t-})\sum_{i=1}^n\left[\big(1+\hat{h}_{t}^i\big)^{q}\frac{g(T-t,Y_t,\bar{H}_{t}^i)}{g(T-t,Y_t,H_{t})}
-(1+\hat{h}_t^i)\right](1-H_t^i)\lambda_i(Y_t,H_t)dt.\nonumber
\end{align}
In the above expression, the function $\nu(t,y,z)$ is defined in \eqref{eq:drifts}. Using the equality 
\[
(1-q)\sum_{i=1}^n\hat{h}_t^i(1-H^i_t)\lambda_i(Y_t,H_t)-(1+\hat{h}_t^i)(1-H_t^i)\lambda_i(Y_t,H_t)=\sum_{i=1}^n[q-1-q(1+\hat{h}_t^i)](1-H_t^i)\lambda_i(Y_t,H_t),
\]
and noticing that $g(t,y,z)$ satisfies the HJB equation~\eqref{eq:hjn2p<0}, the dynamics \eqref{eq:P-dyn-hatXB} follows from \eqref{eq:ito-product2}. 
%


\end{document}